\documentclass{article}

\usepackage{arxiv}

\usepackage[utf8]{inputenc} 
\usepackage[T1]{fontenc}    
\usepackage{lmodern}        
\usepackage{hyperref}       
\usepackage{url}            
\usepackage{booktabs}       
\usepackage{amsfonts}       
\usepackage{nicefrac}       
\usepackage{microtype}      
\usepackage{graphicx}

\title{Power Under Multiplicity Project (PUMP): Estimating Power,
Minimum Detectable Effect Size, and Sample Size When Adjusting for
Multiple Outcomes in Multi-level Experiments}

\author{
    Kristen Hunter \orcidlink{0000-0002-5678-4620}
   \\
    University of New South Wales \\
   \\
  \texttt{\href{mailto:kristen.hunter@unsw.edu.au}{\nolinkurl{kristen.hunter@unsw.edu.au}}} \\
   \And
    Luke Miratrix \orcidlink{0000-0002-0078-1906}
   \\
    Harvard Graduate School of Education \\
   \\
  \texttt{\href{mailto:lmiratrix@g.harvard.edu}{\nolinkurl{lmiratrix@g.harvard.edu}}} \\
   \And
    Kristin Porter
   \\
    K.E. Porter Consulting LLC \\
   \\
  \texttt{\href{mailto:kristin.porter@keporterconsulting.com}{\nolinkurl{kristin.porter@keporterconsulting.com}}} \\
  }

\usepackage{color}
\usepackage{fancyvrb}

\DefineVerbatimEnvironment{Highlighting}{Verbatim}{commandchars=\\\{\}}
\usepackage{framed}
\definecolor{shadecolor}{RGB}{248,248,248}
\newenvironment{Shaded}{\begin{snugshade}}{\end{snugshade}}

\newcommand{\AttributeTok}[1]{\textcolor[rgb]{0.77,0.63,0.00}{#1}}

\newcommand{\CommentTok}[1]{\textcolor[rgb]{0.56,0.35,0.01}{\textit{#1}}}

\newcommand{\ConstantTok}[1]{\textcolor[rgb]{0.00,0.00,0.00}{#1}}

\newcommand{\DecValTok}[1]{\textcolor[rgb]{0.00,0.00,0.81}{#1}}

\newcommand{\FloatTok}[1]{\textcolor[rgb]{0.00,0.00,0.81}{#1}}
\newcommand{\FunctionTok}[1]{\textcolor[rgb]{0.00,0.00,0.00}{#1}}

\newcommand{\NormalTok}[1]{#1}

\newcommand{\OtherTok}[1]{\textcolor[rgb]{0.56,0.35,0.01}{#1}}

\newcommand{\SpecialCharTok}[1]{\textcolor[rgb]{0.00,0.00,0.00}{#1}}

\newcommand{\StringTok}[1]{\textcolor[rgb]{0.31,0.60,0.02}{#1}}

\providecommand{\tightlist}{%
  \setlength{\itemsep}{0pt}\setlength{\parskip}{0pt}}

\newlength{\cslhangindent}
\newlength{\csllabelwidth}
\newlength{\cslentryspacingunit} 
  {}%
  {\par}
\newenvironment{CSLReferences}[2] 
 {
  \setlength{\parindent}{0pt}
  \ifodd #1
  \let\oldpar\par
  \def\par{\hangindent=\cslhangindent\oldpar}
  \fi
  \setlength{\parskip}{#2\cslentryspacingunit}
 }%
 {}
\usepackage{calc}

\usepackage{amsmath}
\usepackage{minitoc}
\usepackage{bm}

\usepackage{orcidlink}
\usepackage{tabularx}
\usepackage{booktabs}
\usepackage{longtable}
\usepackage{array}
\usepackage{multirow}
\usepackage{wrapfig}
\usepackage{float}
\usepackage{colortbl}
\usepackage{pdflscape}
\usepackage{tabu}
\usepackage{threeparttable}
\usepackage{threeparttablex}
\usepackage[normalem]{ulem}
\usepackage{makecell}
\usepackage{xcolor}
\begin{document}
\maketitle

\begin{abstract}
For randomized controlled trials (RCTs) with a single intervention being
measured on multiple outcomes, researchers often apply a multiple
testing procedure (such as Bonferroni or Benjamini-Hochberg) to adjust
\(p\)-values. Such an adjustment reduces the likelihood of spurious
findings, but also changes the statistical power, sometimes
substantially, which reduces the probability of detecting effects when
they do exist. However, this consideration is frequently ignored in
typical power analyses, as existing tools do not easily accommodate the
use of multiple testing procedures. We introduce the \texttt{PUMP}
\texttt{R} package as a tool for analysts to estimate statistical power,
minimum detectable effect size, and sample size requirements for
multi-level RCTs with multiple outcomes. Multiple outcomes are accounted
for in two ways. First, power estimates from \texttt{PUMP} properly
account for the adjustment in \(p\)-values from applying a multiple
testing procedure. Second, as researchers change their focus from one
outcome to multiple outcomes, different definitions of statistical power
emerge. \texttt{PUMP} allows researchers to consider a variety of
definitions of power, as some may be more appropriate for the goals of
their study. The package estimates power for frequentist multi-level
mixed effects models, and supports a variety of commonly-used RCT
designs and models and multiple testing procedures. In addition to the
main functionality of estimating power, minimum detectable effect size,
and sample size requirements, the package allows the user to easily
explore sensitivity of these quantities to changes in underlying
assumptions.
\end{abstract}

\doparttoc
\faketableofcontents

\part{}

\section*{Abstract}

For randomized controlled trials (RCTs) with a single intervention being
measured on multiple outcomes, researchers often apply a multiple
testing procedure (such as Bonferroni or Benjamini-Hochberg) to adjust
\(p\)-values. Such an adjustment reduces the likelihood of spurious
findings, but also changes the statistical power, sometimes
substantially, which reduces the probability of detecting effects when
they do exist. However, this consideration is frequently ignored in
typical power analyses, as existing tools do not easily accommodate the
use of multiple testing procedures. We introduce the \texttt{PUMP}
\texttt{R} package as a tool for analysts to estimate statistical power,
minimum detectable effect size, and sample size requirements for
multi-level RCTs with multiple outcomes. Multiple outcomes are accounted
for in two ways. First, power estimates from \texttt{PUMP} properly
account for the adjustment in \(p\)-values from applying a multiple
testing procedure. Second, as researchers change their focus from one
outcome to multiple outcomes, different definitions of statistical power
emerge. \texttt{PUMP} allows researchers to consider a variety of
definitions of power, as some may be more appropriate for the goals of
their study. The package estimates power for frequentist multi-level
mixed effects models, and supports a variety of commonly-used RCT
designs and models and multiple testing procedures. In addition to the
main functionality of estimating power, minimum detectable effect size,
and sample size requirements, the package allows the user to easily
explore sensitivity of these quantities to changes in underlying
assumptions.

\section{Introduction}
\label{sec:intro}

The \texttt{PUMP} \texttt{R} package fills in an important gap in
open-source software tools to design multi-level randomized controlled
trials (RCTs) with adequate statistical power. With this package,
researchers can estimate statistical power, minimum detectable effect
size (MDES), and needed sample size for multi-level experimental
designs, in which units are nested within hierarchical structures such
as students nested within schools nested within school districts. The
statistical power is calculated for estimating the impact of a single
intervention on multiple outcomes. The package uses a frequentist
framework of mixed effects regression models, which is currently the
prevailing framework for estimating impacts from experiments in
education and other social policy
research.\footnote{Other options include nonparametric or Bayesian methods, but these are less prevalent in applied research \cite{GELMANETAL2012, GelmanHill2007}.}

To our knowledge, none of the existing software tools for power
calculations allow researchers to account for multiple hypothesis tests
and the use of a multiple testing procedure (MTP). MTPs adjust
\(p\)-values to reduce the likelihood of spurious findings when
researchers are testing for effects on multiple outcomes. This
adjustment can result in a substantial change in statistical power,
greatly reducing the probability of detecting effects when they do
exist. Unfortunately, when designing studies, researchers who plan to
test for effects on multiple outcomes and employ MTPs frequently ignore
the power implications of the MTPs.

Also, as researchers change their focus from one outcome to multiple
outcomes, multiple definitions of statistical power emerge
\cite{RN23882, RN23878, RN23881, MTSAS}. The \texttt{PUMP} package
allows researchers to consider multiple definitions of power, selecting
those most suited to the goals of their study. The definitions of power
include:

\begin{itemize}
\tightlist
\item
  \textbf{individual power}: the probability of detecting an effect of a
  particular size (specified by the researcher) or larger for each
  hypothesis test. Individual power corresponds to how power is defined
  when there is focus on a single outcome.
\item
  \textbf{\(1-\)minimal power}: the probability of detecting effects of
  at least a particular size on at least one outcome. Similarly, the
  researcher can consider \textbf{\(d-\)minimal power} for any \(d\)
  less than the number of outcomes, or fractional powers, such as
  \(1/2-\)minimal power.
\item
  \textbf{complete power}: the power to detect effects of at least a
  particular size on \emph{all} outcomes.
\end{itemize}

As noted in \cite{Porter2018}, the prevailing default in many
studies---individual power---may or may not be the most appropriate type
of power. If the researcher's goal is to find statistically significant
estimates of effects on most or all primary outcomes of interest, then
their power may be much lower than anticipated when multiplicity
adjustments are taken into account. On the other hand, if the
researcher's goal is to find statistically significant estimates of
effects on at least one or a small proportion of outcomes, their power
may be much better than anticipated. In both of these cases, by not
accounting for both the challenges and opportunities arising from
multiple outcomes, a researcher may find they have wasted resources,
either by designing an underpowered study that cannot detect the desired
effect sizes, or by designing an overpowered study that had a larger
sample size than necessary. We introduce the \texttt{PUMP} package to
allow for directly answering questions that take multiple outcomes into
account, such as:

\begin{itemize}
\tightlist
\item
  How many schools would I need to detect a given effect on at least
  three of my five outcomes?
\item
  What size effect can I reliably detect on each outcome, given a
  planned MTP across all my outcomes?
\item
  How would the power to detect a given effect change if only half my
  outcomes truly had impact?
\end{itemize}

The methods in the PUMP package build on those introduced in
\cite{Porter2018}. This earlier paper focused only on a single RCT
design and model --- a multisite RCT with the blocked randomization of
individuals, in which effects are estimated using a model with
block-specific intercepts and with the assumption of constant effects
across all units. This earlier paper also did not produce software to
assist researchers in implementing its methods. With this current paper
and with the introduction of the PUMP package, we extend the methodology
to nine additional multi-level RCT designs and models. Also, while
\cite{Porter2018} focused on estimates of power, PUMP goes further to
also estimate MDES and sample size requirements that take multiplicity
adjustments into account.

\texttt{PUMP} extends functionality of the popular PowerUp! \texttt{R}
package (and its related tools in the form of a spreadsheet and Shiny
application), which compute power or MDES for multi-level RCTs with a
single outcome \cite{RN4473}. For a wide variety of RCT designs with a
single outcome, researchers can take advantage of closed-form solutions
and numerous power estimation tools. For example, in education and
social policy research, see \cite{RN4473, RN30153, RN23884, RN24179}.
However, closed-form solutions are difficult or impossible to derive
when a MTP is applied to a setting with multiple outcomes. Instead, we
use a simulation-based approach to achieve estimates of power.

In order to calculate power, the researcher specifies information about
the sample size at each level, the minimum detectable effect size for
each outcome, the level of statistical significance, and parameters of
the data generating distribution. The minimum detectable effect size is
the smallest true effect size the study can detect with the desired
statistical significance level, in units of standard deviations. An
``effect size'' generally refers to the standardized mean difference
effect size, which ``equals the difference in mean outcomes for the
treatment group and control group, divided by the standard deviation of
outcomes across subjects within experimental groups'' \cite{RN27978}.
Researchers often use effect sizes to standardize outcomes so that
outcomes with different scales can be directly compared.

The package includes three core functions:

\begin{itemize}
\tightlist
\item
  \texttt{pump\_power()} for calculating power given a experimental
  design and assumed model, parameters, and minimum detectable effect
  size.
\item
  \texttt{pump\_mdes()} for calculating minimum detectable effect size
  given a target power and sample sizes.
\item
  \texttt{pump\_sample()} for calculating the required sample size for
  achieving a given target power for a given minimum detectable effect
  size.
\end{itemize}

For any of these core functions, the user begins with two main choices.
First, the user chooses the assumed design and model of the RCT. The
\texttt{PUMP} package covers a range of multi-level designs, up to three
levels of hierarchy, that researchers typically use in practice, in
which research units are nested in hierarchical groups. Our power
calculations assume the user will be analyzing these RCTs using
frequentist mixed-effects regression models, containing a combination of
fixed or random intercepts and treatment impacts at different levels, as
we explain in detail in Section \ref{sec:est_power} and in the Technical
Appendix. Second, the user chooses the MTP to be applied. \texttt{PUMP}
supports five common MTPs --- Bonferroni, Holm, single-step and
step-down versions of Westfall-Young, and Benjamini-Hochberg. After
these two main choices, the user must also make a variety of decisions
about parameters of the data generating distribution.

The package also includes functions that allow users to easily explore
power over a range of possible values of parameters. This exploration
encourages the user to determine the sensitivity of estimates to
different assumptions. \texttt{PUMP} also visually displays results.
These additional functions include:

\begin{itemize}
\tightlist
\item
  \texttt{pump\_power\_grid()}, \texttt{pump\_mdes\_grid()}, and
  \texttt{pump\_sample\_grid()} for calculating the given output over a
  range of possible parameter values.
\item
  \texttt{update()} to re-run an existing calculation with a small
  number of parameters updated.
\item
  \texttt{plot()} on \texttt{PUMP}-generated objects to generate plots
  (including grid outputs).
\end{itemize}

The \texttt{PUMP} package is available on CRAN at
\url{https://CRAN.R-project.org/package=PUMP}. The authors of the
\texttt{PUMP} package have also created a web application built with R
Shiny. This web application calls the \texttt{PUMP} package and allows
users to conduct calculations with a user-friendly interface, but it is
less flexible than the package, with a focus on simpler scenarios (e.g.,
10 or fewer outcomes). The app can be found at
\url{https://public.mdrc.org/pump/}.

The remainder of this paper is organized as follows. In Section
\ref{sec:diplomas}, we introduce Diplomas Now, an educational
experiment, to be used as a running example throughout the paper. We
note, however, that the problem of power estimation for multi-level RCTs
is not exclusive to the educational setting. In Section
\ref{sec:mtp_overview}, we provide a summary of the multiple testing
problem. Also in Section \ref{sec:mtp_overview}, we present an overview
of the statistical challenges introduced by multiple hypothesis testing
and how MTPs protect against spurious impact findings. In Section
\ref{sec:est}, we introduce our methodology for estimating power when
taking the use of MTPs into account. This section also briefly discusses
our validation process. Section \ref{sec:choices} discusses the various
choices a user must make when using the package, including the designs
and models, MTPs, and key design and model parameters. Section
\ref{sec:vignette} provides a detailed presentation of the \texttt{PUMP}
package with multiple examples of using the packages functions to
conduct calculations for our education RCT example. Section
\ref{sec:conclusion} is a brief conclusion.

\section{Diplomas Now}
\label{sec:diplomas}

We illustrate our package using an example of a published RCT that
evaluated a secondary school model called Diplomas Now. The Diplomas Now
model is designed to increase high school graduation rates and
post-secondary readiness. Evaluators conducted a RCT comparing schools
who implemented the model to business-as-usual. We refer to this example
throughout this paper to illustrate key concepts and to illustrate the
application of the \texttt{PUMP} package.

The Diplomas Now model, created by three national organizations, Talent
Development, City Year, and Communities In Schools, targets underfunded
urban middle and high schools with many students who are not performing
well academically. The model is designed to be robust enough to
transform high-poverty and high-needs middle and high schools attended
by many students who fall off the path to high school graduation.
Diplomas Now, with MDRC as a partner, was one of the first validation
grants awarded as part of the Investing in Innovation (i3) competition
administered by the federal Department of Education.

We follow the general design of the Diplomas Now evaluation, conducted
by MDRC. The RCT contains three levels (students within schools within
districts) with random assignment at level two (schools). The initial
evaluation, included two cohorts of schools with each cohort
implementing for two years (2011-2013 for Cohort 1 and 2012-2014 for
Cohort 2). The cohorts included 62 secondary schools (both middle and
high schools) in 11 school districts that agreed to participate. Schools
in the active treatment group were assigned to implement the Diplomas
Now model, while the schools in the control group continued their
existing school programs or implemented other reform strategies of their
choosing \cite{DNREPORT}. The MDRC researchers conducted randomization
of the schools within blocks defined by district, school type, and year
of roll-out. After some schools were dropped from the study due to
structural reasons, the researchers were left with 29 high schools and
29 middle schools grouped in 21 random assignment blocks. Within each
block, schools were randomized to the active treatment or
business-as-usual, resulting in 32 schools in the treatment group, and
30 schools in the control group.

The evaluation focused on three categories of outcomes: Attendance,
Behavior, and Course performance, called the ``ABC's,'' with multiple
measures for each category. In addition, the evaluation measured an
overall ABC composite measures of whether a student is above given
thresholds on all three categories. This grouping constitutes 12 total
outcomes of interest. Evaluating each of the 12 outcomes independently
would not be good practice, as the chance of a spurious finding would
not be well controlled. The authors of the MDRC report pre-identified
three of these outcomes as \emph{primary} outcomes before the start of
the study in order to reduce the problem of multiple testing. We, by
contrast, use this example to illustrate what could be done if there was
uncertainty as to which outcomes should be primary. In particular, we
illustrate how to conduct a power analysis to plan a study where one
uses multiple testing adjustment, rather than predesignation, to account
for the multiple outcome problem.

There are different guidelines for how to adjust for groupings of
multiple outcomes in education studies. For example, \cite{RN23748}
recommends organizing primary outcomes into domains, conducting tests on
composite domain outcomes, and applying multiplicity corrections to
composites across domains. The What Works Clearinghouse applies
multiplicity corrections to findings within the same domain rather than
across different domains. We do not provide recommendations for which
guidelines to follow when investigating impacts on multiple outcomes.
Rather, we address the fact that researchers across many domains are
increasingly applying MTPs and therefore need to correctly estimate
power, MDES and sample size requirements accounting for this choice. In
our example, we elect to do a power analysis separately for each of the
three outcome groups of the ABC outcomes to control family-wise error
rather than overall error. This strategy means we adjust for the number
of outcomes within each group independently. For illustration purposes,
we focus on one outcome group, attendance, which we will assume contains
five separate outcomes.

\section{Overview of multiple testing}
\label{sec:mtp_overview}

Our motivating example illustrates that researchers are often interested
in testing the effectiveness of a single intervention on multiple
outcomes. The resulting multiplicity of statistical hypothesis tests can
lead to spurious findings of
effects.\footnote{Testing the effectiveness of an intervention for multiple subgroups, at multiple points in time, or across multiple treatment groups also results in a multiplicity of statistical hypotheses and can also lead to spurious findings of effects, but this is beyond the scope of this paper.}
Multiple testing procedures counteract this problem by adjusting
\(p\)-values for effect estimates; generally, \(p\)-values are adjusted
upward to require a higher burden of proof. When not using a MTP, the
probability of finding false positives increases, sometimes
dramatically, with the number of tests. When using a MTP, this
probability is reduced. Much of the proceeding explanation is borrowed
from or parallels discussion found in \cite{Porter2018}.

We first remind the reader of how raw, or unadjusted, \(p\)-values are
calculated in this setting, before introducing adjustments due to
multiple testing. Consider that a researcher is interested in testing
the impact of an intervention on \(M\) outcomes. In our running example
of the Diplomas Now study, if we had five outcomes in the attendance
group, we would have \(M = 5\). We apply a frequentist hypothesis
testing framework, and frame impacts in terms of effect sizes (\(ES\)).
For outcome \(m\), we can test a null hypothesis of no effect,
\(H_{0_m}: ES_m = 0\), against an alternative hypothesis
\(H_{1_m}: ES_m \neq 0\) for a two-sided tests or \(H_{1_m}: ES_m > 0\)
or \(H_{1_m}: ES_m < 0\) for a one-sided test. A significance test, such
as a two- or one-sided \(t\)-test, would then typically be driven by a
test statistic given by \begin{equation}
t_m = \frac{\widehat{ES}_m}{SE(\hat{ES}_m)},
\end{equation} where \(SE(\hat{ES}_m)\) is the standard error. A raw
\(p\)-value would then be computed as the probability of being at least
as extreme as the one observed, given that the null hypothesis is true.
The term ``raw'' is used to denote unadjusted \(p\)-values, in contrast
to values that have been adjusted using a MTP. For a two-sided test, the
raw \(p\)-value for test \(m\) is \(p_m=2*Pr(T_m \geq |t_m|)\). The
\texttt{PUMP} package allows for either one-sided or two-sided tests,
but we proceed assuming two-sided tests going
forward.\footnote{For a one-sided test, depending on the direction of our alternative hypothesis, the raw $p$-value for test $m$ is computed as $p_m=Pr(T_m \leq t_m)$ or $p_m=Pr(T_m \geq t_m)$.}

When testing a \emph{single} hypothesis under this framework,
``researchers typically specify \(\alpha\), the maximum acceptable
probability of making a Type I error. A Type I error is the probability
of erroneously rejecting the null hypothesis when it is true. The
quantity \(\alpha\) is also referred to as the significance level. If
\(\alpha=0.05\), then the null hypothesis is rejected if the \(p\)-value
is less than \(0.05\)'' \cite{Porter2018}.

In contrast, ``when one tests \emph{multiple} hypotheses under this
framework (such that \(M>1\)) and one conducts a separate test for each
of the hypotheses with \(\alpha=0.05\), there is a \emph{greater} than
\(5\%\) overall chance of a false positive finding in the study. If the
multiple tests are independent, the probability that at least one of the
null hypothesis tests will be erroneously rejected is
\[1-Pr(\text{none of the null hypotheses will be erroneously rejected}) = 1-(1-\alpha)^M.\]
Therefore, if researchers are estimating effects on three outcomes (and
if these outcomes are independent) the probability of at least one false
positive finding is \(1-(1-0.05)^3=0.14\). If the researchers were
instead estimating effects on five independent outcomes, the probability
of at least one false positive finding rises to \(0.23\). This Type I
error inflation for independent outcomes demonstrates the crux of the
multiple testing problem. In practice, however, the multiple outcomes
are usually at least somewhat correlated, which makes the test
statistics correlated and reduces the extent of Type I error inflation.
Nonetheless, any error inflation can still make it problematic to draw
reliable conclusions about the existence of effects above a specified
size'' \cite{Porter2018}.

\subsection{Using MTPs to protect against spurious impact findings}
\label{sec:mtp_use}

As introduced above, multiple testing procedures adjust \(p\)-values to
counteract the multiple testing
problem.\footnote{Alternatively, MTPs can decrease the critical values for rejecting hypothesis tests. For ease of presentation, this paper focuses only on the approach of adjusting $p$-values.}
We next describe how using a MTP protects against false positives.

Considering multiple outcomes presents both challenges and
opportunities. First, we discuss the impact of MTPs on individual power.
The power of an individual hypothesis test is the probability of
correctly rejecting a null hypothesis when the effect is at least a
specified size. We refer to a setting in which the true impact is at
least as large as the desired effect size as a ``false null''
hypothesis, while a ``true null'' is a setting in which the true impact
is zero. In the proceeding explanations, when we refer to rejecting a
null hypothesis or detecting an effect, we assume that we are detecting
an effect of a certain pre-specified size. If \(p\)-values are adjusted
upward, one is less likely to reject true nulls, which reduces the
probability of Type I errors, or false positive findings. At the same
time, MTPs increase the probability of a Type II error, or false
negative findings, when the test fails to reject a false null.
Individual power is \(1 - Pr(\text{Type II error})\), so MTPs have the
tradeoff of reducing Type I errors but also reducing individual power.

Next, we consider the impact of multiple outcomes on other definitions
of power. Applying a MTP reduces power according to all definitions of
power relative to the case when no MTP is applied to adjust
\(p\)-values. However, as discussed previously, having multiple outcomes
also allows for a wider variety of definitions of success. Recall that
\(1-\)minimal power is the probability of detecting an effect on at
least one outcome. Typically, \(1-\)minimal power, even after applying a
MTP, is higher than individual power for a hypothesis test on a single,
pre-specified outcome. Depending on the study, other definitions of
power, such as \(1/2\) or \(1/3\)-minimal power, may or may not have
higher power than the power of a single hypothesis test.

The MTPs that are the focus of this paper have three key features that
affect statistical power: (1) whether the MTP is a familywise procedure
or a false discovery rate procedure; (2) whether the MTP is single-step
or stepwise; and (3) whether the MTP takes the correlation between test
statistics into account. Below we explain each of these features of MTPs
and provide discussion of the new parameter specifications they require
when estimating power.

\subsection{Familywise error rate vs. false discovery error rate}

Familywise procedures ``reframe Type I error as a rate across the entire
set or ``family'' of multiple hypothesis tests. This rate is called the
familywise error rate (FWER) \cite{RN33098}. The FWER is typically set
to the same value as the probability of a Type I error for a single
test, e.g., \(\alpha\). MTPs that control the FWER at \(5\%\) adjust
\(p\)-values in a way that ensures that the probability of at least one
Type I error across the entire set of hypothesis tests is no more than
\(5\%\). The MTPs introduced by Bonferroni \cite{RN24280, RN24281},
\cite{RN24282}, and \cite{RN28696} all control the FWER''
\cite{Porter2018}.

MTPs that control false discovery rate (FDR) take an entirely different
approach to the multiple testing problem. FDR, introduced by
\cite{BenjaminiHochberg1995}, is a less stringent criteria than FWER.
It is the expected proportion of all rejected hypotheses that are
erroneously rejected. As laid out in \cite{Porter2018}, the two-by-two
representation in Table \ref{tab:twobytwo} is often found in articles on
multiple hypothesis testing, and helps to illustrate the difference
between FWER and FDR. Let \(M\) be the total number of tests. Therefore,
we have \(M\) unobserved truths: whether or not each null hypothesis is
true or false. We also have \(M\) observed decisions: whether or not the
null hypotheses were rejected, because the \(p\)-values were less than
\(\alpha\). In Table \ref{tab:twobytwo}, \(U\), \(V\), \(W\), and \(X\)
are four possible scenarios: the numbers of true or false hypotheses not
rejected or rejected. \(M_0\) and \(M_1\) are the unobservable numbers
of true null and false null hypotheses. \(R\) is the number of null
hypotheses that were rejected, and \(M - R\) is the number of null
hypotheses that were not rejected.

\begin{table}[h!]
\centering
\begin{tabular}{l r r r}
\toprule
                                      & \multicolumn{3}{c}{Observed decisions}\\
Unobserved truths                     & Number not rejected     & Number rejected   & Total \\ \midrule
Number of true null hypotheses        & $U$                     & $V$               & $M_0$ \\
Number of false null hypotheses       & $W$                     & $X$               & $M_1$ \\ \hline
Total                                 & $M-R$                   & $R$               & $M$ \\
\bottomrule
\end{tabular}
\caption{Numbers of hypothesis types and decisions.}
  \label{tab:twobytwo}
\end{table}

In Table \ref{tab:twobytwo}, \(V\) is the number of erroneously rejected
null hypotheses, or the number of false positive findings. Therefore,
the FWER is equivalent to \(Pr(V > 0)\), the probability of at least one
false positive finding. As noted in \cite{Porter2018}, ``recall that
Type I error is inflated when testing for effects when no MTPs are
applied. Consider the setting when all the outcomes are independent of
each other. The Type I error is almost \(10\%\) when testing effects on
two independent outcomes and \(23\%\) when testing effects on five
independent outcomes. These Type I error rates both correspond to the
FWER. The goal of MTPs that control the FWER is to bring these
percentages back down to \(5\%\).''

Also in Table \ref{tab:twobytwo}, the FDR is equal to \(E(V/R)\) but is
defined to be \(0\) when \(R=0\), or when no hypotheses are rejected.
``As is frequently noted in the literature (e.g.,
\cite{RN352, RN23748}, the FWER and FDR have different objectives.
Control of the FWER protects researchers from spurious findings and so
may be preferred when even a single false positive could lead to the
wrong conclusion about the effectiveness of an intervention. On the
other hand, the FDR is more lenient with false positives''
\cite{Porter2018}. Researchers may be willing to accept a few false
positives, \(V\), when the total number of rejected hypotheses, \(R\),
is large. Note that under the complete null hypothesis that all \(M\)
null hypotheses are true, the FDR is equal to the FWER. Referring back
to Table \ref{tab:twobytwo}, under the complete null we have \(V = R\),
so \begin{align*}
FDR &= E\left(\frac{V}{R}\right)\\
&=  E\left(\frac{V}{R}\mid R = 0\right) Pr(R = 0) + E\left(\frac{V}{R} \mid R > 0\right)Pr(R > 0) \\
&= 0 \times Pr(R = 0) + 1 \times Pr(R > 0) \\
&= Pr(R > 0) = FWER
\end{align*} However, if any effects truly exist, then FWER \(\geq\)
FDR. As a result of the difference in objective between FWER and FDR, in
the case where there is at least one false null hypothesis (at least one
true effect), a MTP that controls the FDR at \(5\%\) will have a Type I
error rate that is greater than \(5\%\).

A side remark is that MTPs may provide either ``weak control'' or
``strong control'' of the error rate they target. A MTP ``provides weak
control of the FWER or the FDR at level \(\alpha\) if the control can
only be guaranteed when all null hypotheses are true, e.g.~when the
effects on all outcomes are zero. A MTP provides strong control of the
FWER or FDR at level \(\alpha\) if the control is guaranteed when some
null hypotheses are true and some are false, e.g.~when there may be
effects on at least some outcomes. Of course, strong control is
preferred''
\cite{Porter2018}.\footnote{The single-step and step-down Westfall Young MTPs (which we discuss below) always provide at least weak control of the FWER. In order for these procedures to provide strong control of the FWER, they require the assumption of subset pivotality \cite{RN33093}.The distribution of the unadjusted test statistics or $p$-values is said to have subset pivotality if for any subset of null hypotheses, the joint distribution of the test statistics or of the $p-$values for the subset is identical to the distribution under the complete null.
A consequence of this assumption is that the permutation of test statistics or $p$-values can be done under the complete null hypothesis rather than under the unknown partial hypothesis \cite{RN33093}.}

\subsection{Single-step vs. stepwise procedures}
\label{sec:stepwise}

An additional feature of a MTP that affects its statistical power is
whether it is a ``single-step'' or ``stepwise'' procedure. ``Single-step
procedures adjust each \(p\)-value independently of the other
\(p\)-values. For example, the Bonferroni MTP multiplies all raw
\(p\)-values by \(M\). Therefore, one \(p\)-value adjustment does not
depend on other \(p\)-value adjustments, only on the number of tests''
\cite{Porter2018}.

In contrast, ``stepwise procedures first order raw \(p\)-values (or test
statistics), and then adjust according to the order of the tests. The
adjustments depend on the null hypotheses already rejected in previous
steps. For example, the Holm MTP --- the stepwise counterpart to the
Bonferroni MTP --- orders raw \(p\)-values from smallest to largest. The
procedure then multiplies the smallest \(p\)-value by \(M\), the second
smallest \(p\)-value by \(M-1\), and so on. The Holm MTP, like most
other stepwise procedures, also enforces monotonicity: each adjusted
\(p\)-value is greater than or equal to the previous adjusted
\(p\)-value, and enforces that any \(p\)-values is not greater than one.
Overall, stepwise MTPs allow for less adjustment than single-step MTPs
in later steps, and therefore preserve more power (for outcomes in the
later steps). The Bonferroni and Westfall-Young single-step procedure
are single-step; the Holm and Benjamini-Hochberg MTPs and the
Westfall-Young step-down procedure are stepwise'' \cite{Porter2018}.
Note that stepwise procedures may be ``step-down'' or ``step-up,''
referring to whether a procedure begins with the smallest \(p\)-value,
and thus the largest effect size (step-down) or the largest \(p\)-value
(step-up)``.

Due to the dependencies of adjustments in stepwise MTPs, a new
assumption must be considered when estimating power under multiplicity:
the proportion of outcomes on which there are truly impacts, or,
equivalently, the number of false null hypotheses. ``Researchers may be
inclined to assume that there will be effects on all outcomes, as
hypotheses of effects probably drive the selection of outcomes in the
first place.(\ldots) However, if the researchers are incorrect, the
probability of detecting the extant effects can be diminished, sometimes
substantially'' \cite{Porter2018}.

As noted in \cite{Porter2018}, it is important to point out that under
the assumption that some effects are truly null, we must change our
notion of power for \(d-\)minimal powers (e.g., \(1-\)minimal power,
\(1/3-\)minimal power, etc.) and complete power. While individual power
is defined based on the probability of correctly rejecting false nulls,
the definition is loosened here and includes the probability of
erroneous rejections of true nulls. For example, \(1/3-\)minimal power
is defined as the probability of detecting effects on at least \(1/3\)
of the \emph{total outcomes \(M\)}, regardless of the number of outcomes
with true effects. That is, \(1/3-\)minimal power is not defined as the
probability of detecting effects among the \(M\) outcomes on which the
effects truly exist. This reframing of power is necessary for power to
be consistent. If \(d-\)minimal power were defined based on false nulls,
then the value and interpretation would change depending on what
assumption the researcher is making about the number of false nulls,
which is an unknown quantity. For example, with \(M = 5\) outcomes, the
probability of detecting at least one effect would be very different
depending on if we assume all five outcomes are false nulls, or if we
assume only two of them are false nulls. Complete power, which is the
probability of detecting effects on all outcomes, has similar issues. We
define complete power only in the context where all effects are assumed
to be false nulls --- if any outcomes are assumed to be true nulls, then
complete power is undefined.

There is an additional technical note about the calculation of complete
power.\footnote{Complete power has also been referred to as “conjunctive power” \cite{RN33091} and “all pairs power” \cite{RN33097}.}
To calculate complete power, we do not need to adjust the \(p\)-values,
and can instead reject each individual test based on the unadjusted
\(p\)-values. Complete power is the power of the omnibus test
constructed by whether or not we reject all the null hypotheses. This
test was originally introduced as the intersection-union test because
the null hypothesis is expressed as a union and the alternative
hypothesis is expressed as an intersection
\cite{Berger1982, Berger1996}. \cite{Berger1982} showed that if all
the individual tests are level \(\alpha\), the intersection-union test
is also a level \(\alpha\) test. To provide some intuition, we do not
need to adjust \(p\)-values for complete power because it is a special
case where we must reject \emph{all} the hypothesis tests. Thus, there
is no way for the omnibus test to be rejected by chance because of a
favorable configuration \cite{RN23882}. For example, consider if we
have four tests, with two false nulls and two true nulls. If we consider
\(3-\)minimal power, we just need one of the two true nulls to be
rejected by chance alone, and there are two ways for this to occur. For
complete power, there is only one way for us to reject all of the nulls.
The downside of an intersection-union test is that it is conservative:
the FWER is generally less than \(\alpha\). For example, if we have two
independent tests with Type I error \(\alpha\), then if both of are true
nulls, the probability of a Type I error for the omnibus test (the
probability of rejecting both null hypotheses) is \(\alpha^2\)
\cite{Deng2008}.

\subsection{Correlation between test statistics}
\label{sec:corr}

The final feature of a MTP that affects its statistical power is whether
or not it takes into account the correlation between test statistics.
``The Bonferroni and Holm procedures strongly control the FWER in all
cases, even when the test statistics are correlated, but they adjust
\(p\)-values more than is necessary in that case. Along with the
Bonferroni and Holm MTPs, the Benjamin-Hochberg MTP also does not take
correlations into
account.\footnote{The Benjamini-Hochberg procedure was originally shown to control the FDR for independent test statistics. However, \cite{BenjaminiYekutieli2001} showed that it also controls the FDR for true null hypotheses with “positive regression dependence.” This condition is satisfied
for most applications in practice.} In contrast, both of the
Westfall-Young MTPs rely on the estimation of the joint distribution of
test statistics when the complete null hypothesis (that there are not
effects on any of the outcomes) is true. This joint distribution of the
test statistics is estimated from the study's data'' \cite{Porter2018}.
For example, random permutations of the treatment indicator break the
association between treatment status and outcome. Repeating these
permutations a large number of times results in a distribution of test
statistics under the complete null. ``Because the actual data are used
to generate this null distribution, correlations among the test
statistics are captured. Then observed test statistics can be compared
with the distribution of test statistics under the complete null
hypothesis''
\cite{Porter2018}.\footnote{Instead of using test statistics, the Westfall-Young MTPs can alternatively compare raw $p$-values with the estimated joint null distribution of $p$-values.}

The correlation between test statistics is a parameter a researcher must
specify in order to estimate power, MDES or sample size requirements
when using a MTP. When fitting a separate regression model for the
impact on each outcome, the \(\binom{M}{2}\) correlations between test
statistics are equal to the ``pairwise correlations between the
residuals in the \(M\) impact models'' \cite{Porter2018}. Then, ``if
there are no covariates in the impact models or if the \(R^2\)'s of the
covariates are equivalent in all impact models, then the correlations
between the test statistics are equal to the correlations between the
outcomes. However, having different \(R^2\)'s across the impact models
reduces the correlations between the residuals and therefore between
test
statistics.\footnote{For example, one of the multiple outcomes may have a baseline covariate with a high $R^2$ while another may have a baseline covariate with
a smaller $R^2$. Also, block dummies may explain more variation in some outcomes than in others.}
Models of outcomes that are highly correlated are more likely to have
residuals that are highly correlated because baseline covariates will
tend to have similar \(R^2\)'s. The gaps between the correlations
between outcomes and the correlations between residuals --- and
therefore the test statistics --- may be wider for moderately or weakly
correlated outcomes. In any case, the upper bounds of correlations
between the test statistics are the correlations between the outcomes''
\cite{Porter2018}.

\section{Estimating power, MDES and sample size in studies with multiple outcomes}
\label{sec:est}

\subsection{Power estimation approach}
\label{sec:est_power}

We take an innovative simulation-based approach to estimating power, as
introduced in \cite{Porter2018}. This approach is then also applied to
estimate MDES and sample size. In order to estimate power for a single
outcome, we can often use closed-form algebraic expressions, which are
derived from the assumed model. However, with multiple outcomes, finding
such expressions can be quite difficult, or even impossible. In cases
where it is possible to find a closed-form expression, we would need to
find expressions for every design and model, MTP, and definition of
power. Importantly, we would \emph{also} need to find new expressions
for any possible number of outcomes, which quickly becomes an
intractable problem. Furthermore, in some cases, such as
permutation-based procedures like Westfall-Young approaches, a
closed-form solution does not exist. To avoid these complexities, we
rely on simulation to calculate estimated power. The approach outlined
below can estimate power for any scenario.

If we were to rely on a \emph{full} simulation approach, we could use
the following method to estimate power. We introduce this full
simulation approach to provide intuition, but use a simplified and far
less computationally intensive approach in the package, as discussed
below.

\begin{enumerate}
\def\labelenumi{\arabic{enumi}.}
\item
  \emph{Simulate a data sample according to the joint alternative
  hypothesis.} First, we formulate what we will refer to as the
  \emph{joint alternative hypothesis}, which is the set of outcomes we
  assume to have treatment effects above the desired size. We define
  \(\psi_m\) to be the treatment effect for outcome \(m\), with \(M\)
  total outcomes. If we have \(M = 5\) outcomes, as in the Diplomas Now
  study, one possible joint alternative hypothesis is that all outcomes
  have effects above specified sizes:
  \(H_A: \psi_1 > 0.125, \psi_2 > 0.2, \psi_3 > 0.1, \psi_4 > 0.1, \psi_5 > 0.05\).
  Another possible joint alternative hypothesis is one where only the
  first two outcomes have effects above the desired sizes:
  \(H_A: \psi_1 > 0.125, \psi_2 > 0.2, \psi_3 = \psi_4 = \psi_5 = 0\).
  Once our joint alternative hypothesis is specified, we would generate
  simulated data under this hypothesis. To simulate data, we need to
  specify the full set of parameters as mentioned in Section
  \ref{sec:params} that allow for data generation. The Technical
  Appendix contains more details about the assumed data-generating
  process. For example, for the Diplomas Now experiment, we would assume
  a specific data generating process to allow us to simulate synthetic
  students, schools, and districts, including covariates, outcomes, and
  treatment assignment. This process would involve specifying parameter
  values such as \(R^2\), the amount of outcome variation explained by
  covariates at a particular level, and translating these parameter
  choices into data-generating parameters, such as the coefficient
  values for covariates in a linear model.
\item
  \emph{Estimate impacts on the simulated data.} Given simulated data,
  we could fit \(M\) regression models (specified to match the
  experimental design and model assumptions). For the models supported
  by \texttt{PUMP}, the relevant functions would be \texttt{lm()},
  \texttt{lmer()} from the \texttt{lme4} library \cite{lme4}, and
  \texttt{interacted\_linear\_estimators()} from the \texttt{blkvar}
  library.\footnote{This package is currently under development on GitHub; see \url{https://github.com/lmiratrix/blkvar}}
  From the model output we extract the test statistics \(t_m\) for the
  estimated impacts, one statistic for each outcome, along with
  estimated standard errors.
\item
  \emph{Calculate unadjusted \(p\)-values.} The test statistics and
  standard errors would in turn give raw (unadjusted) \(p\)-values. We
  can either calculate these by hand, or use the \(p\)-values routinely
  returned by regression functions. For Diplomas Now we could run a
  regression model of each attendance measure on treatment status and
  student and school covariates, and extract \(p\)-values from the
  regression outputs.
\item
  \emph{Repeat above steps (1 through 3) for a large number of
  iterations.} Denote the number of iterations \texttt{tnum}. Repeating
  steps \(1\)-\(3\) \texttt{tnum} times results in a matrix of
  unadjusted \(p\)-values which we call \(\mathbf{F}\), and is of
  dimension \(tnum \times M\). One row corresponds to one set of
  simulated raw \(p\)-values from regressions for the \(5\) attendance
  outcomes of interest for Diplomas Now.
\item
  \emph{Adjust \(p\)-values.} For each row, corresponding to one
  simulated dataset, the \(M\) raw \(p\)-values corresponding to the
  \(M\) hypothesis tests can be adjusted according to the desired
  multiple testing procedure. This process generates a new matrix
  \(\mathbf{G}\) of adjusted \(p\)-values. For Bonferroni, Holm, and
  Benjamini-Hochberg adjustments, we use the function \texttt{p.adjust}
  in \texttt{R} (found in the \texttt{stats} package). We developed our
  own functions for implementing adjustment using the Westfall-Young
  procedures. One row corresponds to one set of simulated
  \emph{adjusted} \(p\)-values for the \(5\) attendance outcomes of
  interest for Diplomas Now.
\item
  \emph{Calculate hypothesis rejection indicators.} For any MTP, the
  matrix of adjusted \(p\)-values \(\mathbf{G}\) can then be compared
  with a specified value of \(\alpha\) (the default is \(0.05\), but the
  value can be changed by the user). For each row, corresponding to one
  iteration of simulated data, we record whether or not the null
  hypothesis was rejected for each outcome. This process results in a
  new matrix \(\mathbf{H}\), which contains hypothesis rejection
  indicators (still of dimension \(tnum \times M\)). Using
  \(\mathbf{H}\), we can compute all definitions of power.
\item
  \emph{Calculate power.} To compute the different definitions of power:
\end{enumerate}

\begin{itemize}
\item
  \emph{Individual power} for outcome \(m\) is the proportion of the
  \texttt{tnum} rows in which the null hypothesis \(m\) was rejected
  (the mean of column \(m\) of \(\mathbf{H}\)). We would have \(5\)
  different individual power values for Diplomas Now, corresponding to
  each outcome of interest.
\item
  \emph{\(d\)-minimal power} is the proportion of the \(tnum\) rows in
  which at least \(d\) of the \(M\) hypotheses were
  rejected.\footnote{Note that others refer to $1-$minimal power simply as “minimal power” \cite{RN33095, RN23882, MTSAS}, “disjunctive power” \cite{RN33091}, or “any pair” power \cite{RN33097}. \cite{RN23882} use the terminology of “r-power” for what is referred to here as $d-$minimal power for $d>1$.}
  For Diplomas Now, we could consider \(1\)-minimal power through
  \(4\)-minimal power.
\item
  \emph{Complete power} is the proportion of the \(tnum\) rows in which
  all of the null hypotheses were rejected based on the raw \(p\)-values
  rather than adjusted \(p\)-values (based on the matrix \(\mathbf{G}\)
  rather than \(\mathbf{H}\).) We would be interested in complete power
  if we want to evaluate whether Diplomas Now resulted in improvement
  for every single attendance outcome of interest. With \(5\) outcomes,
  this criteria is a relatively strict indicator of success.
\end{itemize}

Above, we outlined a full simulation-based approach for calculating
power. This approach would be computationally intensive because of the
need to generate and analyze a full simulated dataset at each iteration.
We can simplify this process by skipping the simulation of data and
modeling steps. Given an assumed model and correlation structure for the
test statistics, we can directly sample from \(f(t_1, \dots, t_M)\), the
joint alternative distribution of the test statistics. This shortcut
vastly improves both the simplicity and the speed of computation. In
summary, our approach is:

\begin{enumerate}
\def\labelenumi{\arabic{enumi}.}
\tightlist
\item
  \textbf{Generate} draws of \emph{test statistics \(t_1, \dots, t_M\)
  under the joint alternative hypothesis.} This step produces a
  \(tnum \times M\) matrix \(\mathbf{E}\).
\item
  \emph{Calculate unadjusted \(p\)-values.} This produces the matrix
  \(\mathbf{F}\), as in the procedure above.
\item
  \emph{Adjust \(p\)-values.} This produces the matrix \(\mathbf{G}\),
  as in the procedure above.
\item
  \emph{Calculate hypothesis rejection indicators.} This produces the
  matrix \(\mathbf{H}\), as in the procedure above.
\item
  \emph{Calculate power.}
\end{enumerate}

We now describe how to sample from \(f(t_1, \dots, t_M)\) directly.
First, we assume a particular research design and model. In our example
based on the Diplomas Now study, the research design is a \(3\)-level
experiment, with randomization at level \(2\). We plan for analyzing our
data with a linear regression model with fixed intercepts at the
district level, random intercepts at the school level, and a constant
treatment effect across schools and districts. As previously, denote
\(\psi_m\) as the treatment effect for outcome \(m\). We express
treatment effects in terms of effect sizes:
\[ES_m = \frac{\psi_m}{\sigma_{m}}\] where \(\sigma_{m}\) is the
standard deviation of outcome \(Y_m\) in the control group. In order to
calculate power, we also need the standard error of the impact in effect
size units, which we denote as \[Q_m = SE(\hat{ES}_m).\] The quantity
\(Q_m\) is a consequence of the assumed model, the number of units at
different levels, the percent of units treated, the assumed \(R^2\), and
other parameters; our technical appendix shows formulae for \(Q_m\) for
all the designs and models our package supports. In our Diplomas Now
example, \(Q_m\) will be a function of the number of students, schools,
and districts; the proportion of treated units; the number of student
and school covariates; the explanatory power of the student and school
covariates; the proportion of variation in the outcome explained by
schools and districts; and the amount of impact variation relative to
the amount of mean variation. Some parameters, such as the percent of
units treated, will generally be known, while others, such as the
\(R^2\) at different levels, would need to be supplied by the user
through either estimation on pilot data or assumptions based on prior
knowledge.

Given the effect sizes \(ES_m\) and the standard errors \(Q_m\), we can
determine the distribution of the vector of test statistics. When
testing the hypothesis for outcome \(m\), the test statistic for a
\(t\)-test is: \[t_m = \frac{\hat{ES}_m}{\hat{Q}_m}\] with degrees of
freedom \(df\), also defined by the assumed model. Under the alternative
hypothesis for outcome \(m\), \(t_m\) has a \(t\) distribution with
degrees of freedom \(df\) and mean \(ES_m/Q_m\). Finally, in addition to
the parameters above, we also need to choose the correlation matrix
between test statistics \(\rho\) to sample from the joint distribution
of \(f(t_1, \dots, t_M)\). With these distributions specified, we can
calculate \(p\)-values.

Note that this approach of simulating test statistics builds on work by
\cite{RN33089}, who use simulated test statistics to identify critical
values based on the distribution of the maximum test statistics. Their
approach produces the same estimates as the approach described here for
the single-step Westfall-Young MTP. As an alternative to a
simulation-based approach, \cite{RN23882} derived explicit formulae for
\(d-\)minimal powers of stepwise procedures and for complete power of
single-step procedures, but only for \(1\), \(2\), or \(3\) tests. The
approach presented here is more generally applicable, as it can be used
for all MTPs, for any number of tests, and for all definitions of power
discussed in the present paper.

\emph{Remark.} The \(p\)-value adjustment using Westfall-Young
procedures is the most complex correction procedure, so we briefly
outline it here. Similar to above, we first explain a full simulation
approach, and then discuss our simplification. Under a full simulation
approach, we would first generate a single dataset under the joint
alternative hypothesis and calculate a set of \(M\) observed test
statistics. Then, we would permute the single simulated dataset, say
\(B = 3,000\) times, assuming the joint null hypothesis, and calculate
test statistics on each of these permuted datasets. This process
generates an empirical distribution of \(B\) test statistics under the
joint null distribution. Next, we compare the distribution of observed
test statistics to the generated distribution of test statistics under
the joint null distribution to calculate \(p\)-values. We would then
re-generate a new simulated dataset, and repeat the process. If we were
to generate \(tnum = 10,000\) datasets under the joint alternative
hypothesis, for each of these datasets we generate \(B = 3,000\)
permuted datasets under the joint null, so we would have to generate
\(10,000 \times 3,000\) datasets!

When we skip the step of simulating data, then for each iteration \(t\)
in \(1, \dots, tnum\) we first generate a set of \(M\) observed test
statistics from the joint alternative distribution. Then, we draw \(B\)
samples of test statistics under the joint null rather than permuting
the data \(B\) times. Under the null hypothesis, \(t_m\) has a \(t\)
distribution with degrees of freedom \(df\) and mean \(0\). As before,
we then compare the distribution of observed test statistics to the
distribution of test statistics under the joint null distribution to
calculate \(p\)-values. Westfall-Young procedures are computationally
intensive, so the approach of skipping the simulated data step is
particularly helpful here. This approach substantially reduces
computational time by drawing test statistics directly rather than
permutating the data.

\subsection{Determining MDES and sample size}
\label{sec:est_mdes_ss}

Frequently, a researcher's main concern with power is calculating either
the MDES for each outcome in a given study, or determining the necessary
sample size to achieve a target power given a specified set of MDES
values. In Diplomas Now, we might want to know what sample sizes we
would need to detect at least one significant effect across our outcomes
if all the outcomes had a specified effect size (corresponding to
\(1-\)minimal power) and we were planning on using the Holm procedure.

For \texttt{pump\_mdes()} and \texttt{pump\_sample()}, the user provides
a particular target power, say \(80\%\). The method then conducts a
stochastic optimization problem to determine a value (of sample size or
MDES) that is within a specified tolerance of the target power with high
probability. We discuss the algorithm for MDES, although the approach
for determining sample size is the same.

The algorithm first determines an initial range of MDES values that
likely contain the target MDES. This initial range is calculated using
formulae for unadjusted power based on the standard errors and degrees
of freedom. In particular, from \cite{RN4473}, in general the MDES for
a single outcome can be estimated as
\[ MDES = MT_{df} \times SE / \sigma_{m} \] where \(MT_{df}\), the
``multiplier,'' is the sum of two \(t\) statistics with degrees of
freedom \(df\). For one-tailed tests,
\(MT_{df} = t_{\alpha}^\star + t_{1-\beta}^\star\) where \(\alpha\) is
the Type I error rate and \(\beta\) is the desired power. For two-tailed
tests, \(MT_{df} = t_{\alpha/2}^\star + t_{1-\beta}^\star\). We do not
explain the details of the derivations of the multiplier here; for more
details and understanding, see \cite{RN4473} or \cite{RN27978}. These
expressions can be further manipulated to obtain sample size formulae;
see our technical appendix for all formulae used in the package.

We can calculate our initial bounds by manipulating the \(\alpha\) and
\(\beta\) values in the above. First, to calculate the preliminary lower
bound, we apply the formula above as given, assuming individual
unadjusted power will give the smallest MDES; to calculate the
preliminary upper bound, we apply the formula using \(\alpha/M\) to
correspond to a Bonferroni correction. We also adjust \(\beta\) to
account for different power types. For example, if we are interested in
complete power, we need a larger upper bound than for individual power;
in order to have a complete power of \(80\%\), we would need each
outcome to have an individual power of \(\text{0.8}^{(1/M)}\), assuming
independence. If we are interested in minimal power, we must have a
smaller lower bound; in order to have \(1-\)minimal power of \(80\%\),
each outcome would only need to have individual power of
\(1 - (1 - \text{0.8})^{(1/M)}\). We ignore correlation in the setting
of the initial bounds; the bounds do not need to be strict, given the
adaptive nature of the subsequent search.

Once the initial range is established, we use \texttt{pump\_power()}
with the complete array of design parameters, including the correlation
between test statistics, to obtain rough (using a small \texttt{tnum},
or number of simulation trials) estimates of power for five initial
values across this range. We then fit a scaled logistic curve to these
five points, and identify where the curve crosses the desired power
level. After fitting an initial curve, we iterate, repeatedly
calculating power for the targeted point and using the result to update
the logistic curve model. At any point, if the current fitted curve's
range does not contain the target power, the algorithm extrapolates
beyond the initial bounds for the next step. With each iteration we
increase \texttt{tnum} to increase precision as we narrow in on the
final answer; with each update to our estimated power curve, we weigh
the collection of observations by their precisions (determined by
corresponding \texttt{tnum} value). Once a test point achieves the
target power to within tolerance, we conduct an additional simulation
check using a high number of replicates to verify the proposed answer is
within a specified tolerance of the target power; if it is not, we
continue the iterative search. The default tolerance is \(1\%\), so
given a target power of \(80\%\), we stop when we find a MDES that gives
an estimated power between \(79\%\) and \(81\%\).

In practice, due to the monotonic nature of the logistic functional
form, our algorithm generally converges fairly rapidly. However, in
certain corner cases the algorithm may fail to converge on a value
within tolerance. For more information on applying the search algorithm,
see the sample size vignette.

\subsection{Package Validation}

We completed extensive validation checks to ensure our power calculation
procedures are correct. First, we compared our power estimates in
scenarios with only one outcome, \(M = 1\), to those from the
\texttt{PowerUpR} package. Without a multiple testing procedure
adjustment, our estimates match. Second, in order to validate our
estimates under multiplicity, we followed the full simulation approach
outlined above, in Section \ref{sec:est_power}. The simulation approach
involves generating many iterations of full datasets according to the
assumed design and model, calculating \(p\)-values, and calculating an
empirical estimate of power. Using a binomial distribution we
constructed Monte Carlo confidence intervals for the power estimates
from the full simulation approach. Then, we validated that the
\texttt{PUMP} estimates fall within these confidence intervals.

A more detailed explanation of the validation procedure can be found in
the Appendix, and full validation code and results are in a
supplementary GitHub repository
\href{https://github.com/MDRCNY/pump_validate}{pump\_validate}. For some
scenarios, we have some apparent discrepancies from PowerUp, but these
result from different modeling choices. For example, for certain models
PowerUp assumes the intraclass correlation is zero, while we allow for
nonzero values. These discrepancies are noted in the appendix.

\section{User choices}
\label{sec:choices}

\subsection{Designs and models}
\label{sec:d_m}

When planning a study, the researcher first has to identify the design
of the experiment, including the number of levels, and the level at
which randomization occurs. These decisions can be a mix of the
realities of the context (e.g., the treatment must be applied at the
school level, and students are naturally nested in schools, making for a
cluster randomization), or deliberate (e.g., the researcher groups
similar schools to block their experiment in an attempt to improve
power). Second, based on the design and the inferential goals of the
study, the researchers chooses an assumed model, including whether
intercepts and treatment effects should be treated as constant, fixed,
or random. For the same experimental design, the analyst can sometimes
choose from a variety of possible models, and these two decisions should
be kept conceptually separated from each other.

\emph{The design.} The \texttt{PUMP} package supports designs with one,
two, or three levels, with randomization occurring at any level. For
example, a design with two levels and randomization at level one is a
blocked design (or equivalently a multisite experiment), where level two
forms the blocks (blocks being groups of units, some of which are
treated and some not). Ideally, the blocks in a trial will be groups of
relatively homogenous units, but frequently they are a consequence of
the units being studied (e.g., evaluations of college supports, with
students, the units, nested in colleges, the blocks). A design with two
levels and randomization at level two is commonly called a cluster
design (e.g., a collection of schools, with treatment applied to a
subset of the schools, with outcomes at the student level); here the
schools are the clusters, with a cluster being a collection of units
which is entirely treated or entirely not. We can also have both
blocking and clustering: randomizing schools within districts, creating
a series of cluster-randomized experiments, would be a blocked (by
district), cluster-randomized experiment, with randomization at level
two.

\emph{The model.} Given a design, the researcher can select a model via
a few modeling choices. In particular the researcher has to decide, for
each level beyond the first, about the intercepts and the treatment
impacts:

\begin{itemize}
\tightlist
\item
  Whether level two and level three intercepts are:

  \begin{itemize}
  \tightlist
  \item
    fixed: we have a separate intercept for each unit.
  \item
    random: we have a separate intercept for each unit as above, but
    model the collection of intercepts as Normally distributed, allowing
    for partial pooling.
  \end{itemize}
\item
  Whether level two and level three treatment effects are:

  \begin{itemize}
  \tightlist
  \item
    constant: we model all units within a group as having the same
    single average impact.
  \item
    fixed: we allow each block or cluster within a level to have its own
    individual estimated impact (we can only do this if we have treated
    and control units within said block or cluster).
  \item
    random: we allow variation as with fixed, but model the collection
    of treatment impacts as Normally distributed around a grand mean
    mean impact. This is implicitly allowing for the sample as being
    representative of a larger super-population, in terms of treatment
    impact estimation.
  \end{itemize}
\end{itemize}

We denote the research design by \(d\), followed by the number of levels
and randomization level, so \texttt{d3.1} is a three level design with
randomization at level one. The model is denoted by \(m\), followed by
the level and the assumption for the intercepts, either \(f\) or \(r\)
and then the assumption for the treatment impacts, \(c\), \(f\), or
\(r\). For example, \texttt{m3ff2rc} means at level \(3\), we assume
fixed intercepts and fixed treatment impacts, and at level two we assume
random intercepts and constant treatment impacts. The full design and
model are specified by concatenating these together,
e.g.~\texttt{d2.1\_m3fc}. The Diplomas Now model, for example, is
\texttt{d3.2\_m3fc2rc}.

The full list of supported design and model combinations is below. The
user can see the list by calling \texttt{pump\_info()}, which provides
the designs and models, MTPs, power definitions, and model parameters.
We also include the corresponding names from the PowerUP! package where
appropriate. For more details about each combination of design and
model, see the Technical Appendix.

\begin{table}
\centering
\begin{tabular}{lllll}
\toprule
d\_m & Design & Model & PowerUp & Params\\
\midrule
d1.1\_m1c & d1.1 & m1c & n/a & R2.1\\
d2.1\_m2fc & d2.1 & m2fc & bira2\_1c & R2.1, ICC.2\\
d2.1\_m2ff & d2.1 & m2ff & bira2\_1f & R2.1, ICC.2\\
d2.1\_m2fr & d2.1 & m2fr & bira2\_1r & R2.1, ICC.2, omega.2\\
d2.1\_m2rr & d2.1 & m2rr & n/a & R2.1, ICC.2, omega.2\\
\addlinespace
d2.2\_m2rc & d2.2 & m2rc & cra2\_2r & R2.1, R2.2, ICC.2\\
d3.1\_m3rr2rr & d3.1 & m3rr2rr & bira3\_1r & R2.1, ICC.2, omega.2, ICC.3, omega.3\\
d3.2\_m3ff2rc & d3.2 & m3ff2rc & bcra3\_2f & R2.1, R2.2, ICC.2, ICC.3\\
d3.2\_m3fc2rc & d3.2 & m3fc2rc & n/a & R2.1, R2.2, ICC.2, ICC.3\\
d3.2\_m3rr2rc & d3.2 & m3rr2rc & bcra3\_2r & R2.1, R2.2, ICC.2, ICC.3, omega.3\\
\addlinespace
d3.3\_m3rc2rc & d3.3 & m3rc2rc & cra3\_3r & R2.1, R2.2, ICC.2, R2.3, ICC.3\\
\bottomrule
\end{tabular}
\end{table}

\subsection{Multiple testing procedures}

The supported multiple testing procedures were covered in more detail in
Section \ref{sec:mtp_overview}. Here we provide a review of the multiple
testing procedures supported by the \texttt{PUMP} package:

\begin{itemize}
\tightlist
\item
  \emph{Bonferroni}: adjusts \(p\)-values by multiplying them by \(M\)
  to ensure strong control of the FWER. Bonferroni is a simple
  procedure, but the most conservative.
\item
  \emph{Holm}: a step-down version of Bonferroni. Starting from smallest
  to largest, \(p\)-values are sequentially adjusted by different
  multipliers. Holm is less conservative than Bonferroni for larger
  \(p\)-values.
\item
  \emph{Benjamini-Hochberg}: A sequential, step-up procedure that
  controls the FDR. Using the BH method, only null hypotheses with
  \(p\)-values below a certain threshold are rejected, where the
  threshold is determined by the number of tests and the level
  \(\alpha\).
\item
  \emph{Single-step Westfall-Young}: A permutation-based procedure for
  controlling the FWER, which directly takes into account the joint
  correlation structure of the outcomes. In the single-step approach,
  all outcomes are adjusted by using the permuted distribution of the
  minimum \(p\)-value. Although Westfall-Young procedures are less
  conservative while still protecting against false discoveries, they
  are computationally very intensive.
\item
  \emph{Step-down Westfall-Young}: A similar approach to the single-step
  procedure, except that outcomes are adjusted sequentially from
  smallest to largest according to the permuted distributions of the
  corresponding sequential \(p\)-values.
\end{itemize}

For a more detailed explanation of each MTP, see Appendix A of
\cite{Porter2018}.

The following table from \cite{Porter2018} summarizes the important
features for each of the MTPs supported by \texttt{PUMP}.

\begin{table}[h!]
\centering
\begin{tabular}{l r r r}
\toprule
Procedure                             & Control & Single-step or stepwise & Accounts for correlation \\ \midrule
Bonferroni (BF)                       & FWER    & single-step             & No                       \\
Holm (HO)                             & FWER    & stepwise                & No\\
Westfall-Young Single-step (WY-SS)    & FWER    & single-step             & Yes\\
Westfall-Young Step-down (WY-SD)      & FWER    & stepwise                & Yes\\
Benjamini-Hochberg (BH)               & FDR     & stepwise                & No \\
\bottomrule
\end{tabular}
\caption{Summary of MTP procedures.}
  \label{tab:mtp}
\end{table}

\subsection{Model parameters}
\label{sec:params}

The table below shows the parameters that influence \(Q_m\), the
standard error, for different designs and models.

\begin{table}
\centering
\begin{tabular}{ll}
\toprule
Parameter & Description\\
\midrule
nbar & harmonic mean of level 1 units per
                     level 2 unit (students per school)\\
J & harmonic mean of number of level 2 
                      units per level 3 unit (schools per district)\\
K & number of level 3 units (districts)\\
Tbar & proportion of units assigned to treatment\\
numCovar.1 & number of level 1 (individual) covariates\\
\addlinespace
numCovar.2 & number of level 2 (school) covariates\\
numCovar.3 & number of level 3 (district) covariates\\
R2.1 & percent of variation explained by level 1 covariates\\
R2.2 & percent of variation explained by level 2 covariates\\
R2.3 & percent of variation explained by level 3 covariates\\
\addlinespace
ICC.2 & level 2 intraclass correlation\\
ICC.3 & level 3 intraclass correlation\\
omega.2 & ratio of variance of level 2 average impacts to
                      level 2 random intercepts\\
omega.3 & ratio of variance of level 3 average impacts to
                      level 3 random intercepts\\
\bottomrule
\end{tabular}
\end{table}

A few parameters warrant more explanation.

\begin{itemize}
\item
  The quantity \(\text{ICC}\) is the unconditional Intraclass
  Correlation, and gives a measure of variation at different levels of
  the model. For each outcome, the ICC for each level is defined as the
  ratio of the variance at that level divided by the overall variance of
  the individual outcomes. The ICC includes the variation due to
  covariates.
\item
  For each outcome, the quantity omega (\(\omega\)) for each level is
  the ratio between impact variation at that level and variation in
  intercepts (including covariates) at that level. It is a measure of
  treatment impact heterogeneity.
\item
  The \(R^2\) expressions are the percent of variation at a particular
  level predicted by covariates specific to that level. For simplicity
  we assume covariates at a level are group mean centered, so only
  covariates at a particular level explain variance at that level.
\end{itemize}

For precise formulae of these expressions, see the Technical Appendix,
which outlines the assumed data-generating process, and the resulting
expressions for \(\text{ICC}\), \(\omega\), and \(R^2\).

In addition to design parameters, there are additional parameters that
control the precision of the power estimates themselves:

\begin{itemize}
\tightlist
\item
  \texttt{tnum} is the number of test statistics generated in order to
  estimate power. A larger number of test statistics results in greater
  computation time, but also a more precise estimate of power. Note that
  the \texttt{pump\_mdes()} and \texttt{pump\_sample()} have multiple
  \texttt{tnum} parameters controlling the precision of the search.
\item
  \texttt{B} is the number of Westfall-Young permutations. Again, there
  is a tradeoff between precision and computation time.
\item
  \texttt{parallel.WY.cores} specifies the number of cores to use for
  parallel computation of the Westfall-Young Step-Down procedure, which
  is the most computationally intensive. The default of \texttt{1} does
  not result in parallel computation. Parallelization is done using
  \texttt{parApply} from the \texttt{parallel} package.
\end{itemize}

\section{Using the PUMP package}
\label{sec:vignette}

In this section, we illustrate how to use the \texttt{PUMP} package,
using our example motivated by the Diplomas Now study. Given the study's
design, we ask a natural initial question: What size of impact could we
reasonably detect after using a MTP to adjust \(p\)-values to account
for our multiple outcomes?

We mimic the planning process one might use for planning a study similar
to Diplomas Now (e.g., if we were planning a replication trial in a
slightly different context). To answer this question we therefore first
have to decide on our experimental design and modeling approach. We also
have to determine values for the associated design parameters that
accompany these choices. In the following sections we walk through
selecting these parameters (sample size, control variables, intraclass
correlation coefficients, impact variation, and correlation of
outcomes). We calculate MDES for the resulting context and determine how
necessary sample sizes change depending on what kind of power we desire.
We finally illustrate some sensitivity checks, looking at how MDES
changes as a function of rho, the correlation of the test statistics.

\subsection{Establishing needed design parameters}

To conduct power, MDES, and sample size calculations, we first specify
the design, sample sizes, analytic model, and level of statistical
significance. We also must specify parameters of the data generating
distribution that match the selected design and model. All of these
numbers have to be determined given resource limitations, or estimated
using prior knowledge, pilot studies, or other sources of information.

We next discuss selection of all needed design parameters and modeling
choices. For further discussion of selecting these parameters see, for
example \cite{RN27978} and \cite{RN4473}. For discussion in the
multiple testing context, especially with regards to the overall power
measures such as \(1-\)minimal or complete power, see
\cite{Porter2018}; the findings there are general, as they are a
function of the final distribution of test statistics. The key insight
is that power is a function of only a few summarizing elements: the
individual-level standard errors, the degrees of freedom, and the
correlation structure of the test statistics. Once we have these
elements, regardless of the design, we can proceed.

\emph{Analytic model.} We first need to specify how we will analyze our
data; this choice also determines which design parameters we will need
to specify. Following the original Diplomas Now report, we plan on using
a multi-level model with fixed effects at level three, a random
intercept at level two, and a single treatment coefficient. We represent
this model as ``m3fc2rc.'' The ``3fc'' means we are including block
fixed effects, and not modeling any treatment impact variation at level
three. The ``2rc'' means random intercept and no modeled variation of
treatment within each block (the ``c'' is for ``constant''). We note
that the Diplomas Now report authors call their model a ``two-level''
model, but this is not quite aligned with the language of this package.
In particular, fixed effects included at level two are actually
accounting for variation at level three; we therefore identify their
model as a three level model with fixed effects at level three.

\emph{Sample sizes.} We assume equal size randomization blocks and
schools, as is typical of most power analysis packages. For our context,
this gives about three schools per randomization block; we can later do
a sensitivity check where we increase and decrease this to see how power
changes. The Diplomas Now report states there were 14,950 students,
yielding around 258 students per school. Normally we would use the
geometric means of schools per randomization block and students per
school as our design parameters, but that information is not available
in the report. We assume 50\% of the schools are treated; our
calculations will be approximate here in that we could not actually
treat exactly 50\% in small and odd-sized blocks.

\emph{Control variables.} We next need values for the \(R^2\) of the
possible covariates. The report does not provide these quantities, but
it does mention covariate adjustment in the presentation of the model.
Given the types of outcomes we are working with, it is unlikely that
there are highly predictive individual-level covariates, but our prior
year school-average attendance measures are likely to be highly
predictive of corresponding school-average outcomes. We thus set
\(R^2_1 = 0.1\) and \(R^2_2 = 0.5\). We assume five covariates at level
one and three at level two; this decision, especially for level one,
usually does not matter much in practice, unless sample sizes are very
small (the number of covariates along with sample size determine the
degrees of freedom for our planned tests).

\emph{ICCs.} We also need a measure of where variation occurs: the
individual, the school, or the randomization block level. We capture
this with Intraclass Correlation Coefficients (ICCs), one for level two
and one for level three. ICC measures specify overall variation in
outcome across levels: e.g., do we see relatively homogeneous students
within schools that are quite different, or are the schools generally
the same with substantial variation within them? We typically would
obtain ICCs from pilot data or external reports on similar data. We here
specify a level-two ICC of 0.05, and a level-three ICC of 0.40. We set a
relatively high level three ICC as we expect our school type by district
blocks to isolate variation; in particular we might believe middle and
high school attendance rates would be markedly different.

\emph{Correlation of outcomes.} We finally need to specify the number
and relationship among our outcomes and associated test-statistics. For
illustration, we select attendance as our outcome group. We assume we
have five different attendance measures. The main decision regarding
outcomes is the correlation of our test statistics. As a rough proxy, we
use the correlation of the outcomes at the level of randomization; in
our case this would be the correlation of school-average attendance
within block. We believe the attendance measures would be fairly
related, so we select \texttt{rho\ =\ 0.40} for all pairs of outcomes.
This value is an estimate, and we strongly encourage exploration of
different values of this correlation choice as a sensitivity check for
any conducted analysis. Selecting a candidate rho is difficult, and will
be new for those only familiar with power analyses of single outcomes;
we need to more research in the field, both empirical and theoretical,
to further guide this choice.

If the information were available, we could specify different values for
the design parameters such as the \(R^2\)s and \(ICC\)s for each
outcome, if we thought they had different characteristics; for
simplicity we do not do this here. The \texttt{PUMP} package also allows
specifying different pairwise correlations between the test statistics
of the different outcomes via a matrix of \(\rho\)s rather than a single
\(\rho\); also for simplicity, we do not do that here.

Once we have established initial values for all needed parameters, we
first conduct a baseline calculation, and then explore how MDES or other
quantities change as these parameters change.

\subsection{Calculating MDES}

We now have an initial planned design, with a set number of schools and
students. But is this a large enough experiment to reliably detect
reasonably sized effects? To answer this question we calculate the
minimal detectable effect size (MDES), given our planned analytic
strategy, for our outcomes.

To identify the MDES of a given setting we use the \texttt{pump\_mdes}
method, which conducts a search for a MDES that achieves a target level
of power. The MDES depends on all the design and model parameters
discussed above, but also depends on the type of power and target level
of power we are interested in. For example, we could determine what size
effect we can reliably detect on our first outcome, after multiplicity
adjustment. Or, we could determine what size effects we would need
across our five outcomes to reliably detect an impact on at least one of
them. We set our goal by specifying the type (\texttt{power.definition})
and desired power (\texttt{target.power}).

Here, for example, we find the MDES if we want an 80\% chance of
detecting an impact on our first outcome when using the Holm procedure:

\begin{Shaded}
\begin{Highlighting}[]
\NormalTok{m }\OtherTok{\textless{}{-}} \FunctionTok{pump\_mdes}\NormalTok{(}
  \AttributeTok{d\_m =} \StringTok{"d3.2\_m3fc2rc"}\NormalTok{,      }\CommentTok{\# choice of design and analysis strategy}
  \AttributeTok{MTP =} \StringTok{"HO"}\NormalTok{,                 }\CommentTok{\# multiple testing procedure}
  \AttributeTok{target.power =} \FloatTok{0.80}\NormalTok{,          }\CommentTok{\# desired power}
  \AttributeTok{power.definition =} \StringTok{"D1indiv"}\NormalTok{, }\CommentTok{\# power type}
  \AttributeTok{M =} \DecValTok{5}\NormalTok{,                        }\CommentTok{\# number of outcomes}
  \AttributeTok{J =} \DecValTok{3}\NormalTok{,                        }\CommentTok{\# number of schools per block}
  \AttributeTok{K =} \DecValTok{21}\NormalTok{,                       }\CommentTok{\# number districts}
  \AttributeTok{nbar =} \DecValTok{258}\NormalTok{,                   }\CommentTok{\# average number of students per school}
  \AttributeTok{Tbar =} \FloatTok{0.50}\NormalTok{,                  }\CommentTok{\# prop treated}
  \AttributeTok{alpha =} \FloatTok{0.05}\NormalTok{,                 }\CommentTok{\# significance level}
  \AttributeTok{numCovar.1 =} \DecValTok{5}\NormalTok{,               }\CommentTok{\# number of covariates at level 1}
  \AttributeTok{numCovar.2 =} \DecValTok{3}\NormalTok{,               }\CommentTok{\# number of covariates at level 2}
  \AttributeTok{R2.1 =} \FloatTok{0.1}\NormalTok{, }\AttributeTok{R2.2 =} \FloatTok{0.7}\NormalTok{,       }\CommentTok{\# explanatory power of covariates for each level}
  \AttributeTok{ICC.2 =} \FloatTok{0.05}\NormalTok{, }\AttributeTok{ICC.3 =} \FloatTok{0.4}\NormalTok{,    }\CommentTok{\# intraclass correlation coefficients}
  \AttributeTok{rho =} \FloatTok{0.4}\NormalTok{ )                   }\CommentTok{\# how correlated outcomes are}
\end{Highlighting}
\end{Shaded}

The results are easily made into a nice table via the \texttt{knitr}
\texttt{kable()} command:

\begin{Shaded}
\begin{Highlighting}[]
\NormalTok{knitr}\SpecialCharTok{::}\FunctionTok{kable}\NormalTok{( m, }\AttributeTok{digits =} \DecValTok{3}\NormalTok{, }\AttributeTok{booktabs =} \ConstantTok{TRUE}\NormalTok{,}
              \AttributeTok{position =} \StringTok{"h!"}\NormalTok{, }\AttributeTok{caption =} \StringTok{"MDES Estimate"}\NormalTok{ ) }\SpecialCharTok{\%\textgreater{}\%}
\NormalTok{  kableExtra}\SpecialCharTok{::}\FunctionTok{kable\_styling}\NormalTok{( }\AttributeTok{position =} \StringTok{"center"}\NormalTok{ )}
\end{Highlighting}
\end{Shaded}

\begin{table}[h!]

\caption{\label{tab:unnamed-chunk-3}MDES Estimate}
\centering
\begin{tabular}[t]{lrr}
\toprule
MTP & Adjusted.MDES & D1indiv.power\\
\midrule
HO & 0.106 & 0.797\\
\bottomrule
\end{tabular}
\end{table}

The answers \texttt{pump\_mdes()} gives are approximate as we are
calculating them via monte carlo simulation. To control accuracy, we can
specify a tolerance (\texttt{tol}) of how close the estimated power
needs to be to the desired target along with the number of iterations in
the search sequence (via \texttt{start.tnum}, \texttt{tnum}, and
\texttt{final.tnum}). The search will stop when the estimated power is
within \texttt{tol} of \texttt{target.power}, as estimated by
\texttt{final.tnum} iterations. Lower tolerance and higher \texttt{tnum}
values will give more exact results (and take more computational time).

Changing the type of power is straightforward: for example, to identify
the MDES for \(1-\)minimal power (i.e., what effect do we have to assume
across all observations such that we will find at least one significant
result with 80\% power?), we simply update our result with our new power
definition:

\begin{Shaded}
\begin{Highlighting}[]
\NormalTok{m2 }\OtherTok{\textless{}{-}} \FunctionTok{update}\NormalTok{( m, }\AttributeTok{power.definition =} \StringTok{"min1"}\NormalTok{ )}
\end{Highlighting}
\end{Shaded}

\begin{verbatim}
#> mdes result: d3.2_m3fc2rc d_m with 5 outcomes
#>   target min1 power: 0.80
#>  MTP Adjusted.MDES min1.power   SE
#>   HO    0.08144353    0.79075 0.01
#>  (10 steps in search)
\end{verbatim}

The \texttt{update()} method can replace any number of arguments of the
prior call with new ones, making exploration of different scenarios very
straightforward.\footnote{The update() method re-runs the underlying call of pump\_mdes(), pump\_sample(), or pump\_power() with the revised set of design parameters. You can even change which call to use via the type parameter.}
Our results show that if we just want to detect at least one outcome
with 80\% power, we can reliably detect an effect of size \(0.08\)
(assuming all five outcomes have effects of at least that size).

When estimating power for multiple outcomes, it is important to consider
cases where some of the outcomes in fact have null, or very small,
effects, to hedge against circumstances such as one of the outcomes not
being well measured. One way to do this is to set two of our outcomes to
no effect with the \texttt{numZero} parameter:

\begin{Shaded}
\begin{Highlighting}[]
\NormalTok{m3 }\OtherTok{\textless{}{-}} \FunctionTok{update}\NormalTok{( m2, }\AttributeTok{numZero =} \DecValTok{2}\NormalTok{ )}
\end{Highlighting}
\end{Shaded}

\begin{verbatim}
#> mdes result: d3.2_m3fc2rc d_m with 5 outcomes
#>   target min1 power: 0.80
#>  MTP Adjusted.MDES min1.power SE
#>   HO    0.09050718     0.8065  0
#>  (7 steps in search)
\end{verbatim}

The MDES goes up, as expected: when there are not effects on some
outcomes, there are fewer good chances for detecting an effect.
Therefore, an increased MDES (for the nonzero outcomes) is required to
achieve the same level of desired power (80\%). Below we provide a
deeper dive into the extent to which \texttt{numZero} can effect power
estimates.

\subsection{Determining necessary sample size}

The MDES calculator tells us what we can detect given a specific design.
We might instead want to ask how much larger our design would need to be
in order to achieve a desired MDES. In particular, we might want to
determine the needed number of students per school, the number of
schools, or the number of blocks to detect an effect of a given size.
The \texttt{pump\_sample} method will search over any one of these.

Assuming we have three schools per block, we first calculate how many
blocks we would need to achieve a MDES of 0.10 for \(1-\)minimal power
(this answers the question of how big of an experiment do we need in
order to have an 80\% chance of finding at least one outcome
significant, if all outcomes had a true effect size of 0.10):

\begin{Shaded}
\begin{Highlighting}[]
\NormalTok{smp }\OtherTok{\textless{}{-}} \FunctionTok{pump\_sample}\NormalTok{(}
  \AttributeTok{d\_m =} \StringTok{"d3.2\_m3fc2rc"}\NormalTok{,}
  \AttributeTok{MTP =} \StringTok{"HO"}\NormalTok{,}
  \AttributeTok{typesample =} \StringTok{"K"}\NormalTok{,}
  \AttributeTok{target.power =} \FloatTok{0.80}\NormalTok{, }\AttributeTok{power.definition =} \StringTok{"min1"}\NormalTok{, }\AttributeTok{tol =} \FloatTok{0.01}\NormalTok{,}
  \AttributeTok{MDES =} \FloatTok{0.10}\NormalTok{, }\AttributeTok{M =} \DecValTok{5}\NormalTok{, }\AttributeTok{nbar =} \DecValTok{258}\NormalTok{, }\AttributeTok{J =} \DecValTok{3}\NormalTok{,}
  \AttributeTok{Tbar =} \FloatTok{0.50}\NormalTok{, }\AttributeTok{alpha =} \FloatTok{0.05}\NormalTok{, }\AttributeTok{numCovar.1 =} \DecValTok{5}\NormalTok{, }\AttributeTok{numCovar.2 =} \DecValTok{3}\NormalTok{,}
  \AttributeTok{R2.1 =} \FloatTok{0.1}\NormalTok{, }\AttributeTok{R2.2 =} \FloatTok{0.7}\NormalTok{, }\AttributeTok{ICC.2 =} \FloatTok{0.05}\NormalTok{, }\AttributeTok{ICC.3 =} \FloatTok{0.40}\NormalTok{, }\AttributeTok{rho =} \FloatTok{0.4}\NormalTok{ )}
\end{Highlighting}
\end{Shaded}

\begin{verbatim}
#> sample result: d3.2_m3fc2rc d_m with 5 outcomes
#>   target min1 power: 0.80
#>  MTP Sample.type Sample.size min1.power   SE
#>   HO           K          15      0.798 0.01
#>  (4 steps in search)
\end{verbatim}

We would need 15 blocks, rather than the originally specified 21, giving
45 total schools in our study, to achieve 80\% \(1-\)minimal power.

We recommend checking MDES and sample-size calculators, as the
estimation error combined with the stochastic search can give results a
bit off the target in some cases. A check is easy to do; simply run the
found design through \texttt{pump\_power()}, which directly calculates
power for a given scenario, to see if we recover our originally targeted
power (we can use \texttt{update()} and set the type to \texttt{power}
to pass all the design parameters automatically). When we do this, we
can also increase the number of iterations to get more precise estimates
of power, as well:

\begin{Shaded}
\begin{Highlighting}[]
\NormalTok{p\_check }\OtherTok{\textless{}{-}} \FunctionTok{update}\NormalTok{( smp, }\AttributeTok{type =} \StringTok{"power"}\NormalTok{, }\AttributeTok{tnum =} \DecValTok{50000}\NormalTok{,}
                   \AttributeTok{long.table =} \ConstantTok{TRUE}\NormalTok{ )}
\end{Highlighting}
\end{Shaded}

\begin{table}[h!]

\caption{\label{tab:unnamed-chunk-7}Power table}
\centering
\begin{tabular}[t]{lrr}
\toprule
power & None & HO\\
\midrule
individual outcome 1 & 0.7 & 0.53\\
individual outcome 2 & 0.7 & 0.52\\
individual outcome 3 & 0.7 & 0.53\\
individual outcome 4 & 0.7 & 0.53\\
individual outcome 5 & 0.7 & 0.53\\
\addlinespace
mean individual & 0.7 & 0.53\\
1-minimum &  & 0.81\\
2-minimum &  & 0.64\\
3-minimum &  & 0.51\\
4-minimum &  & 0.39\\
\addlinespace
complete &  & 0.33\\
\bottomrule
\end{tabular}
\end{table}

When calculating power directly, we get power for all the implemented
definitions of power applicable to the design.

In the above, the first five rows are the powers for rejecting each of
the five outcomes---they are (up to simulation error) the same since we
are assuming the same MDES and other design parameters for each. The
``mean individual'' is the mean individual power across all outcomes.
The first column is power without adjustment, and the second has our
power with the listed \(p\)-value adjustment.

The next rows show different multi-outcome definitions of power. In
particular, \texttt{1-minimum} shows the chance of rejecting at least
one hypotheses. The \texttt{complete} row shows the power to reject all
hypotheses; it is only defined if all outcomes are specified to have a
non-zero
effect.\footnote{The package does not show power for these without adjustment for multiple testing, as that power would be grossly inflated and meaningless.}

We can look at a power curve of our \texttt{pump\_sample()} call to
assess how sensitive power is to our level two sample
size:\footnote{The points on the plots show the evaluated simulation trials, with larger points corresponding to more iterations and greater precision.}

\begin{Shaded}
\begin{Highlighting}[]
\FunctionTok{plot}\NormalTok{( smp )}
\end{Highlighting}
\end{Shaded}

\begin{center}\includegraphics{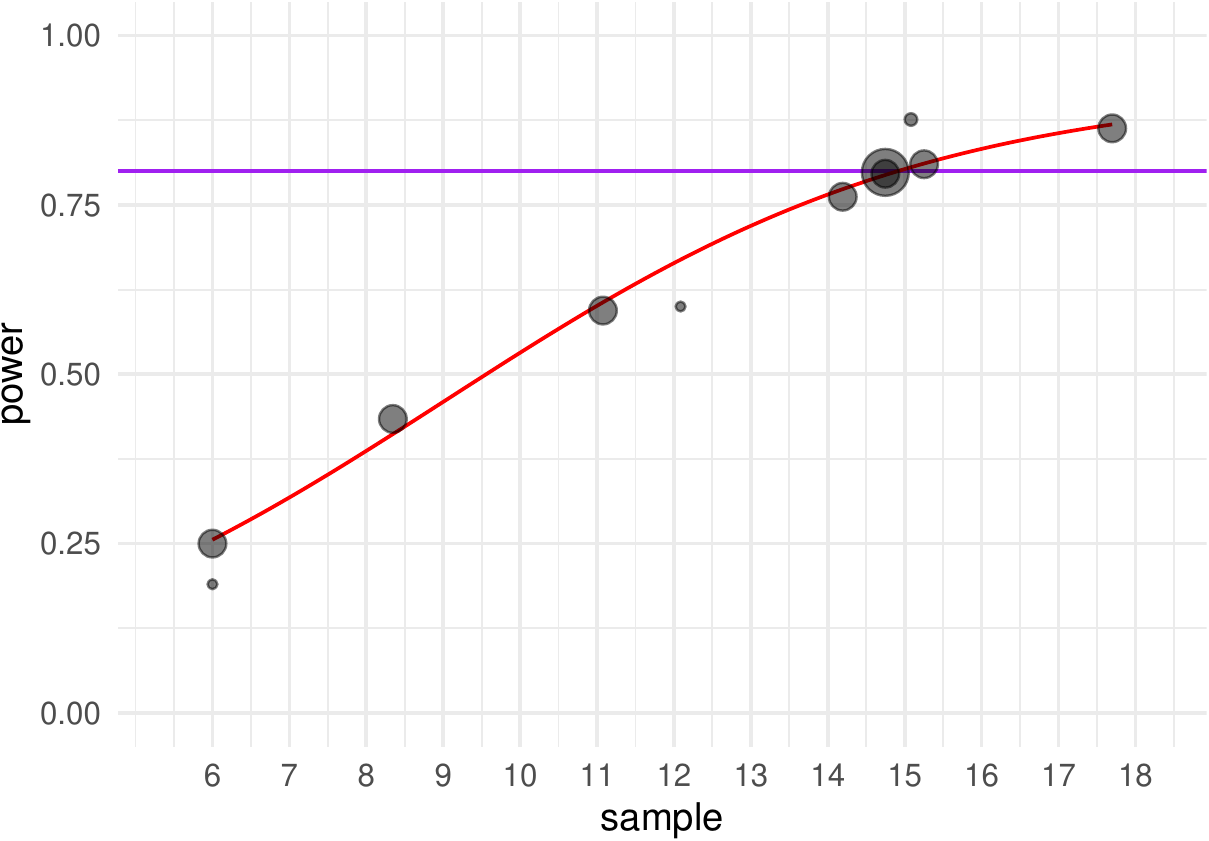} \end{center}

Though increasing \texttt{tnum} is useful for checking the power
calculation, it also increases computation time. Thus, for future
calculations we save a call with the default \texttt{tnum} to reduce
computation time.

\begin{Shaded}
\begin{Highlighting}[]
\NormalTok{pow }\OtherTok{\textless{}{-}} \FunctionTok{update}\NormalTok{( p\_check, }\AttributeTok{tnum =} \DecValTok{10000}\NormalTok{ )}
\end{Highlighting}
\end{Shaded}

\emph{Remark.} In certain settings, a wide range of sample sizes may
result in very similar levels of power. In this case, the algorithm may
return a sample size that is larger than necessary. This pattern does
not occur for the sample size at the highest level of the hierarchy, and
only for occurs for sample sizes at lower levels of the hierarchy;
e.g.~for \texttt{nbar} for all models, and for \texttt{nbar} and
\texttt{J} for three level models. In addition, due to the nature of the
search algorithm, occasionally the algorithm may not converge. For a
more detailed discussion of these challenges, see the package sample
size vignette.

\subsection{Comparing adjustment procedures}

It is easy to rerun the above using the Westfall-Young Stepdown
procedure (this procedure is much more computationally intensive to
run), or other procedures of interest. Alternatively, simply provide a
list of procedures you wish to compare. If you provide a list, the
package will re-run the power calculator for each item on the list; this
can make the overall call computationally intensive. Here we obtain
power for our scenario using Bonferroni, Holm and Westfall-Young
adjustments, and plot the results using the default \texttt{plot()}
method:

\begin{Shaded}
\begin{Highlighting}[]
\NormalTok{p2 }\OtherTok{\textless{}{-}} \FunctionTok{update}\NormalTok{( pow,}
              \AttributeTok{MTP =} \FunctionTok{c}\NormalTok{( }\StringTok{"BF"}\NormalTok{, }\StringTok{"HO"}\NormalTok{, }\StringTok{"WY{-}SD"}\NormalTok{ ),}
              \AttributeTok{tnum =} \DecValTok{5000}\NormalTok{,}
              \AttributeTok{parallel.WY.cores =} \DecValTok{2}\NormalTok{ )}
\FunctionTok{plot}\NormalTok{( p2 )}
\end{Highlighting}
\end{Shaded}

\begin{center}\includegraphics{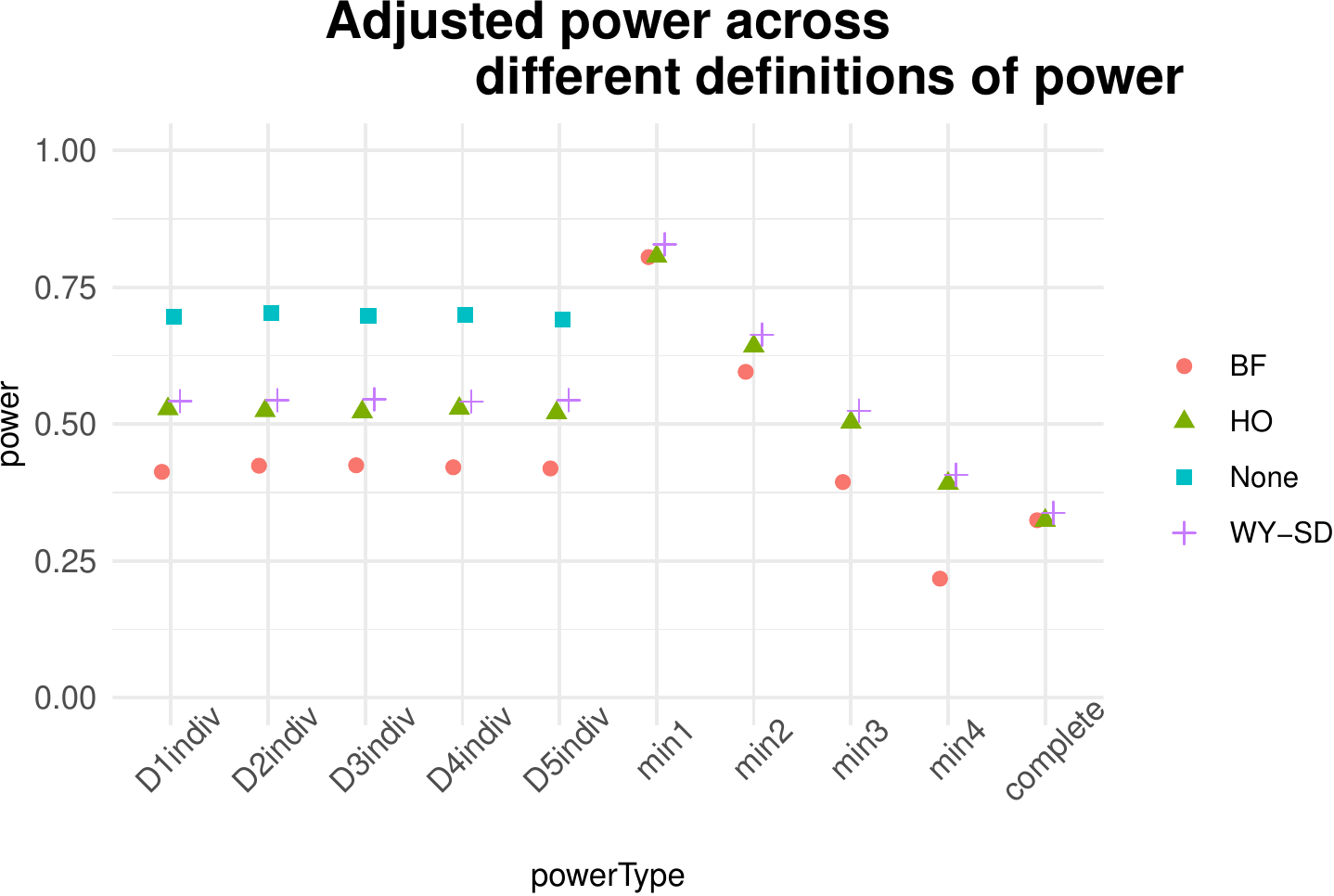} \end{center}

To speed up computation, we set an additional option,
\texttt{parallel.WY.cores}, to the number of desired cores to
parallelize the computation. We also reduce \texttt{tnum} to decrease
computation time.

The more sophisticated (and less conservative) adjustment exploits the
correlation in our outcomes (\texttt{rho\ =\ 0.4}) to provide higher
individual power. Note, however, that we do not see elevated rates for
\(1-\)minimal power. Accounting for the correlation of the test
statistics when adjusting \(p\)-values can drive some power (individual
power) up, but on the flip side \(1-\)power can be driven down as the
lack of independence between tests gives fewer chances for a significant
result. See \cite{Porter2018} for further discussion; while the paper
focuses on the multisite randomized trial context, the lessons learned
there apply to all designs as the only substantive differences between
different design and modeling choices is in how we calculate the
unadjusted distribution of their test statistics.

\subsection{Exploring sensitivity to design parameters}

Within the pump package we have two general ways of exploring design
sensitivity. The first is with \texttt{update()}, which allows for
quickly generating a single alternate scenario. To explore sensitivity
to different design parameters more systematically, use the
\texttt{grid()} functions, which calculate power, mdes, and sample size
for all combinations of a set of passed parameter values. There are two
main differences between the two approaches. First, \texttt{update()}
allows for different values of a parameter for the different outcomes;
the \texttt{grid} approach, by contrast, is more limited in this regard,
and assumes the same parameter value across different outcomes. Second,
the \texttt{grid} functions are a powerful tool for systematically
exploring many possible combinations, while \texttt{update()} only
allows the user to explore one value at a time.

We first illustrate the \texttt{update()} approach, and then turn to
illustrating \texttt{grid()} across three common areas of exploration:
Intraclass Correlation Coefficients (ICCs), the correlation of test
statistics, and the assumed number of non-zero effects. The last two are
particularly important for multiple outcome contexts.

\subsubsection{Exploring power with update()}

Update allows for a quick change of some of the set of parameters used
in a prior call; we saw \texttt{update()} used several times above. As a
further example, here we examine what happens if the ICCs are more
equally split across levels two and three:

\begin{Shaded}
\begin{Highlighting}[]
\NormalTok{p\_b }\OtherTok{\textless{}{-}} \FunctionTok{update}\NormalTok{( pow, }\AttributeTok{ICC.2 =} \FloatTok{0.20}\NormalTok{, }\AttributeTok{ICC.3 =} \FloatTok{0.25}\NormalTok{ )}
\FunctionTok{print}\NormalTok{( p\_b )}
\CommentTok{\#\textgreater{} power result: d3.2\_m3fc2rc d\_m with 5 outcomes}
\CommentTok{\#\textgreater{}   MTP D1indiv D2indiv D3indiv D4indiv D5indiv indiv.mean   min1   min2   min3}
\CommentTok{\#\textgreater{}  None  0.2474  0.2421  0.2353  0.2395  0.2431    0.24148     NA     NA     NA}
\CommentTok{\#\textgreater{}    HO  0.0991  0.0960  0.0938  0.0967  0.0927    0.09566 0.2635 0.1157 0.0579}
\CommentTok{\#\textgreater{}    min4 complete}
\CommentTok{\#\textgreater{}      NA       NA}
\CommentTok{\#\textgreater{}  0.0279   0.0215}
\CommentTok{\#\textgreater{}  0.000 \textless{}= SE \textless{}= 0.002}
\end{Highlighting}
\end{Shaded}

We immediately see that our assumption of substantial variation in level
three matters a great deal for power.

When calculating power for a given scenario, it is also easy to vary
many of our design parameters by outcome. For example, if we thought we
had better predictive covariates for our second outcome, we might try:

\begin{Shaded}
\begin{Highlighting}[]
\NormalTok{p\_d }\OtherTok{\textless{}{-}} \FunctionTok{update}\NormalTok{( pow,}
               \AttributeTok{R2.1 =} \FunctionTok{c}\NormalTok{( }\FloatTok{0.1}\NormalTok{, }\FloatTok{0.3}\NormalTok{, }\FloatTok{0.1}\NormalTok{, }\FloatTok{0.2}\NormalTok{, }\FloatTok{0.2}\NormalTok{ ),}
               \AttributeTok{R2.2 =} \FunctionTok{c}\NormalTok{( }\FloatTok{0.4}\NormalTok{, }\FloatTok{0.8}\NormalTok{, }\FloatTok{0.3}\NormalTok{, }\FloatTok{0.2}\NormalTok{, }\FloatTok{0.2}\NormalTok{ ) )}
\FunctionTok{print}\NormalTok{( p\_d )}
\CommentTok{\#\textgreater{} power result: d3.2\_m3fc2rc d\_m with 5 outcomes}
\CommentTok{\#\textgreater{}   MTP D1indiv D2indiv D3indiv D4indiv D5indiv indiv.mean   min1   min2  min3}
\CommentTok{\#\textgreater{}  None  0.4362  0.8541  0.3891   0.349  0.3407    0.47382     NA     NA    NA}
\CommentTok{\#\textgreater{}    HO  0.2469  0.6552  0.2153   0.191  0.1887    0.29942 0.7155 0.3782 0.213}
\CommentTok{\#\textgreater{}    min4 complete}
\CommentTok{\#\textgreater{}      NA       NA}
\CommentTok{\#\textgreater{}  0.1226   0.0878}
\CommentTok{\#\textgreater{}  0.001 \textless{}= SE \textless{}= 0.002}
\end{Highlighting}
\end{Shaded}

Notice how the individual powers are heavily impacted. The \(d\)-minimal
powers naturally take the varying outcomes into account as we are
calculating a joint distribution of test statistics that will have the
correct marginal distributions based on these different design parameter
values.

After several \texttt{update()}s, we may lose track of where we are; to
find out, we can always check details with \texttt{print\_design()} or
\texttt{summary()}:

\begin{Shaded}
\begin{Highlighting}[]
\FunctionTok{summary}\NormalTok{( p\_d )}
\CommentTok{\#\textgreater{} power result: d3.2\_m3fc2rc d\_m with 5 outcomes}
\CommentTok{\#\textgreater{}   MDES vector: 0.1, 0.1, 0.1, 0.1, 0.1}
\CommentTok{\#\textgreater{}   nbar: 258  J: 3    K: 15   Tbar: 0.5}
\CommentTok{\#\textgreater{}   alpha: 0.05    }
\CommentTok{\#\textgreater{}   Level:}
\CommentTok{\#\textgreater{}     1: R2: 0.1 / 0.3 / 0.1 / 0.2 / 0.2 (5 covariates)}
\CommentTok{\#\textgreater{}     2: R2: 0.4 / 0.8 / 0.3 / 0.2 / 0.2 (3 covariates)    ICC: 0.05   omega: 0}
\CommentTok{\#\textgreater{}     3:   fixed effects   ICC: 0.4    omega: 0}
\CommentTok{\#\textgreater{}   rho = 0.4}
\CommentTok{\#\textgreater{}   MTP D1indiv D2indiv D3indiv D4indiv D5indiv indiv.mean   min1   min2  min3}
\CommentTok{\#\textgreater{}  None  0.4362  0.8541  0.3891   0.349  0.3407    0.47382     NA     NA    NA}
\CommentTok{\#\textgreater{}    HO  0.2469  0.6552  0.2153   0.191  0.1887    0.29942 0.7155 0.3782 0.213}
\CommentTok{\#\textgreater{}    min4 complete}
\CommentTok{\#\textgreater{}      NA       NA}
\CommentTok{\#\textgreater{}  0.1226   0.0878}
\CommentTok{\#\textgreater{}  0.001 \textless{}= SE \textless{}= 0.002}
\CommentTok{\#\textgreater{}  (tnum = 10000)}
\end{Highlighting}
\end{Shaded}

Using update allows for targeted comparison of major choices, but if we
are interested in how power changes across a range of options, we can do
this more systematically with the \texttt{grid()} functions, as we do
next.

\subsubsection{Exploring the impact of the ICC}

We above saw that the ICC does impact power considerably. We next extend
this evaluation by exploring a range of options for both level two and
three ICCs, so we can assess whether our power is sufficient across a
set of plausible values. The \texttt{update\_grid()} call makes this
straightforward: we pass our baseline scenario along with lists of
parameters to additionally explore:

We can then easily visualize the variation in min1 power by calling
\texttt{plot()} on the object.

\begin{Shaded}
\begin{Highlighting}[]
\NormalTok{grid }\OtherTok{\textless{}{-}} \FunctionTok{update\_grid}\NormalTok{( pow,}
            \AttributeTok{ICC.2 =} \FunctionTok{seq}\NormalTok{( }\DecValTok{0}\NormalTok{, }\FloatTok{0.3}\NormalTok{, }\FloatTok{0.05}\NormalTok{ ),}
            \AttributeTok{ICC.3 =} \FunctionTok{seq}\NormalTok{( }\DecValTok{0}\NormalTok{, }\FloatTok{0.60}\NormalTok{, }\FloatTok{0.20}\NormalTok{ ) )}

\FunctionTok{plot}\NormalTok{( grid, }\AttributeTok{power.definition =} \StringTok{"min1"}\NormalTok{ )}
\end{Highlighting}
\end{Shaded}

\begin{center}\includegraphics{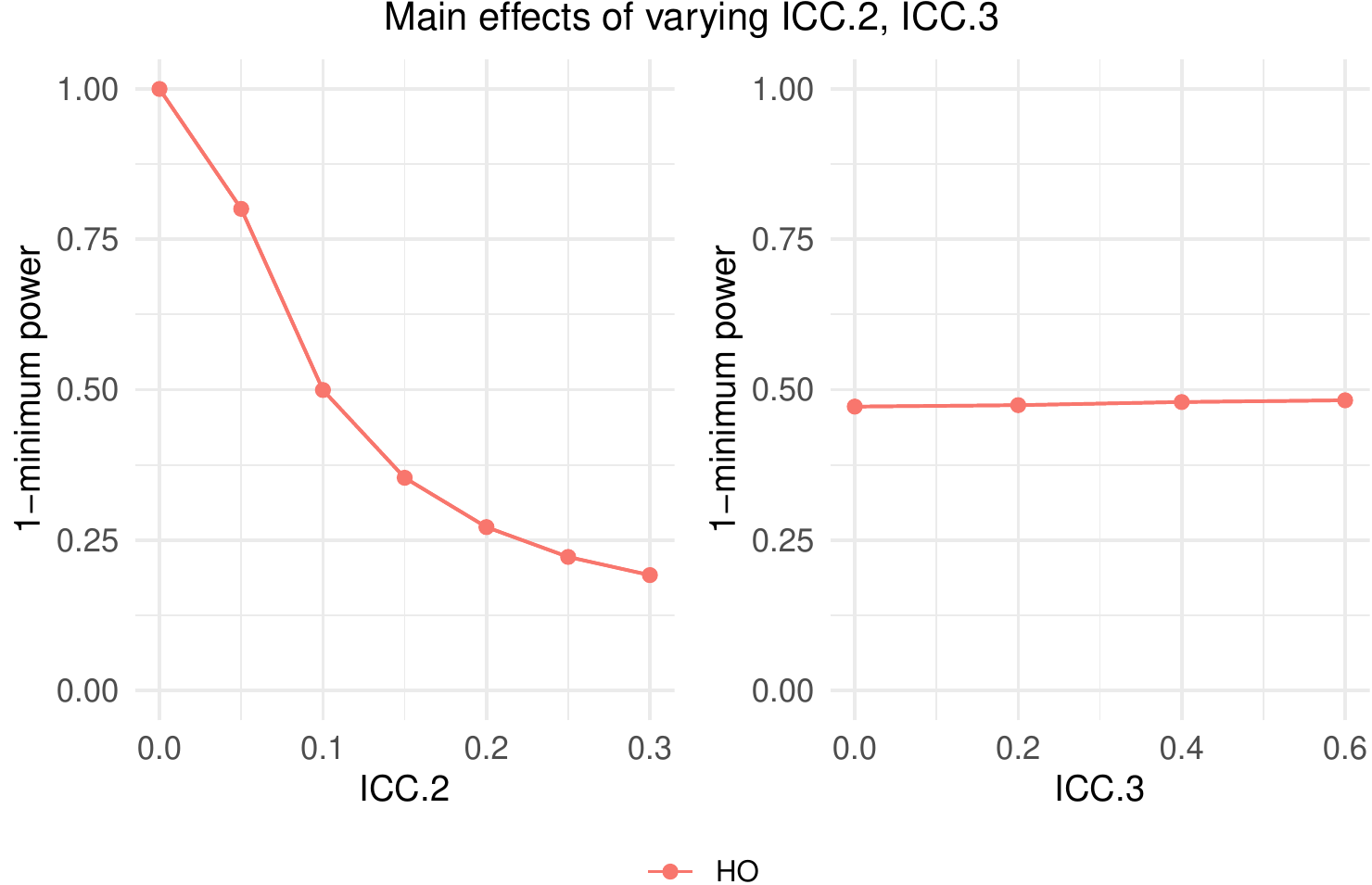} \end{center}

Note that in addition to \texttt{update\_grid()}, there are also base
functions \texttt{pump\_power\_grid()}, \texttt{pump\_mdes\_grid()}, and
\texttt{pump\_sample\_grid()}.

We see that higher ICC.2 radically reduces power to detect anything and
ICC.3 does little. To understand why, we turn to our standard error
formula for this design and model: \[
\begin{aligned}
SE( \hat{\tau} ) = \sqrt{
\frac{\text{ICC}_{2}(1 - R^2_{2})}{\bar{T}(1 - \bar{T}) JK} +
\frac{(1-\text{ICC}_{2} - \text{ICC}_{3})(1-R^2_{1})}{\bar{T}(1 - \bar{T}) J K\bar{n}} } .
\end{aligned}
\] In the above, the \(\bar{n} = 258\) students per group makes the
second term very small compared to the first, regardless of the ICC.3
value. The first term, however, is a direct scaling of ICC.2; changing
it will change the standard error, and therefore power, a lot. All
provided designs and models implemented in the package are discussed,
along with corresponding formula such as these, in our technical
supplement accompanying this paper and package.

For grid searches we recommend reducing the number of permutations, via
\texttt{tnum}, to speed up computation. As \texttt{tnum} shrinks, we
will get increasingly rough estimates of power, but even these rough
estimates can help us determine trends.

The \texttt{grid()} functions provide easy and direct ways of exploring
how power changes as a function of the design parameters. We note,
however, that in order to keep syntax simple, they do not allow
different design parameters, including MDES, by outcome. This is to keep
package syntax simpler. When faced with contexts where it is believed
that these parameters do vary, we recommend using average values for the
broader searches, and then double-checking a small set of potential
final designs with the \texttt{pump\_power()} method.

\subsubsection{Exploring the impact of rho}

The correlation of test statistics, \(\rho\), is a critical parameter
for how power will play out across the multiple tests. For example, with
Westfall-Young, we saw that the correlation can improve our individual
power, as compared to Bonferroni. We might not know what will happen to
\(2-\)minimal power, however: on one hand, correlated statistics make
individual adjustment less severe, and on the other correlation means we
succeed or fail all together. We can explore this question relatively
easily by letting \texttt{rho} vary as so:

\begin{Shaded}
\begin{Highlighting}[]
\NormalTok{gridRho }\OtherTok{\textless{}{-}} \FunctionTok{update\_grid}\NormalTok{( pow,}
              \AttributeTok{MTP =} \FunctionTok{c}\NormalTok{( }\StringTok{"BF"}\NormalTok{, }\StringTok{"WY{-}SD"}\NormalTok{ ),}
              \AttributeTok{rho =} \FunctionTok{seq}\NormalTok{( }\DecValTok{0}\NormalTok{, }\FloatTok{0.9}\NormalTok{, }\AttributeTok{by =} \FloatTok{0.15}\NormalTok{ ),}
              \AttributeTok{tnum =} \DecValTok{500}\NormalTok{,}
              \AttributeTok{B =} \DecValTok{3000}\NormalTok{ )}
\end{Highlighting}
\end{Shaded}

We then plot our results.

\begin{Shaded}
\begin{Highlighting}[]
\FunctionTok{plot}\NormalTok{( gridRho )}
\end{Highlighting}
\end{Shaded}

\begin{center}\includegraphics{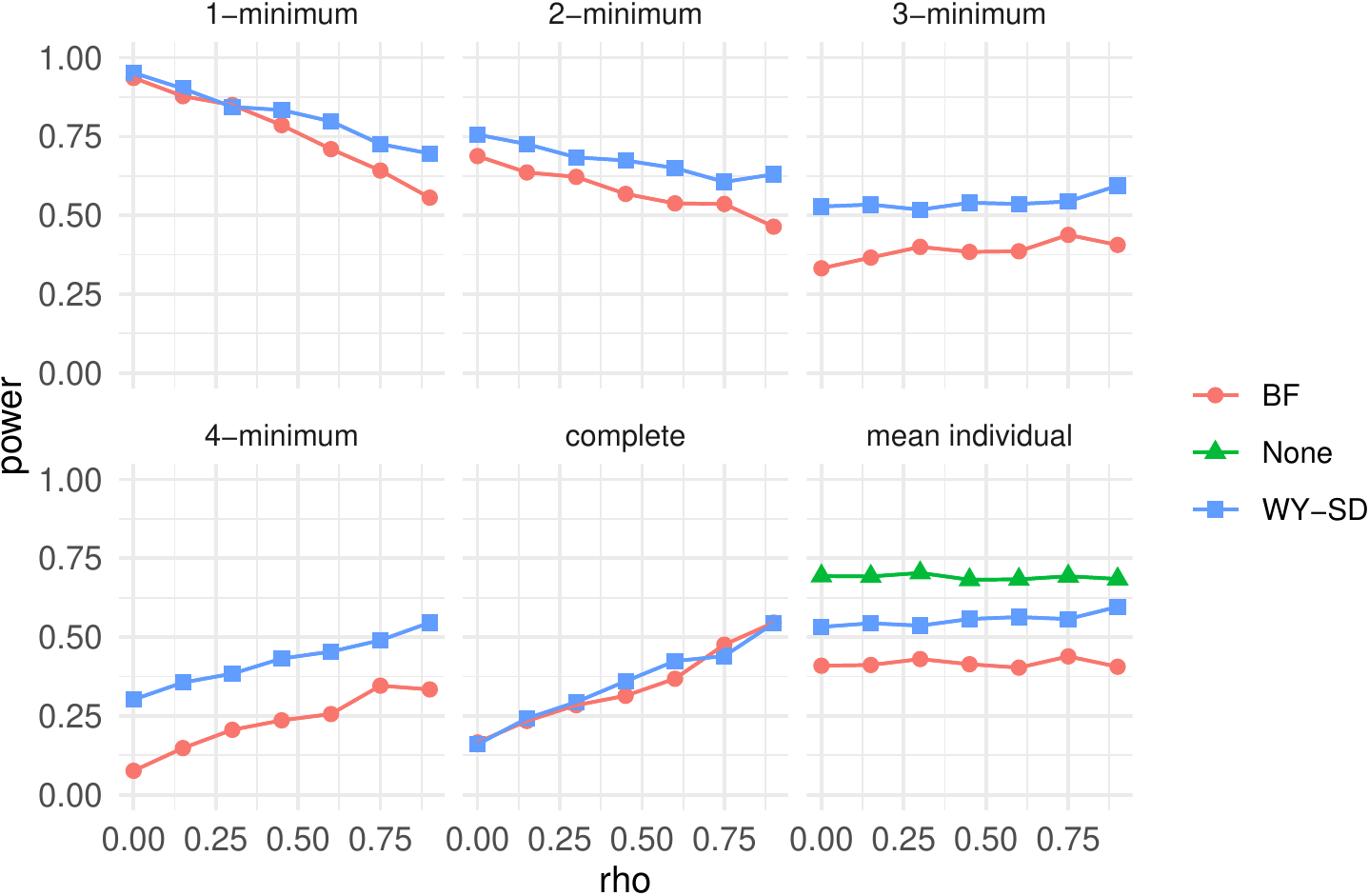} \end{center}

First, we see the benefit of the Westfall-Young single-step procedure is
minimal, as compared to Bonferroni. Second, the impact on individual
adjustment is flat, as anticipated. Third, across a very broad range of
rho, we maintain good \(1-\)minimal power. Complete power climbs as
correlation increases, and \(2-\)minimal power is generally unchanged.

\subsubsection{Exploring the impact of null outcomes}

We finally explore varying the number of outcomes with no effects. This
exploration is an important way to hedge a design against the
possibility that some number of the identified outcomes are measured
poorly, or are simply not impacted by treatment. We use a grid search,
varying the number of outcomes that have no treatment impact via the
\texttt{numZero} design parameter:

\begin{Shaded}
\begin{Highlighting}[]
\NormalTok{gridZero }\OtherTok{\textless{}{-}} \FunctionTok{update\_grid}\NormalTok{( pow,}
                           \AttributeTok{numZero =} \DecValTok{0}\SpecialCharTok{:}\DecValTok{4}\NormalTok{,}
                         \AttributeTok{M =} \DecValTok{5}\NormalTok{ )}
\FunctionTok{plot}\NormalTok{( gridZero, }\AttributeTok{nrow =} \DecValTok{1}\NormalTok{ )}
\end{Highlighting}
\end{Shaded}

\begin{center}\includegraphics{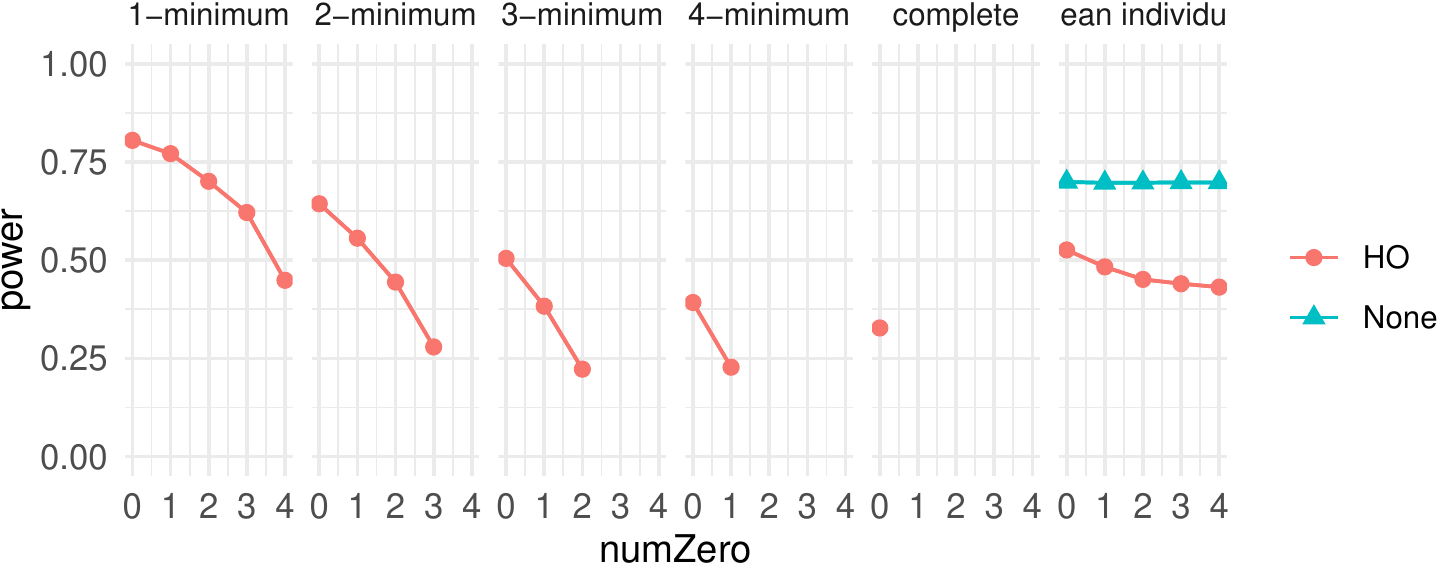} \end{center}

There are other ways of exploring the impact of weak or null effects on
some outcomes. In particular, the \texttt{pump\_power()} and
\texttt{pump\_sample()} methods allow the researcher to provide an MDES
vector with different values for each outcome, including 0s for some
outcomes. The \texttt{grid()} functions, by contrast, take a single MDES
value for the non-null outcomes, with a separate specification of how
many of the outcomes are 0. (This single value plus \texttt{numZero}
parameter also works with \texttt{pump\_power()} if desired.)

\section{Conclusion}
\label{sec:conclusion}

We introduce the power under multiplicity project (\texttt{PUMP})
\texttt{R} package, which estimates power for multi-level randomized
control trials with multiple outcomes. \texttt{PUMP} allows users to
estimate power, MDES, and sample size requirements for a wide variety of
commonly used RCT designs and models across different definitions of
power and applying different MTPs. The functionality of \texttt{PUMP}
fills an important gap, as existing tools do not allow researchers to
conduct power, MDES or sample size calculations when applying a MTP. An
online interface is also available at
\url{https://public.mdrc.org/pump/}.

The main advantage of the \texttt{PUMP} package is to provide easily
accessible estimation procedures so that users can properly account for
power when making adjustments for multiple hypothesis testing. However,
one of the additional strengths of the package is the ease with which a
user can explore the impact of different designs, models, and
assumptions on power, MDES or sample size. Even if a user is only
interested in a single outcome, \texttt{PUMP} provides useful
functionality for more robust power calculations. A user can and should
try a range of parameter values to determine the sensitivity of the
power of their study to different assumptions; this package simplifies
that process.

In addition to this paper, there is a variety of supporting information.

\begin{itemize}
\tightlist
\item
  The package is available on CRAN,
  \url{https://CRAN.R-project.org/package=PUMP}.
\item
  The code is available on GitHub, \url{https://github.com/MDRCNY/PUMP}.
\item
  The Technical Appendix contains detailed information about each design
  and model, the assumed data generating process, and understanding
  parameters such as \(ICC\) and \(\omega\). It is a useful reference
  not just for users of the package, but also as a general summary of
  multi-level models.
\item
  The package has an additional vignette on understanding sample size
  calculations, which present unique challenges.
\item
  The package has supplementary functions that allow a user to simulate
  data from multi-level models. Although these functions are not
  directly related to the power calculations, we provide them as a
  potentially useful tool. A short vignette explains these functions.
\item
  The code and results for validating the package are in a separate
  repository, \url{https://github.com/MDRCNY/pump_validate}.
\end{itemize}

\section*{Acknowledgements}

We acknowledge the Diplomas Now team at MDRC. Development of this
package was supported by a grant from the Institute of Education
Sciences (R305D170030). We would like to thank members of the Harvard
CARES lab for their feedback on the manuscript.

\hypertarget{refs}{}
\begin{CSLReferences}{0}{0}
\end{CSLReferences}

\bibliographystyle{unsrt}
\bibliography{refs.bib}

\section*{Appendix: Validation}

In order to validate that our power estimates are working as intended,
we compared three different methods of estimating power:

\begin{itemize}
\tightlist
\item
  \texttt{PUMP}
\item
  \texttt{PowerUpR}
\item
  Monte Carlo simulations
\end{itemize}

We compute values from \texttt{PowerUpR} for individual, unadjusted
power, as \texttt{PowerUpR} only provides power estimates in that
setting. For all other types of power definitions and adjustments, we
compare \texttt{PUMP} to the estimated power obtained from full Monte
Carlo simulations. We follow the simulation approach outlined in detail
in Section \ref{sec:est_power}. After repeatedly simulating data and
calculating \(p\)-values, we calculate power and a 95\% confidence
interval, assuming a conservative standard error estimate of
\(\sqrt{0.25/S}\).

To validate the estimates, we first check that the \texttt{PUMP} and
\texttt{PowerUpR} estimates match. In some settings we expect some
discrepancies between these values because \texttt{PUMP} has different
assumptions than \texttt{PowerUpR} for certain models. For details about
differences between \texttt{PUMP} and \texttt{PowerUpR} assumptions, see
the Technical Appendixs. Second, we check that the \texttt{PUMP}
estimate is within the Monte Carlo confidence interval.

We also validate MDES and sample size calculations. For MDES, we choose
one default scenario for each design and model, then input the
already-calculated individual power and see if the output MDES is the
same as the original input MDES. Similarly, for sample size validation,
we input the already-calculated individual power and see if the output
sample size (\texttt{nbar}, \texttt{J}, and \texttt{K} depending on
design) is the same as the original input sample size.

\textbf{Simulation parameters}

In order to validate that the method works in a wide range of scenarios,
we vary the following parameters. For most scenarios, we vary only one
parameter at a time. Thus, to test varying \(\rho\), we set
\(\rho = 0.2\) with all other parameters being set to the default
values, and try another scenario with \(\rho = 0.8\) with all other
parameters being set to the default values. Table \ref{tab:val_params}
shows the default parameter values, and the other values we try to test
out varying that parameter.

\begin{table}
\centering
\begin{tabular}{l l l}
\toprule
Parameter               & Default                & Other values \\ \midrule
school size $\bar{n}$   & 50                     & 75, 100           \\
$R^2$                   & 0.1                    & 0.6               \\
$\rho$                  & 0.5                    & 0.2, 0.8          \\
MDES                    & (0.125, 0.125, 0.125)  & (0.125, 0, 0)     \\
ICC                     & 0.2                    & 0.7               \\
$\omega$                & 0.1                    & 0.8               \\
\bottomrule
\end{tabular}
\label{tab:val_params}
\caption{Validation parameters}
\end{table}

We do not vary:

\begin{itemize}
\tightlist
\item
  \texttt{M\ =\ 3}
\item
  \texttt{J} and \texttt{K} are fixed for each scenario
\item
  Scalar grand mean of control outcomes
\item
  Correlations between school random effects and impacts
\item
  \texttt{rho} informs all correlations; we keep the same correlation
  between covariates, residuals, impacts, random effects for all levels
  and across all outcomes.
\end{itemize}

\textbf{Validation results}

\begin{figure}[h!]
\centering
  \includegraphics[width=6in]{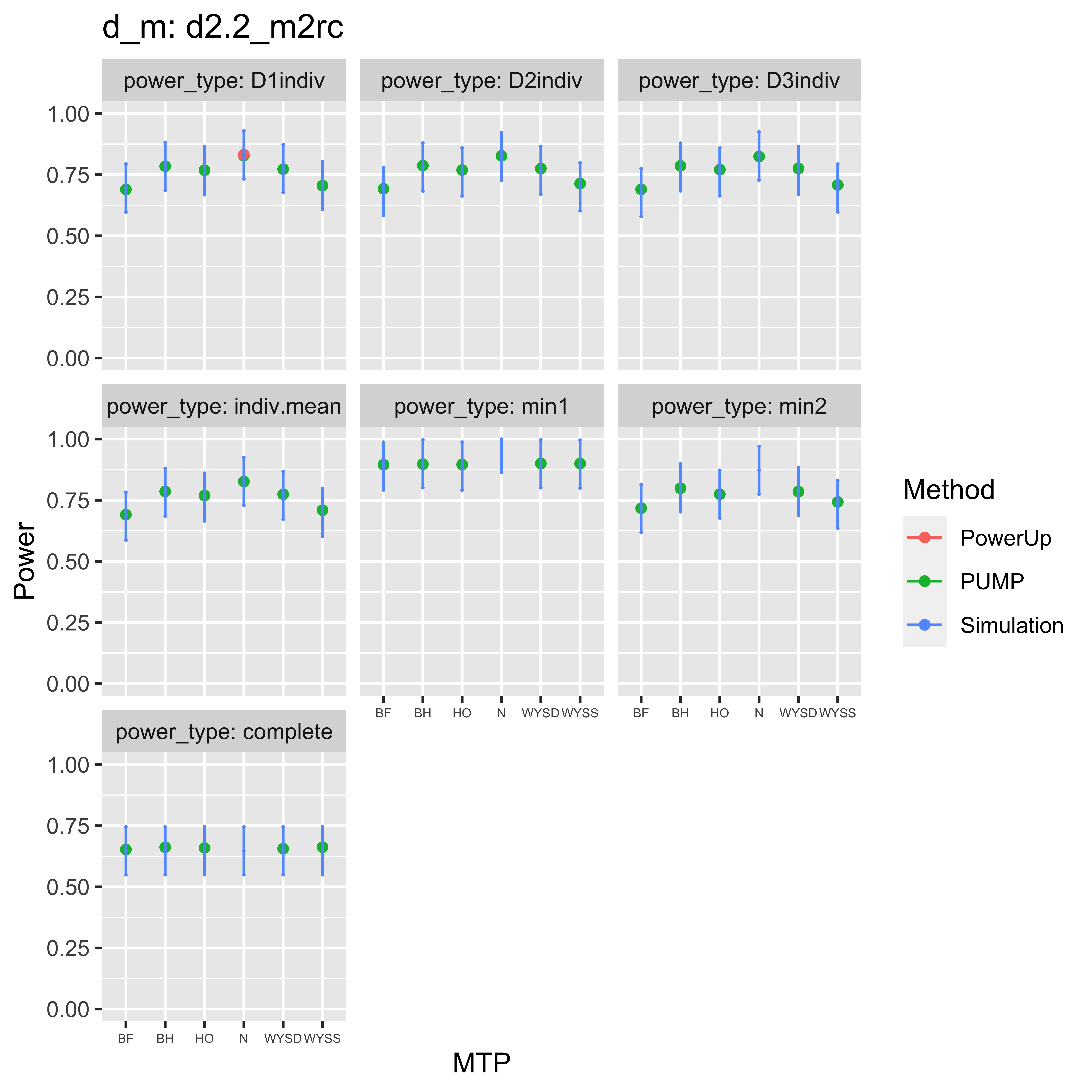}
  \label{fig:validate}
  \caption{Validation plot}
\end{figure}

Figure \ref{fig:validate} is an example of a graph we use for
validation. The green dots are \texttt{PUMP} estimate of power, the red
dot is the \texttt{PowerUpR} estimate of power, and the 95\% confidence
intervals based on the Monte Carlo simulations are shown in blue. To
validate that \texttt{PUMP} produces the expected result, we want to see
the red and green points match, and for the red point to be within the
blue intervals. Figure \ref{fig:validate} shows the results across
different types of power and different MTPs. We repeat this plot for a
variety of different parameter values for each design and model.

Next, we validate MDES and sample size calculations. We put in our found
power, and then see if the \texttt{pump\_mdes()} function returns the
MDES we originally plugged in to achieve this power. In Table
\ref{tab:mdes}, the first column shows the calculated MDES, the middle
column is the power we plugged into the calculation, and the last column
shows the MDES that we are targeting. Thus, ideally we want the first
and last columns to match.

\begin{table}[h!]
\centering
\begin{tabular}{lrrr}
\toprule
MTP & Adjusted MDES & D1indiv Power & Target MDES\\
\midrule
Bonferroni & 0.122 & 0.447 & 0.125\\
BH & 0.127 & 0.578 & 0.125\\
Holm & 0.125 & 0.540 & 0.125\\
\bottomrule
\end{tabular}
\label{tab:mdes}
\caption{MDES validation}
\end{table}

Similarly, we validate our sample size calculations. Using our found
power, we see if \texttt{pump\_sample()} returns the original sample
size. In Table \ref{tab:ss}, we are targeting a sample size of
\(J = 20\).

\begin{table}[h!]
\centering
\begin{tabular}{llrr}
\toprule
MTP & Sample.type & Sample.size & D1indiv.power\\
\midrule
Bonferroni & J & 21 & 0.500\\
BH & J & 21 & 0.580\\
Holm & J & 20 & 0.544\\
\bottomrule
\end{tabular}
\label{tab:ss}
\caption{Sample size validation}
\end{table}

\clearpage \newpage
\appendix
\addcontentsline{toc}{section}{Technical Appendix}
\setcounter{tocdepth}{3}
\part{Technical Appendix}
\parttoc

\section{Introduction}
\label{sec:app_intro}

Our package allows for power calculations across a range of common
scenarios that a user might select. Each scenario is categorized by two
choices. First, the user chooses the planned experimental design (e.g.,
clustered data, with randomization within cluster). Second, the user
makes a choice of the planned analytic model they would use for the
data, once they had it (e.g.~a multilevel model with random impacts). To
provide a concrete example, throughout we assume an education setting
where we have students (level 1) nested within schools (level 2) nested,
in the three-level case, within districts (level 3).

The \emph{design} is characterized by the number of levels of nesting
(e.g., students in schools, no districts, would be two levels) and the
level of randomization (e.g., randomization of schools would be
randomization at level two).

The \emph{model} choices are a bit more complex, and we discuss them in
detail below. In particular, for each model we support, we create a
taxonomy by noting what modeling choice is used at each level of the
model, and how covariates are used.

The outline of this appendix is as follows. In the introduction, we
provide notation, a taxonomy for models, explain user-set parameters,
and outline the power estimation strategy. The bulk of the appendix then
provides detailed information about each supported scenario, including
the assumed model and standard error formula. We then describe the data
generating process that we used for the package validation; this section
might also be useful for readers who wish to understand the models and
assumptions more deeply. The final two sections provide explicit formula
linking the user-specified parameters to the full set of parameters used
for data generation, followed by derivations of these formula.

\subsection{Model taxonomy}

To define our models, we first assume a set of observed quantities,
shown in Table\textasciitilde{}\ref{tab:observed_param}, such as sample
sizes, the outcomes, and covariates. For all notation, we use \(i\) to
index level 1 (individuals), \(j\) to index level 2 (schools), \(k\) to
index level 3 (districts), and \(m\) to index outcomes.

\begin{table}[h]
\centering
\begin{tabularx}{0.8\textwidth}{p{1.5cm} | p{10cm}}
Param                                           & Description \\ \hline
$M$                                             & Number of outcomes \\
$J$                                             & Number of level 2 units in each level 3 group (assumed constant across level 3 groups) \\
$K$                                             & Number of level 3 units \\
$\bar{n}$                                       & Number of level 1 units (assumed constant across level 2 groups) \\
$\bar{T}$                                       & Proportion of the sample that is assigned to the treatment group (assumed constant across groups) \\
$N$                                             & Total number of units $N =\sum_{k=1}^{K} \sum_{j=1}^{J}  \bar{n}$ \\ \hline
$S_{id}$                                        & Categorical variable indicating the membership of individual $i$ to a level 2 group\\
$D_{id}$                                        & Categorical variable indicating the membership of individual $i$ to a level 3 group\\ \hline
$Y_{ijmk}(0)$                                   & Potential outcome for unit $i$ in level 2 group $j$ in level 3 group $k$ for outcome $m$ given no treatment\\
$Y_{ijmk}(1)$                                   & Potential outcome for unit $i$ in level 2 group $j$ in level 3 group$k$ for outcome $m$ given treatment\\ \hline
$V_{km}$                                        & Level 3 covariates \\
$g_{3,m}$                                       & Number of level 3 covariates for outcome $m$\\
$X_{jkm}$                                       & Level 2 covariates \\
$g_{2,m}$                                       & Number of level 2 covariates for outcome $m$\\
$C_{ijkm}$                                      & Level 1 covariates \\
$g_{1,m}$                                       & Number of level 1 covariates for outcome $m$\\
\end{tabularx}
\caption{Observed quantities\label{tab:observed_param}}
\end{table}

We also assume a set of unobserved, latent parameters, shown in
Table\textasciitilde{}\ref{tab:latent_param}. These parameters include
intercepts, impacts, and coefficients on covariates.

\begin{table}[h]
\centering
\begin{tabularx}{0.8\textwidth}{p{1.5cm} | p{10cm}}
Param                                           & Description \\ \hline
$\Xi_{0,m}$                                     & Grand mean outcome under no treatment across level 3 units for outcome $m$\\
$\Xi_{1,m}$                                     & Grand mean impact across level 3 units for outcome $m$\\
$\mu_{0,km}$                                    & Grand mean outcome under no treatment across level 2 units in level 3 unit $k$ for outcome $m$\\
$\mu_{1,km}$                                    & Grand mean impact across level 2 units in level 3 unit $k$ for outcome $m$\\
$\theta_{0,jkm}$                                & Mean outcome under no treatment for level 2 unit $j$ in level 3 unit $k$ for outcome $m$\\
$\psi_{1,jkm}$                                  & Mean impact for level 2 unit $j$ in level 3 unit $k$ for outcome $m$\\ \hline
$w_{0,km}$                                      & Level 3/District intercepts\\
$w_{1,km}$                                      & Level 3/District impacts \\
$\eta^2_{0,m}$                                  & Variance of level 3 random effects for outcome $m$ \\
$\eta^2_{1,m}$                                  & Variance of level 3 impacts for outcome $m$ (cross-district treatment heterogeneity) \\
$\xi_m$                                         & Coefficient of level 3 covariates $V_{km}$ \\ \hline
$u_{0,jkm}$                                     & Level 2/School intercepts\\
$u_{1,jkm}$                                     & Level 2/School impacts \\
$\tau^2_{0,m}$                                  & Variance of level 2 random effects for outcome $m$ \\
$\tau^2_{1,m}$                                  & Variance of level 2 impacts for outcome $m$ (cross-school treatment heterogeneity) \\
$\delta_{m}$                                    & Coefficient of level 2 covariates $X_{jkm}$ \\\hline
$r_{ijkm}$                                      & Level 1/Individual intercepts \\
$\sigma^2_{m}$                                  & Variance of individual/level 1 residuals \\
$\gamma_{m}$                                    & Coefficient of individual/level 1 covariates $C_{ijkm}$ \\
\end{tabularx}
\caption{Latent parameters\label{tab:latent_param}}
\end{table}

Now, we can create a model taxonomy, based on modeling choices for
intercepts and impacts. In particular we determine for each level:

\begin{itemize}
\item Whether the level 2 and level 3 intercepts are:
\begin{itemize}
\item fixed ($u_{0,jkm}$ and $w_{0,jkm}$ are fixed effects)
\item random ($u_{0,jkm}$ and $w_{0,jkm}$ are considered to be Normally distributed, allowing for partial pooling)
\end{itemize}
\item Whether the level 2 and level 3 treatment effects are:
\begin{itemize}
\item constant, e.g. all units are modeled as having a single average impact ($u_{1,jkm}= 0$ or $w_{1,km} = 0$)
\item fixed, e.g. each unit has an individual estimated impact ($u_{1,jkm}$ are fixed effects constrained to have mean 0), with an additional mean impact.
\item random ($u_{1,jkm}$ and $w_{1,km}$ are Normally distributed around a mean impact)
\end{itemize}
\end{itemize}

In addition, for each level the user can plan to adjust for baseline
covariates, unless there are fixed effects at that level or below. Note
that in some models, PowerUp! includes treatment by covariate
interactions, allowing for, in principle, heterogeneous treatment
effects correlated with said covariates. We do not allow for this, as
including treatment by covariate interactions adds complexity with
estimation of average treatment effects, and is unlikely to help with
the precision of an overall average impact estimate. We view covariate
by treatment interactions as primarily for modeling treatment effect
heterogeneity, which is not the goal of this package or project. When
this difference between PowerUp! and our approach occurs, it is noted.

For users familiar with PowerUP, Table\textasciitilde{}\ref{tab:powerup}
provides a reference for translating notation between this document and
PowerUpR!

\begin{table}[ht!]
\centering
\begin{tabularx}{0.8\textwidth}{p{1.5cm} | p{1.75cm} | p{8cm}}
PowerUp             & PUMP                          & Description \\ \hline
$\beta_{0j}$        & $\theta_{0,jkm}$              & Mean outcome under no treatment for school $j$ in district $k$\\
$\beta_{1j}$        & $\psi_{1,jkm}$                    & Mean impact for school $j$ in district $k$ \\ \hline
$X_{ij}$            & $C_{ijkm}$                    & Individual covariates \\ 
$\beta_{2j}$        & $\gamma_{m}$                  & Coefficient vector for individual covariates $C_{ijkm}$ \\ \hline
$\gamma_{00}$       & $\mu_{0,km}$                  & Grand mean outcome under no treatment across schools in district $k$ \\
$\gamma_{10}$       & $\mu_{1,km}$                  & Grand mean impact across schools in district $k$\\
$W_{jk}$            & $X_{jkm}$                     & School covariates \\
$\gamma_{01k}$      & $\delta_{m}$                  & Coefficient vector for school covariates $X_{jkm}$ \\
$\mu_{0j}$          & $u_{0,jkm}$                   & School intercepts\\
$\mu_{1j}$          & $u_{1,jkm}$                   & School impacts\\
$\tau^2_{2|W}$      & $\tau^2_{0,m}$                    & Variance of school random effects \\
$\tau^2_{2}$        & $\tau^2_{0,m} + \delta_{m}^2$ & Overall variance of schools \\
$\tau^2_{T2|W}$     & $\tau^2_{1,m}$                    & Variance of school impacts \\
$\rho_2$            & $\text{ICC}_2$                    & Intraclass correlation (unconditional) for level 2 \\
$\omega_2$          & $\omega_2$                    & Ratio of variation of impacts to residuals     for level 2 \\
$\tau_{2T2}$        & $\boldsymbol{\kappa}^{w}$         & Correlations between school random effects and impacts \\ \hline
$\xi_{000}$         & $\Xi_{0,m}$                   & Grand mean outcome under no treatment across districts \\
$\xi_{100}$         & $\Xi_{1,m}$                   & Grand mean impact across districts\\
$V_{k}$             & $V_{km}$                      & District covariates \\
$\xi_{001}$         & $\xi_{m}$                     & Coefficient vector for district covariates $V_{km}$ \\
$\zeta_{00}$        & $w_{0,km}$                    & District intercepts\\
$\zeta_{10}$        & $w_{1,km}$                    & District impacts\\
$\tau^2_{3|V}$      & $\eta^2_{0,m}$                    & Variance of district random effects \\
$\tau^2_{3}$        & $\eta^2_{0,m} + \xi_{m}^2$        & Overall variance of districts\\
$\tau^2_{T3|V}$     & $\eta^2_{1,m}$                    & Variance of district impacts\\
$\tau_{3T3}$        & $\boldsymbol{\kappa}^{w}$         & Correlations between district random effects and impacts \\
\end{tabularx}
\caption{Correspondence with PowerUpR!\label{tab:powerup}}
\end{table}

\subsection{Scenario naming convention}

We denote the research design by \(d\), followed by the number of levels
and randomization level, so \texttt{d3.1} is a \(3\)-level design with
randomization at level \(1\). The model is denoted by \(m\), followed by
the level and the assumption for the intercepts, either \(f\) or \(r\)
and then the assumption for the treatment impacts, \(c\), \(f\), or
\(r\). For example, m3ff2rc means at level \(3\), we assume fixed
intercepts and fixed treatment impacts, and at level \(2\) we assume
random intercepts and constant treatment impacts. The full design and
model are specified by concatenating these together, e.g.~d2.1\_m3fc.

Examples:

\begin{itemize}
\item d2.1\_m2rr: 2 level, individual assignment, level 2 random intercept and random treatment effect.  Corresponds to PowerUP! blocked\_i1\_2r.
\item d3.2\_m3ff2rc: 3 level, level 2 assignment, level 3 fixed intercepts and fixed treatment effects, level 2 random intercepts and constant treatment effects. Corresponds to PowerUP! blocked\_c2\_3f.
\end{itemize}

Table\textasciitilde{}\ref{tab:names} shows the list of supported
scenarios and their corresponding names in PowerUp!

\begin{table}[h!]
\centering
\begin{tabular}{l | l}
\textbf{PowerUpR!}  & \textbf{PUMP} \\ \hline
n/a                 & d1.1\_m1c \\
bira2\_1c           & d2.1\_m2fc \\
bira2\_1f           & d2.1\_m2ff \\
bira2\_1r           & d2.1\_m2fr \\
bira3\_1r           & d3.1\_m3rr2rr \\
cra2\_2r            & d2.2\_m2rc \\
cra3\_3r            & d3.3\_m3rc2rc \\
bcra3\_2f           & d3.2\_m3ff2rc \\
n/a                 & d3.2\_m3fc2rc \\
bcra3\_2r           & d3.2\_m3rr2rc \\
\end{tabular}
\caption{Scenarios: designs and models\label{tab:names}}
\end{table}

\subsection{Derived parameters}

To calculate power, a user must choose assumed values for some of the
latent parameters. However, for certain parameters, the user may instead
have more intuition about likely values of functions of these
parameters, rather than the parameters themselves. For example, rather
than choosing the value of the coefficient for a level 3 covariate
(\(\xi_m\)), the user sets \(R_{3,m}^2\), the amount of level three
variation explained by covariates. These derived parameters, which are
functions of unobserved parameters, are listed in
Table\textasciitilde{}\ref{tab:derived_param}.

\begin{table}[h]
\centering
\begin{tabularx}{0.8\textwidth}{p{1.5cm} | p{10cm}}
Param               & Description                                                                               \\ \hline
$ES_{m}$            & treatment impact in effect size units                                                     \\
$\text{ICC}_{3,m}$  & level 3 (district) intraclass correlation                                                 \\
$\omega_{3,m}$      & ratio of variation of district impacts to district intercepts                             \\
$\text{ICC}_{2,m}$  & level 2 (school) intraclass correlation                                                   \\
$\omega_{2,m}$      & ratio of variation of school impacts to school intercepts                                 \\
$R_{3,m}^2$         & percent of district variation explained by level 3 (district) covariates  $V_{km}$            \\
$R_{2,m}^2$         & percent of school variation explained by level 2 (school) covariates  $X_{jkm}$           \\
$R_{1,m}^2$         & percent of individual variation explained by level 1 (individual) covariates $C_{ijkm}$   \\
\end{tabularx}
\caption{Derived parameters\label{tab:derived_param}}
\end{table}

We now provide further clarification on these derived parameters. For a
more detailed discussion of these expressions, see
Section\textasciitilde{}\ref{sec:tune}.

To keep the formula for intraclass correlation coefficients (ICCs) and
\(R^2\) terms simple and clear, we assume all covariates are unit
variance and group-mean centered. In particular, group-mean centering
means covariates only explain variation at their level; in practice, raw
lower level covariates can explain variation in higher levels, if they
systematically differ by group. For using the formula, however,
researchers are welcome to input total \(R^2\) values that include the
full explanatory power of all covariates for a given level.

The \(\text{ICC}\)s are unconditional Intraclass Correlations, meaning
they include the variation explained by covariates. Because we assume
all covariates are group-mean centered and have unit variance, we get
the pairing structure of terms in the equations below.

\begin{align}
\text{ICC}_{3,m} &= \frac{Var(\mu_{0,km})}{ Var(Y_{ijkm}(0))} = \frac{\xi^2_m + \eta_{0,m}^2}{\left(\xi_m^2 +  \eta^2_{0,m}\right) + \left(\delta_m^2  + \tau^2_{0,m}\right) + \left(\gamma_m^2 + \sigma^2_m\right)}\\
\text{ICC}_{2,m} &= \frac{Var(\theta_{0,jkm} \mid \mu_{0,km})}{ Var(Y_{ijkm}(0))} = \frac{\delta_m^2  + \tau_{0,m}^2}{ \left(\xi_m^2 +  \eta^2_{0,m}\right) + \left(\delta_m^2  + \tau^2_{0,m}\right) + \left(\gamma_m^2 + \sigma^2_m\right)}
\end{align}

The quantity \(\omega\) is the ratio between the variation in average
impact of a unit and the variation in the control-side mean of a unit.

\begin{align}
\omega_{3,m} &=  \frac{Var(\mu_{1,jkm})}{Var(\mu_{0,km})} = \frac{\eta^2_{1,m}}{\xi^2_m + \eta_{0,m}^2}\\
\omega_{2,m} &=  \frac{Var(\psi_{1,jkm} \mid \mu_{1,km})}{Var(\theta_{0,jkm} \mid \mu_{0,km})} = \frac{\tau^2_{1,m}}{\delta_m^2 + \tau^2_{0,m}}
\end{align}

The \(R^2\) expressions are the percent of variation at a particular
level predicted by covariates. The group-mean centering makes these
formula only involve covariates at the same level as the \(R^2\); this
is a simplification of convenience. The standard error formula for the
models in the remainder of the document are general, however, and a user
would not need to group-mean center or rescale any covariate in
practice.

\begin{align}
R_{3,m}^2 &= 1 - \frac{Var(w_{0,km})}{Var(\mu_{0,km})} = 1 - \frac{\eta^2_{0,m}}{\xi_m^2 + \eta^2_{0,m}} \\
R_{2,m}^2 &= 1 - \frac{Var(u_{0,jkm})}{Var(\theta_{0,jkm} \mid D_{id})} = 1 - \frac{\tau^2_{0,m}}{\delta_m^2 + \tau^2_{0,m}}\\
R^2_{1,m} &= 1 - \frac{Var(r_{ijkm})}{Var( Y_{ijkm}(0) \mid S_{id}, D_{id})} = 1 - \frac{ \sigma^2_m }{ \gamma_m^2 + \sigma^2_m }
\end{align}

\subsection{Power estimation strategy}

The same strategy is followed for all designs. First, we lay out a model
for our outcomes, \(Y_{ijkm}\). Next, we calculate the standard error of
the average treatment effect estimate, \(\hat{\psi}_m\). When expressing
the estimated treatment effect as an effect size, the standard error is
given by: \begin{align}\label{eqn:qm}
Q_m \equiv \text{SE}\left(\hat{\text{ES}}_m\right) &= \text{SE}\left(\frac{\hat{\psi}_m}{\mbox{VAR}}\right) = \frac{1}{VAR} \text{SE}\left(\hat{\psi}_m\right),
\end{align} where \(VAR\) is some ``Index Variation'\,' that we are
measuring our impacts against.

When analyzing actual data, we would, to estimate \(Q_m\), plug in known
values for \(\bar{T}\), \(J\), and \(\bar{n}\). Any other parameters are
replaced by sample estimates. Then, when testing the \(m^{\text{th}}\)
null hypothesis, \(\text{ES}_m = 0\), the test statistic for a
\(t\)-test is given by
\begin{align}t_m \equiv \frac{\hat{\text{ES}}_m}{\hat{Q}_m}.\end{align}
When the null is true, \(t_m\) follows a \(t\) distribution with mean
\(0\) and degrees of freedom \(\text{df}_m\), which depends on the
design and model.

For power calculations we calculate, given our assumptions on the design
and selected model, a reasonable value for \(Q_m\). We can then
calculate the power to detect an impact expressed in effect size units.

From the power formulas, we can also calculate MDES and sample size
requirements. From \cite{RN4473}, in general the MDES can be estimated
as

\[ MDES = MT_{df} \times \text{SE} / \text{VAR} \]

where \(MT_{df}\) is known as the multiplier and is the sum of two t
statistics based on degrees of freedom \(df\). For one-tailed tests,
\(MT_{df} = t_{\alpha} + t_{1-\beta}\) where \(\alpha\) is the type I
error rate and \(\beta\) is the desired power. For two-tailed tests,
\(MT_{df} = t_{\alpha/2} + t_{1-\beta}\). For more details, see
\cite[page 31]{RN4473} or \cite[page 22]{RN27978}. Manipulating this
expression then results in sample size formulae.

\paragraph{A note on effect sizes.}

In describing the standard error of our estimators in terms of effect
size, we need to carefully identify what we mean by an ``effect
size.'\,' We commonly think of an effect size as the size of an impact
relative to some reference amount of variation. If the reference amount
of variation is different, then the effect size, for the same absolute
effect, will also be different. This concern of what the denominator is
can create some tension regarding some of the power formula, as we will
note in the following sections.

In particular, the effect size formula can use either the variation in
\emph{overall} control group, or just the within-group variation only.
Overall variation includes the student variation within each site, but
also how the sites vary from each other. Within-group variation is just
this latter component. We argue that overall variation is more natural.
We also believe all the formula should use the same definition of effect
size.

Where this is most obviously a concern is with fixed effect regression.
In particular, with overall variation, if we increase the ICC at level 2
or level 3, then there is less variation (relative to the reference
variation) in level 1; thus an increased ICC will increase power for
fixed effect regression. This is simply the realized the gains of a
blocked experiment. If the effect size is calculated relative to
within-group variation, however, this gain is not seen. We will note how
this plays out explicitly in the following sections. An alternate
approach is to have the \(R^2\) measure include variance explained by
the fixed effects. To make the formula more directly comparable, we do
not take this route, but one can obtain the same results by selecting an
appropriate \(R^2\) and then setting the ICC to 0.

\newpage
\section{Scenarios}

\subsection{d1.1 designs: 1 level, randomization at level 1}

This is the classic individually randomized experiment where we allocate
some fraction of a single set of units to treatment.

The randomization scheme is simple random sampling:

\begin{verbatim}
T.x <- randomizr::simple_ra(N = nbar, prob = Tbar)
\end{verbatim}

\subsubsection{Constant effects (d1.1\_m1c)}

\textbf{PowerUp name:} Not applicable.

\textbf{Design:} 1-level design, randomization at level 1.

\textbf{Model:} constant intercepts, constant treatment effects, no
school or district covariates.

The model for estimating impacts on outcome \(m\) is given by:
\begin{align}
Y_{ijkm} &= \psi_{1,jkm} T_{ijk} + \theta_{0,jkm} + \sum_{p=1}^{g_{1,m}} \gamma_{mp} C_{ijkmp} + r_{ijkm}
\end{align}

The standard error of the treatment effect estimate is:
\begin{align} Q_m = \sqrt{\frac{(1-R^2_{1,m})}{\bar{T}(1 - \bar{T}) J \bar{n}}} . \end{align}

The degrees of freedom are:
\begin{align}\text{df}_m = J \bar{n} - g_{1,m} - 1.\end{align}

\paragraph{Sample size formula.}

The sample size formulas is: \begin{align}
\bar{n} &= \left(\frac{MT_{df}}{MDES}\right)^2 \left(\frac{1-R^2_{1,m}}{ \bar{T} (1 - \bar{T})} \right) .
\end{align}

\paragraph{Code syntax.}

The R model is

\begin{verbatim}
Yobs ~ 1 + T.x + C.ijk
\end{verbatim}

\newpage
\subsection{d2.1 designs: 2 levels, randomization at level 1}

This section of designs comprise what are usually referred to as
\emph{multisite experiments}. In a multisite experiment, we have a
collection of sites (here, schools) and are able to randomize the
individuals within each site into treatment and control. This allows for
estimating an average impact for each site, in principle. That being
said, we are usually interested in estimating some overall summary of
impacts across all our sites. These are also called blocked experiments,
especially if the sites are viewed as fixed.

Critically, there are four different estimands we might consider: the
average impact for persons vs.~impact for sites, and the average impact
of the sample we have vs.~the average impact of the population where the
sample came from. When sites are equal sized, a common assumption for
power calculations, the site and person average will be the same. We
therefore ignore it here. For finite vs.~super-population, we have to be
more careful. Some estimation strategies target a finite-population
estimand. In this document, the ones that do are d2.1\_m2fc and
d2.1\_m2ff. The d2.1\_m2fc estimation strategy does because it assumes a
constant treatment impact; given this assumption, there is no
uncertainty due to the sample itself as all samples have the same
average impact by assumption. The d2.1\_m2ff estimation strategy allows
each school to have an individually estimated impact, but due to using
fixed effects rather than random, it is evaluating the sample at hand.
See \cite{MiratrixWeiss2020} for further, in-depth, discussion.
Estimators that target the super-population need to take any uncertainty
of the sample being representative of the super-population into account.
Here, the one that does this is d2.1\_m2fr, with a model of each school
having an average impact drawn from some random distribution.

Regardless of the model used to analyze these data, the randomization
scheme is the same. It is simple random sampling within each school,
with proportion \(\bar{T}\) units assigned to treatment in each school.
In R, we could randomize this way as so:

\begin{verbatim}
T.x <- randomizr::block_ra( S.id, prob = Tbar )
\end{verbatim}

\subsubsection{Constant effects (d2.1\_m2fc)}

\textbf{PowerUp name:} bira2\_1c

\textbf{Design:} 2-level design, randomization at level 1 (blocked).

\textbf{Model:} fixed intercepts, constant treatment effect, no school
covariates.

When we assume constant effects, each school has its own fixed intercept
for the control outcome, and the treatment effect is modeled as constant
across schools. We can also call this a fixed effects, constant
treatment model \cite{MiratrixWeiss2020}. This model allows some
schools to have higher average outcomes than others (allowed for with
the fixed effects), but assumes the treatment impact is the same.

The model for estimating impacts on outcome \(m\) is given by:
\begin{align}
Y_{ijkm} &= \psi_{1,m} T_{ijk} + \theta_{0,jkm} + \sum_{p=1}^{g_{1,m}} \gamma_{mp} C_{ijkmp} + r_{ijkm}\\
\nonumber \theta_{0,jkm} &= \mu_{0,km} + u_{0,jkm}
\end{align} with reduced form: \begin{align}
Y_{ijkm} &= \psi_{1,m} T_{ijk} + \mu_{0,km} + \sum_{p=1}^{g_{1,m}} \gamma_{mp} C_{ijkmp} + u_{0,jkm} + r_{ijkm}
\end{align} and distributions: \begin{align}
r_{ijkm} &\sim N\left(0, \sigma^2_m\right).
\end{align}

The standard error formula we use is

\begin{align} Q_m = \sqrt{\frac{(1-\text{ICC}_{2,m})(1-R^2_{1,m})}{\bar{T}(1 - \bar{T}) J \bar{n}}} . \end{align}

See below for important details on this specific formula, and how it
differs from PowerUp!

The degrees of freedom for our impact estimate are
\begin{align}\text{df}_m = J \bar{n} - g_{1,m} - J - 1.\end{align}

\paragraph{Sample size formula.}

The sample size formulas are: \begin{align}
J &= \left(\frac{MT_{df}}{MDES}\right)^2 \left(\frac{(1-\text{ICC}_{2,m})(1-R^2_{1,m})}{\bar{n} \bar{T} (1 - \bar{T})} \right)\\
\bar{n} &= \left(\frac{MT_{df}}{MDES}\right)^2 \left(\frac{(1-\text{ICC}_{2,m})(1-R^2_{1,m})}{J \bar{T} (1 - \bar{T})} \right) .
\end{align}

The constant effects model means that we assume no treatment variation
across our sites, i.e.,:

\begin{itemize}
\item $\omega_{2,m} = 0$
\end{itemize}

\paragraph{Difference from PowerUP.}

The PowerUp formula in effect assumes

\begin{itemize}
\item $\text{ICC}_{2,m} = 0$,
\end{itemize}

although this can be motivated differently; see below.

\paragraph{Code syntax.}

The R model is

\begin{verbatim}
Yobs ~ 1 + T.x + C.ijk + S.id
\end{verbatim}

\paragraph{PowerUp! Differences.}

PowerUp assumes there is no \(\text{ICC}_{2,m}\) term while we allow for
it. This can be viewed as within (PowerUp!) vs.~overall (this work)
effect size metrics.

\paragraph{Remark on effect sizes.}

The standard error of the treatment effect estimate \emph{not} in effect
size units is (this taken from the PowerUp! documentation):
\begin{align}SE( \hat{\psi}_m ) = \sqrt{\frac{1}{\bar{T}(1 - \bar{T}) J \bar{n}}} \cdot \sigma_m .\end{align}

To convert this to an effect size, we need to scale by overall
variation. Unfortunately, under a fixed effect model, there is no
natural way to express this as we have not parameterized how the
individual site intercepts, the \(\delta_{0,jkm}\), vary. PowerUp!
therefore indexes by within group variation, which is \begin{align*}
Var( Y_{ijkm}(0) | S_{id} ) &= \frac{ \sigma^2_m }{ 1 - R^2_{1,m} } \end{align*}
using the formula for \(R^2_{1,m}\), capturing the predictive power of
our individual-level covariates on the outcomes within a given school,
of
\[ R^2_{1,m} = 1 - \frac{ \sigma^2_m }{ Var( Y_{ijkm}(0) |  S_{id} )} . \]
If we divide the above \(SE(\hat{\psi}_m)\) formula by
\(\sigma^2_m/(1 - R^2_{1,m})\) we get the reported standard error
formula for \(Q_m\) of \[ 
\tilde{Q}_m = \sqrt{\frac{1-R^2_{1,m}}{\bar{T}(1 - \bar{T}) J \bar{n}}} ,
\] with the tilde denoting that these effect size units are in terms of
within-school variation, which is not often done. Equivalently, this is
assuming the blocks are all homogeneous, which both goes counter to the
design principles of blocking and also is known to generally not hold
when evaluating schools. If we want the more classic effect size indexed
by cross-site variation, we need to go further.

Assume we have an \(\text{ICC}_{2,m}\), an assumed measure of how much
overall (control-side) variation is at the school level:
\[ \text{ICC}_{2,m} = 1 - \frac{ Var( Y_{ijkm}(0) |  S_{id} ) ) }{ Var( Y_{ijkm}(0)) } . \]
This ICC is even defined for a finite sample, if we view the above as
comparing the emperical (pooled) within-group variation to full
variation. Rearranging this gives
\(Var( Y_{ijkm}(0) ) = Var( Y_{ijkm}(0) | S_{id} ) /(1 - \text{ICC}_{2,m})\).

We can then plug this and the \(R^2_{1,m}\) formula together to get
\[ Var( Y_{ijkm}(0) ) = \frac{ \sigma^2_m }{ 1 - R^2_{1,m} } \cdot \frac{1}{1- \text{ICC}_{2,m}} .\]
If we use this expression to scale our SE formula, we finally obtain our
formula listed above.

\newpage 
\subsubsection{Fixed effects (d2.1\_m2ff)}

\textbf{PowerUp name:} bira2\_1f

\textbf{Design:} 2-level design, randomization at level 1 (blocked).

\textbf{Model:} fixed intercepts, fixed treatment effects, no school
covariates.

The constant effects model assumes treatment is the same for each block.
If it is not, and the blocks are different sizes or have different
proportions of units treated, the constant effects estimator is
precision-weighted and can thus be biased. Some may instead choose to
allow each school to have its own estimated impact, with a second
averaging step where we calculate an overall site-average of the site
specific impact estimates.

We do this by interacting our site fixed effects with treatment. Now
each school has its own fixed intercept for the control outcome, and
each school also has its own fixed coefficient for the treatment effect.
We can also call this a fixed effects with interactions model
\cite{MiratrixWeiss2020}.

In practice, the power calculations for this model will be the same as
for constant effects, unless we allow for block size variation or
variable proportion treated.

The model for estimating impacts on outcome \(m\) is given by:
\begin{align}
Y_{ijkm} &= \psi_{1,jkm} T_{ijk} + \theta_{0,jkm} + \sum_{p=1}^{g_{1,m}} \gamma_{mp} C_{ijkmp} + r_{ijkm}\\
\nonumber \theta_{0,jkm} &= \mu_{0,km} + u_{0,jkm}\\
\nonumber \psi_{1,jkm} &= \mu_{1,km} + u_{1,jkm}
\end{align} with reduced form: \begin{align}
Y_{ijkm} &= \left(\mu_{1,km} + u_{1,jkm}\right) T_{ijk} + \mu_{0,km} + \sum_{p=1}^{g_{1,m}} \gamma_{mp} C_{ijkmp} + u_{0,jkm} + r_{ijkm}
\end{align} and distributions: \begin{align}
r_{ijkm} &\sim N\left(0, \sigma^2_m\right).
\end{align} The standard error of the treatment effect estimate (and
therefore the sample size formula) are all the same as in the constant
effects model, i.e.~for SE we have:

\begin{align} Q_m = \sqrt{\frac{(1-\text{ICC}_{2,m})(1-R^2_{1,m})}{\bar{T}(1 - \bar{T}) J \bar{n}}} . \end{align}

However, the degrees of freedom are different due to the additional
interaction terms we need to estimate:
\begin{align}\text{df}_m = J \bar{n} - g_{1,m} - 2J.\end{align}

\paragraph{Sample size formula.}

The sample size formulas are: \begin{align}
J &= \left(\frac{MT_{df}}{MDES}\right)^2 \left(\frac{(1-\text{ICC}_{2,m})(1-R^2_{1,m})}{\bar{n} \bar{T} (1 - \bar{T})} \right)\\
\bar{n} &= \left(\frac{MT_{df}}{MDES}\right)^2 \left(\frac{(1-\text{ICC}_{2,m})(1-R^2_{1,m})}{J \bar{T} (1 - \bar{T})} \right) .
\end{align}

\paragraph{Difference from PowerUP.}

Note that the PowerUp formula assumes

\begin{itemize}
\item $\text{ICC}_{2,m} = 0$
\end{itemize}

\paragraph{Code syntax.}

The R model is

\begin{verbatim}
Yobs ~ 0 + T.x:S.id - T.x + C.ijk
\end{verbatim}

The overall treatment effect is then the average of the
\texttt{T.x:S.id} interaction terms.

\paragraph{PowerUp! Differences.}

Just as with the constant model, PowerUp assumes there is no
\(\text{ICC}_{2,m}\) term while we allow for it. This can be viewed as
within (PowerUp!) vs.~overall (this work) effect size metrics.

\newpage 
\subsubsection{Random effects (d2.1\_m2fr or d2.1\_m2rr)}

\textbf{PowerUp name:} bira2\_1r

\textbf{Design:} 2-level design, randomization at level 1 (blocked).

\textbf{Model:} random intercepts, random treatment effect, school
covariates for intercept. PowerUp! also includes interaction terms of
treatment and school covariates to allow for modeling treatment effect
heterogeneity; we do not include this.

If we are interested in generalizing from our sample to a
superpopulation, we may wish to view the sample of schools themselves as
representative of something larger. Then, if some schools have different
average impacts than other schools, we have to account for the
possibility that our sample of schools has an overall average impact
different from the target population. We can account for this additional
uncertainty with a random effects model that has a random effect for the
school-level average impacts.

The classic random effects model gives each school both a random
intercept for the control average outcome (the intercept), and a random
coefficient for the treatment effect. This is also known as the RIRC
model: random intercept, random coefficient. Recently, researchers also
use a variant of this model, the Fixed Intercept, Random Coefficient
(FIRC) model to account for concerns such as varying proportions of
units treated in different schools. For power calculations, they have
the same performance (they are also similar in practice; see
\cite{MiratrixWeiss2020}).

For RIRC, the model for estimating impacts on outcome \(m\) is given by:
\begin{align}
Y_{ijkm} &= \psi_{1,jkm} T_{ijk} + \theta_{0,jkm} + \sum_{p=1}^{g_{1,m}} \gamma_{mp} C_{ijkmp} + r_{ijkm}\\
\nonumber \theta_{0,jkm} &= \mu_{0,km} + \sum_{r=1}^{g_{2,m}} \delta_{mr} X_{jkmr} + u_{0,jkm}\\
\nonumber \psi_{1,jkm} &= \mu_{1,km} + u_{1,jkm}
\end{align} with reduced form: \begin{align}
Y_{ijkm} &= \left(\mu_{1,km} + u_{1,jkm}\right) T_{ijk} + \mu_{0,km} \\
\nonumber & + \sum_{r=1}^{g_{2,m}} \delta_{mr} X_{jkmr} + \sum_{p=1}^{g_{1,m}} \gamma_{mp} C_{ijkmp} + u_{0,jkm} + r_{ijkm}
\end{align} and random effect and residual distributions of:
\begin{align}
\begin{pmatrix} u_{0, jkm} \\ u_{1,jkm}\\ \end{pmatrix} &\sim
N\left(\begin{pmatrix} 0 \\ 0\\ \end{pmatrix}, \begin{pmatrix} \tau^2_{0,m} & \kappa^u_{mm} \tau_{0,m} \tau_{1,m} \\ \kappa^u_{mm} \tau_{1,m} \tau_{0,m} & \tau^2_{1,m} \\ \end{pmatrix}\right) \\
\nonumber r_{ijkm} &\sim N\left(0, \sigma^2_m\right).
\end{align}

For FIRC, we only have the random effects model on the \(u_{1,jkm}\),
and have fixed effects for the \(u_{0,jkm}\). For RIRC, we assume
bivariate Normal effects with variances \(\tau_{0,m}^2\) and
\(\tau_{1,m}^2\) and correlation \(\kappa_{mm}^u\). The correlation
structure \(\kappa_{mm}^u\) does not heavily impact the distribution of
the final test statistic.

We make an important note. In PowerUp!, they assume that school and
district covariates also influence the treatment impact:
\[ \psi_{1,jkm} = \mu_{1,km} + \sum_{r=1}^{g_{2,m}} \phi_{mr} X_{jkmr} u_{1,jkm}\]

but we do not make this assumption. The result of this is that we
assume, in their notation, that \(R_{2T}^2=0\), where \(R_{2T}^2\) is
the percent of treatment variation explained by level 2 covariates; we
are exploiting none of the cross-site impact heterogeneity. This
assumption affects the first term in the standard error formula below.

The standard error of the treatment effect estimate is given by:
\begin{align}Q_m = \sqrt{\frac{\text{ICC}_{2,m} \omega_{2,m}}{J} + \frac{(1-\text{ICC}_{2,m})(1-R^2_{1,m})}{\bar{T}(1 - \bar{T}) J \bar{n}}}.\end{align}

Note that this formula is simply the formula for d2.1\_m2fc with an
additional term of \(\text{ICC}_{2,m} \omega_{2,m} / J\). This term
captures the additional uncertainty from extrapolating from our sample
to the super-population. \(Q_m\) with this model, therefore, will be
larger than the prior models to the extent that the schools differ in
terms of their impact variation (the \(\text{ICC}_{2,m} \omega_{2,m}\)
term is simply the variation in the random impact terms scaled by our
overall variation).

The degrees of freedom are
\begin{align}\text{df}_m = J - g_{1,m} - 1.\end{align}

\paragraph{Sample size formula.}

The sample size formulas are: \begin{align}
J &= \left(\frac{MT_{df}}{MDES}\right)^2 \left(\text{ICC}_{2,m} \omega_{2,m} + \frac{(1-\text{ICC}_{2,m})(1-R^2_{1,m})}{\bar{T}(1 - \bar{T}) \bar{n}} \right)\\
\bar{n} &= \frac{(1-\text{ICC}_{2,m})(1-R^2_{1,m})}{\bar{T}(1 - \bar{T})\left(J \left(\frac{MT_{df}}{MDES}\right)^{-2} - \text{ICC}_{2,m} \omega_{2,m}\right)}
\end{align}

\paragraph{Code syntax.}

The R model is, for RIRC,

\begin{verbatim}
Yobs ~ 1 + T.x + X.jk + C.ijk + (1 + T.x | S.id)
\end{verbatim}

For FIRC it is

\begin{verbatim}
Yobs ~ 0 + T.x + X.jk + C.ijk + S.id + (0 + T.x | S.id)
\end{verbatim}

\paragraph{PowerUp! Differences.}

PowerUp allows for school covariates to influence the treatment impact,
while we do not allow for this. In PowerUp terms, we assume
\(R^2_{2T} = 0\).

\newpage 
\subsection{d2.2 designs: 2 levels, randomization at level 2}

These are commonly called cluster randomized experiments, with the
schools being the clusters.

The randomization scheme is a simple random sample of \(J\bar{T}\)
schools are assigned to treatment:

\begin{verbatim}
T.x <- randomizr::cluster_ra( S.id, prob = Tbar )
\end{verbatim}

\subsubsection{Random effects (d2.2\_m2rc)}

\textbf{PowerUp name:} cra2\_2r

\textbf{Design:} 2-level design, randomization at level 2 (clusters).

\textbf{Model:} random intercepts, constant treatment effect for all
schools, school covariates for intercept.

The model for estimating impacts on outcome \(m\) is given by:
\begin{align}
Y_{ijkm} &=  \theta_{0,jkm} + \sum_{p=1}^{g_{1,m}} \gamma_{mp} C_{ijkmp} + r_{ijkm}\\
\nonumber \theta_{0,jkm} &= \mu_{0,km} + \psi_{1,m} T_{jk} + \sum_{r=1}^{g_{2,m}} \delta_{mr} X_{jkmr} + u_{0,jkm}
\end{align} with reduced form: \begin{align}
Y_{ijkm} &= \psi_{1,m} T_{jk} + \mu_{0,km} + \sum_{r=1}^{g_{2,m}} \delta_{mr} X_{jkmr} + \sum_{p=1}^{g_{1,m}} \gamma_{mp} C_{ijkmp} + u_{0,jkm} + r_{ijkm}
\end{align} and distributions: \begin{align}
u_{0,jkm} &\sim N\left(0, \tau^2_{0,m}\right)\\
\nonumber r_{ijkm} &\sim N\left(0, \sigma^2_m\right).
\end{align}

The standard error of the treatment effect estimate is given by:
\begin{align}
Q_m = \sqrt{\frac{\text{ICC}_{2,m}(1 - R^2_{2,m})}{\bar{T}(1 - \bar{T}) J} + \frac{(1-\text{ICC}_{2,m})(1-R^2_{1,m})}{\bar{T}(1 - \bar{T}) J \bar{n}}}.\end{align}
The degrees of freedom are
\begin{align}\text{df}_m = J - g_{1,m} - 2.\end{align}

The constant effects model means that we assume no treatment variation
across our sites, i.e.:

\begin{itemize}
\item $\omega_{2,m} = 0$
\end{itemize}

\paragraph{Sample size formula.}

\begin{align}
J &= \left(\frac{MT_{df}}{MDES}\right)^2 \left(\frac{ \bar{n} \text{ICC}_{2,m}(1-R^2_{2,m}) + (1-\text{ICC}_{2,m})(1-R^2_{1,m})}{\bar{T}(1 - \bar{T}) \bar{n}} \right)\\
\bar{n} &= \frac{(1-\text{ICC}_{2,m})(1-R^2_{1,m})}{\bar{T}(1 - \bar{T}) J \left(\frac{MT_{df}}{MDES}\right)^{-2} - \text{ICC}_{2,m}(1-R^2_{2,m})  }
\end{align}

\paragraph{Code syntax.}

The R model is

\begin{verbatim}
Yobs ~ 1 + T.x + X.jk + C.ijk + (1 | S.id)
\end{verbatim}

\newpage 
\subsection{d3.1 designs: 3 levels, randomization at level 1}

In these designs we have schools nested in districts, and students
nested in schools. The only difference here, as compared to blocked
individual randomization with two levels, is the third level of
district. Since we are randomizing at the student level, this will only
impact how we think about where variation is in terms of our effect size
units.

In this context, if we are interested in the finite-sample impacts,
other than for calculating our reference variation for effect sizes, the
districts do not matter. We can simply use the prior two level fixed
effect designs if we lump district variation into the
\(\text{ICC}_{2,m}\) terms. In particular, one could use d2.1\_m2ff or
d2.1\_m2fc for the three level case by just entering
\(\text{ICC}_{2,m} + \text{ICC}_{3,m}\) in for \(\text{ICC}_{2,m}\). In
fact, we cannot have district random or fixed effects given school-level
fixed effects due to collinearity.

The randomization scheme is: simple random sampling occurs within each
school, with proportion \(\bar{T}\) units assigned to treatment in each
school.

\begin{verbatim}
T.x <- randomizr::block_ra( S.id, prob = Tbar )
\end{verbatim}

\subsubsection{Random effects (d3.1\_m3rr2rr)}

\textbf{PowerUp name:} cra3\_3r

\textbf{Design:} 3-level design, randomization at level 1 (blocked).

\textbf{Model:} random intercepts for district, random treatment effects
for district, random intercepts for school, random effects for schools,
school and district covariates for intercepts. Powerup also allows for
school and district covariates for cross-site impact heterogeneity.

The model for estimating impacts on outcome \(m\) is given by:
\begin{align}
Y_{ijkm} &= \psi_{1,jkm} T_{ijk} + \theta_{0,jkm} + \sum_{p=1}^{g_{1,m}} \gamma_{mp} C_{ijkmp} + r_{ijkm}\\
\nonumber \theta_{0,jkm} &= \mu_{0,km} + \sum_{r=1}^{g_{2,m}} \delta_{mr} X_{jkmr} + u_{0,jkm}\\
\nonumber \psi_{1,jkm} &= \mu_{1,km} + u_{1,jkm} \\
\nonumber \mu_{0,km}  &= \Xi_{0,m} + \sum_{s=1}^{g_{3,m}} \xi_{ms} V_{kms} + w_{0,km}\\
\nonumber \mu_{1,km}  &= \Xi_{1,m} + w_{1,km}
\end{align} with reduced form: \begin{align}
Y_{ijkm} &= \left(\Xi_{1,jkm} + w_{1,km} + u_{1,jkm}\right) T_{ijk} + \Xi_{0,km} \\
\nonumber & + \sum_{s=1}^{g_{3,m}} \xi_{ms} V_{kms} + \sum_{r=1}^{g_{2,m}} \delta_{mr} X_{jkmr} + \sum_{p=1}^{g_{1,m}} \gamma_{mp} C_{ijkmp}\\
 \nonumber &+ w_{0,km} + u_{0,jkm} + r_{ijkm}
\end{align} and distributions: \begin{align}
\begin{pmatrix} u_{0, jkm} \\ u_{1,jkm}\\ \end{pmatrix} &\sim
N\left(\begin{pmatrix} 0 \\ 0\\ \end{pmatrix}, \begin{pmatrix} \tau^2_{0,m} & \kappa^u_{mm} \tau_{0,m} \tau_{1,m} \\ \kappa^u_{mm} \tau_{1,m} \tau_{0,m} & \tau^2_{1,m} \\ \end{pmatrix}\right) \\
\nonumber \begin{pmatrix} w_{0, km} \\ w_{1,km}\\ \end{pmatrix} &\sim
N\left( \begin{pmatrix} 0 \\ 0\\ \end{pmatrix}, \begin{pmatrix} \eta^2_{0,m} & \kappa^w_{mm} \eta_{0,m} \eta_{1,m} \\ \kappa^w_{mm} \eta_{1,m} \eta_{0,m} & \eta^2_{1,m} \\ \end{pmatrix}\right) \\
\nonumber r_{ijkm} &\sim N\left(0, \sigma^2_m\right).
\end{align}

Similar to the two-level blocked model, in PowerUp! they further assume
that school and district covariates also influence the treatment impact
\begin{align*}
\psi_{1,jkm} &= \mu_{1,km} + \sum_{r=1}^{g_{2,m}} \phi_{mr} X_{jkmr} u_{1,jkm}\\
\mu_{1,jkm} &= \xi_{1,m} + \sum_{s=1}^{g_{3,m}} \zeta_{mr} V_{kms} w_{1,km}\\
\end{align*} but we do not make this assumption.

The standard error of the treatment effect estimate is given by:
\begin{align}Q_m = \sqrt{
\frac{\text{ICC}_{3,m} \omega_{3,m}}{K} +
\frac{\text{ICC}_{2,m} \omega_{2,m}}{JK} +
\frac{(1-\text{ICC}_{2,m} - \text{ICC}_{3,m})(1-R^2_{1,m})}{\bar{T}(1 - \bar{T}) JK\bar{n}}
}.\end{align} The degrees of freedom are
\begin{align}\text{df}_m = K - 1.\end{align}

This is a very conservative degrees of freedom.

\paragraph{Sample size formula.}

\begin{align}
K &= \left(\frac{MT_{df}}{MDES}\right)^2 \left(\text{ICC}_{3,m} \omega_{3,m} + \frac{\text{ICC}_{2,m} \omega_{2,m}}{J} + \frac{(1-\text{ICC}_{2,m}-\text{ICC}_{3,m})(1-R^2_{1,m})}{\bar{T}(1 - \bar{T}) J \bar{n}} \right)\\
J &= \frac{(1-\text{ICC}_{2,m}-\text{ICC}_{3,m})(1-R^2_{1,m}) + \bar{T}(1 - \bar{T}) \bar{n}\text{ICC}_{2,m} \omega_{2,m}}{\bar{T}(1 - \bar{T}) \bar{n}\left(K \left(\frac{MT_{df}}{MDES}\right)^{-2} - \text{ICC}_{3,m} \omega_{3,m}\right)}\\
\bar{n} &= \frac{(1-\text{ICC}_{2,m}-\text{ICC}_{3,m})(1-R^2_{1,m})}{\bar{T}(1 - \bar{T})\left(JK \left(\frac{MT_{df}}{MDES}\right)^{-2} - J\text{ICC}_{3,m} \omega_{3,m} - \text{ICC}_{2,m} \omega_{2,m}\right)}
\end{align}

\paragraph{Code syntax.}
\begin{verbatim}
Yobs ~ 1 + T.x + V.k + X.jk + C.ijk + (1 + T.x | S.id) + (1 + T.x | D.id) 
\end{verbatim}

\paragraph{PowerUp! Differences.}

PowerUp allows for school and district covariates to influence the
treatment impact, while we do not allow for this. In PowerUp terms, we
assume \(R^2_{3T} = 0\) and \(R^2_{2T} = 0\). This also impacts our
degrees of freedom formula, which is \(\text{df}_m = K - 1\) instead of
\(\text{df}_m = K - g_{3,m} - 1\).

\newpage 
\subsection{d3.2 designs: 3 levels, randomization at level 2}

These are commonly called blocked, cluster-randomized experiments. You
find these if, for example, schools are randomized within a set of
districts, or teachers are randomized within a set of schools (with
students as outcomes in both cases).

The randomization scheme is: simple random sampling occurs within each
district, with \(J \bar{T}\) schools assigned to treatment in each
district. In R we have:

\begin{verbatim}
T.x <- randomizr::block_and_cluster_ra( blocks = D.id, clusters = S.id, prob = Tbar )
\end{verbatim}

\subsubsection{Fixed effects (d3.2\_m3ff2rc)}

\textbf{PowerUp name:} bcra3\_2f

\textbf{Design:} 3-level design, randomization at level 2 (blocked
cluster).

\textbf{Model:} fixed intercepts for districts, fixed treatment effects
for districts, random intercepts for schools, constant effects for
schools within a district, school covariates for intercept.

The model for estimating impacts on outcome \(m\) is given by:
\begin{align}
Y_{ijkm} &=  \theta_{0,jkm} + \sum_{p=1}^{g_{1,m}} \gamma_{mp} C_{ijkmp} + r_{ijkm}\\
\nonumber \theta_{0,jkm} &= \mu_{0,km} + \psi_{1,jkm} T_{jk} + \sum_{r=1}^{g_{2,m}} \delta_{mr} X_{jkmr} + u_{0,jkm}\\
\nonumber \mu_{0,km}  &= \Xi_{0,m}  + w_{0,km}\\
\nonumber \psi_{1,km} &= \Xi_{1,m} + w_{1,km}
\end{align} with reduced form: \begin{align}
Y_{ijkm} &= \left(\Xi_{1,m} + w_{1,km} \right) T_{jk} + \Xi_{0,m} + \sum_{r=1}^{g_{2,m}} \delta_{mr} X_{jkmr} + \sum_{p=1}^{g_{1,m}} \gamma_{mp} C_{ijkmp}\\
\nonumber &+ w_{0,km} + u_{0,jkm} + r_{ijkm}
\end{align} and distributions: \begin{align}
u_{0,jkm} &\sim N\left(0, \tau^2_{0,m}\right)\\
\nonumber r_{ijkm} &\sim N\left(0, \sigma^2_m\right).
\end{align}

The standard error of the treatment effect estimate is given by:
\begin{align}
Q_m = \sqrt{
\frac{\text{ICC}_{2,m}(1 - R^2_{2,m})}{\bar{T}(1 - \bar{T}) JK} +
\frac{(1-\text{ICC}_{2,m} - \text{ICC}_{3,m})(1-R^2_{1,m})}{\bar{T}(1 - \bar{T}) J K\bar{n}} }.\end{align}
The degrees of freedom are
\begin{align}\text{df}_m = K( J - 2) - g_{2,m}.\end{align}

This model assumes: no variation of impacts within schools, and no
variation at the district level.

\begin{itemize}
\item $\omega_{2,m} = 0$
\end{itemize}

\paragraph{Difference from PowerUP.}

Note that the PowerUp formula assumes

\begin{itemize}
\item $\text{ICC}_{3,m} = 0$
\end{itemize}

\paragraph{Sample size formula.}

\begin{align}
K &= \left(\frac{MT_{df}}{MDES}\right)^2 \left( \frac{\text{ICC}_{2,m} (1-R_{2,m}^2)}{\bar{T}(1 - \bar{T}) J} + \frac{(1-\text{ICC}_{2,m} - \text{ICC}_{3,m})(1-R^2_{1,m})}{\bar{T}(1 - \bar{T}) J \bar{n}} \right)\\
J &= \frac{\bar{n}\text{ICC}_{2,m} (1-R_{2,m}^2) + (1-\text{ICC}_{2,m} - \text{ICC}_{3,m})(1-R^2_{1,m})}{\bar{n} \bar{T}(1 - \bar{T}) K \left(\frac{MT_{df}}{MDES}\right)^{-2} } \\
\bar{n} &= \frac{(1-\text{ICC}_{2,m} - \text{ICC}_{3,m})(1-R^2_{1,m})}{\bar{T}(1 - \bar{T})J K \left(\frac{MT_{df}}{MDES}\right)^{-2} -  \text{ICC}_{2,m} (1-R_{2,m}^2)}
\end{align}

\paragraph{Code syntax.}

The R model is

\begin{verbatim}
Yobs ~ 0 + T.x * D.id - T.x + X.jk + C.ijk + (1 | S.id)
\end{verbatim}

The overall treatment effect is then the average of the T.x interaction
terms.

\newpage
\subsubsection{Random effects (d3.2\_m3rr2rc)}

\textbf{PowerUp name:} bcra3\_2r

\textbf{Design:} 3-level design, randomization at level 2 (blocked
cluster).

\textbf{Model:} random intercepts for districts, random treatment effect
for districts, random intercepts for schools, constant effects for
schools within a district, school and district covariates for intercept.
Powerup also allows for district covariates for treatment effects.

The model for estimating impacts on outcome \(m\) is given by:
\begin{align}
Y_{ijkm} &=  \theta_{0,jkm} + \sum_{p=1}^{g_{1,m}} \gamma_{mp} C_{ijkmp} + r_{ijkm}\\
\nonumber \theta_{0,jkm} &= \mu_{0,km} + \sum_{r=1}^{g_{2,m}} \delta_{mr} X_{jkmr} + u_{0,jkm}\\
\nonumber \mu_{0,km}  &= \Xi_{0,m} + \psi_{1,km} T_{k} + \sum_{s=1}^{g_{3,m}} \xi_{ms} V_{kms} + w_{0,km} \\
\nonumber \psi_{1,jkm} &= \Xi_{1,m} + w_{1,km}
\end{align} with reduced form: \begin{align}
Y_{ijkm} &= \left(\Xi_{1,m} + w_{1,km}\right) T_{jk} + \Xi_{0,m}\\
\nonumber & + \sum_{s=1}^{g_{3,m}} \xi_{ms} V_{kms} + \sum_{r=1}^{g_{2,m}} \delta_{mr} X_{jkmr} + \sum_{p=1}^{g_{1,m}} \gamma_{mp} C_{ijkmp}\\
\nonumber &+ w_{0,km} + u_{0,jkm} + r_{ijkm}
\end{align} and distributions: \begin{align}
u_{0,jkm} &\sim N\left(0, \tau^2_{0,m}\right)\\
\nonumber \begin{pmatrix} w_{0, km} \\ w_{1,km}\\ \end{pmatrix} &\sim
N\left(\begin{pmatrix} 0 \\ 0\\ \end{pmatrix}, \begin{pmatrix} \eta^2_{0,m} & \kappa^w_{mm} \eta_{0,m} \eta_{1,m} \\ \kappa^w_{mm} \eta_{1,m} \eta_{0,m} & \eta^2_{1,m} \\ \end{pmatrix}\right) \\
\nonumber r_{ijkm} &\sim N\left(0, \sigma^2_m\right).
\end{align}

Similar to other blocked models model, in PowerUp! they further assume
that district covariates also influence the treatment impact
\begin{align*}
\mu_{1,jkm} &= \xi_{1,m} + \sum_{s=1}^{g_{3,m}} \zeta_{mr} V_{kms} w_{1,km}
\end{align*} but we do not make this assumption.

The standard error of the treatment effect estimate is given by:
\begin{align}
Q_m = \sqrt{
\frac{\text{ICC}_{3,m} \omega_{3,m}}{K} +
\frac{\text{ICC}_{2,m}(1 - R^2_{2,m})}{\bar{T}(1 - \bar{T}) J K } +
\frac{(1-\text{ICC}_{2,m} - \text{ICC}_{3,m})(1-R^2_{1,m})}{\bar{T}(1 - \bar{T}) J K\bar{n}} }.\end{align}
The degrees of freedom are \begin{align}\text{df}_m = K - 1.\end{align}

Parameter assumptions

\begin{itemize}
\item $\omega_{2,m} = 0$
\end{itemize}

\paragraph{PowerUp! Differences.}

PowerUp allows for district covariates to influence the treatment
impact, while we do not allow for this. In PowerUp terms, we assume
\(R^2_{3T} = 0\). This also impacts our degrees of freedom formula,
which is \(\text{df}_m = K - 1\) instead of
\(\text{df}_m = K - g_{3,m} - 1\).

\paragraph{Sample size formula.}

\begin{align}
K &= \left(\frac{MT_{df}}{MDES}\right)^2 \left( \text{ICC}_{3,m} \omega_3  +  \frac{\text{ICC}_{2,m} (1-R_{2,m}^2)}{\bar{T}(1 - \bar{T}) J} + \frac{(1-\text{ICC}_{2,m}-\text{ICC}_{3,m})(1-R^2_{1,m})}{\bar{T}(1 - \bar{T}) J \bar{n}} \right)\\
J &=  \frac{\bar{n}\text{ICC}_{2,m} (1-R_{2,m}^2) + (1-\text{ICC}_{2,m}-\text{ICC}_{3,m})(1-R^2_{1,m})}{\bar{n} \bar{T}(1 - \bar{T})\left( K \left(\frac{MT_{df}}{MDES}\right)^{-2} -  \text{ICC}_{3,m} \omega_3\right)}\\
\bar{n} &= \frac{(1-\text{ICC}_{2,m}-\text{ICC}_{3,m})(1-R^2_{1,m})}{\bar{T}(1 - \bar{T})J \left(K \left(\frac{MT_{df}}{MDES}\right)^{-2} -  \text{ICC}_{3,m}\omega_{3,m}\right) -  \text{ICC}_{2,m} (1-R_{2,m}^2)}
\end{align}

\paragraph{Code syntax.}

The R model is

\begin{verbatim}
Yobs ~ 1 + T.x + V.k + X.jk + C.ijk + (1 | S.id) + (1 + T.x | D.id)
\end{verbatim}

\newpage 
\subsection{d3.3 designs: 3 levels, randomization at level 3}

These designs have randomization at the top level. They are cluster
randomized, but we can model the nesting structure within cluster.

The randomization scheme is: simple random sampling occurs across
districts, with \(K\bar{T}\) districts assigned to treatment.

\begin{verbatim}
T.x <- randomizr::cluster_ra( D.id, prob = Tbar )
\end{verbatim}

\subsubsection{Random effects (d3.3\_m3rc2rc)}

\textbf{PowerUp name:} cra3\_3r

\textbf{Design:} 3-level design, randomization at level 3 (cluster).

\textbf{Model:} random intercepts for districts, constant treatment
effects for districts, random intercepts for schools, constant treatment
effects for schools, school and district covariates for intercept.

The model for estimating impacts on outcome \(m\) is given by:
\begin{align}\label{eqn:bi12c_model}
Y_{ijkm} &=  \theta_{0,jkm} + \sum_{p=1}^{g_{1,m}} \gamma_{mp} C_{ijkmp} + r_{ijkm}\\
\nonumber \theta_{0,jkm} &= \mu_{0,km} + \sum_{r=1}^{g_{2,m}} \delta_{mr} X_{jkmr} + u_{0,jkm}\\
\nonumber \mu_{0,km}  &= \Xi_{0,m} + \psi_{1,m} T_{k} + \sum_{s=1}^{g_{3,m}} \xi_{ms} V_{kms} + w_{0,km}
\end{align} with reduced form: \begin{align}
Y_{ijkm} &= \psi_{1,m} T_{k} + \Xi_{0,m} + \sum_{s=1}^{g_{3,m}} \xi_{ms} V_{kms} + \sum_{r=1}^{g_{2,m}} \delta_{mr} X_{jkmr} + \sum_{p=1}^{g_{1,m}} \gamma_{mp} C_{ijkmp}\\
\nonumber &+ w_{0,km} + u_{0,jkm} + r_{ijkm}
\end{align} and distributions: \begin{align}
u_{0,jkm} &\sim N\left(0, \tau^2_{0,m}\right)\\
\nonumber w_{0,jkm} &\sim N\left(0, \eta^2_{0,m}\right)\\
\nonumber r_{ijkm} &\sim N\left(0, \sigma^2_m\right).
\end{align}

The standard error of the treatment effect estimate is given by:
\begin{align}
Q_m = \sqrt{
\frac{\text{ICC}_{3,m}(1 - R^2_{3,m})}{\bar{T}(1 - \bar{T}) K} +
\frac{\text{ICC}_{2,m}(1 - R^2_{2,m})}{\bar{T}(1 - \bar{T}) J K } +
\frac{(1-\text{ICC}_{2,m} - \text{ICC}_{3,m})(1-R^2_{1,m})}{\bar{T}(1 - \bar{T}) J K\bar{n}} }.\end{align}
The degrees of freedom are
\begin{align}\text{df}_m = K - g_{3,m} - 2.\end{align}

The constant effects model means that we assume no treatment variation
across our sites, i.e.:

\begin{itemize}
\item $\omega_{2,m} = 0$
\item $\omega_{3,m} = 0$
\end{itemize}

\paragraph{Sample size formula.}

\begin{align}
K &= \left(\frac{MT_{df}}{MDES}\right)^2 \left( \frac{\text{ICC}_{3,m}(1-R_{3,m}^2)}{\bar{T}(1 - \bar{T})}  + \frac{\text{ICC}_{2,m} (1-R_{2,m}^2)}{\bar{T}(1 - \bar{T}) J} + \frac{(1-\text{ICC}_{2,m}-\text{ICC}_{3,m})(1-R^2_{1,m})}{\bar{T}(1 - \bar{T}) J \bar{n}} \right)\\
J&=   \frac{\bar{n}\text{ICC}_{2,m} (1-R_{2,m}^2) + (1-\text{ICC}_{2,m}-\text{ICC}_{3,m})(1-R^2_{1,m})}{\bar{n}\left(\bar{T}(1 - \bar{T}) K \left(\frac{MT_{df}}{MDES}\right)^{-2} - \text{ICC}_{3,m}(1-R_{3,m}^2)\right)} \\
\bar{n} &= \frac{(1-\text{ICC}_{2,m}-\text{ICC}_{3,m})(1-R^2_{1,m})}{\bar{T}(1 - \bar{T})JK \left(\frac{MT_{df}}{MDES}\right)^{-2} - J\text{ICC}_{3,m}(1-R_{3,m}^2) - \text{ICC}_{2,m} (1-R_{2,m}^2)}
\end{align}

\paragraph{Code syntax.}

The R model is

\begin{verbatim}
Yobs ~ 1 + T.x + V.k + X.jk + C.ijk + (1 | S.id) + (1 | D.id)
\end{verbatim}

\newpage
\section{The data generating process}

We now discuss the assumed data generating process (DGP), indexed by
parameters directly tied to the structural equations we use. The data
generating process is done in the following stages, outlined below.

\subsection{Determine DGP parameters}
\label{sec:dgp_param}

We have already discussed most of the required parameters in
Section\textasciitilde{}\ref{sec:app_intro}. However, there are a few
additional parameters required to generate data that do not directly
feed into our equations for single-outcome power or MDES, related to
correlations, show in Table\textasciitilde{}\ref{tab:corr_param}.

The parameters used in this section need to be picked based on desired
aggregate relationships of the full data. See the next section for how
to translate parameters such as ICC to the DGP parameters

\begin{table}[ht!]
\centering
\begin{tabularx}{0.8\textwidth}{p{1.5cm} | p{10cm}}
Param                                           & Description \\ \hline
$\boldsymbol{\rho}^V$                                   & Correlation matrix of district covariates $\boldsymbol{V}_{k\cdot}$ \\
$\boldsymbol{\rho}^{w_0}$                               & Correlation matrix of district random effects $\boldsymbol{w}_{0,k\cdot}$ \\
$\boldsymbol{\rho}^{w_1}$                               & Correlation matrix of district impacts $\boldsymbol{w}_{1,k\cdot}$\\
$\boldsymbol{\kappa}^{w}$                       & Non-symmetric matrix of correlations between district random effects and impacts, composed of entries $\{\kappa_{m,m^\prime}^{w}\} = Corr(w_{0,km}, w_{1,km^\prime}$) \\ \hline
$\boldsymbol{\rho}^X$                                   & Correlation matrix of school covariates $\boldsymbol{X}_{jk\cdot}$\\
$\boldsymbol{\rho}^{u_0}$                               & Correlation matrix of school random effects $\boldsymbol{u}_{0,jk\cdot}$\\
$\boldsymbol{\rho}^{u_1}$                               & Correlation matrix of school impacts $\boldsymbol{u}_{1,jk\cdot}$\\
$\boldsymbol{\kappa}^{u}$                       & Non-symmetric matrix of correlations between school random effects and impacts, composed of entries $\{\kappa_{m,m^\prime}^{u}\} = Corr(u_{0,jkm}, u_{1,jkm^\prime}$) \\ \hline
$\boldsymbol{\rho}^C$                                   & Correlation matrix of individual covariates $\boldsymbol{C}_{ijk\cdot}$\\
$\boldsymbol{\rho}^r$                                   & Correlation matrix of individual residuals $\boldsymbol{r}_{ijk\cdot}$
\end{tabularx}
\label{tab:corr_param}
\caption{Correlation parameters}
\end{table}

\subsection{Generate level 3 (district) data}

\subsubsection{Level 3 covariates}

Each outcome has its own district-level covariates, \(V_{km}\) with
\(k = 1, \ldots, K\) and \(m = 1, \ldots, M\). We have \(E(V_{km}) = 0\)
and \(Var(V_{km}) = 1\). We assume a correlation between covariates, so
we define \(\boldsymbol{\rho}^V\) is a \(M \times M\) symmetric
correlation matrix, with \(\rho^V_{ij}\) is the value in row \(i\) and
column \(j\) of the matrix \(\boldsymbol{\rho}^V\).

\[ \left(
\begin{array}{c}
V_{k1}  \\
\vdots  \\
V_{km}
\end{array}\right)
\sim
N\left[\left(
\begin{array}{c}
0       \\
\vdots  \\
0
\end{array}\right),\left(
\begin{array}{ccc}
1               & \cdots    & \rho^V_{1M}   \\
\vdots          & 1         & \vdots    \\
\rho^V_{M1}     & \cdots    & 1
\end{array}
\right)\right].\]

\subsubsection{Level 3 outcomes}
\label{sec:level3_outcomes}

Let \(\mu_{0,km}\) be the grand mean outcome under no treatment for
district \(k\), and \(\mu_{1,km}\) be the grand mean impact across
schools for district \(k\). \begin{align}
\mu_{0,km}  &= \Xi_{0,m} + \xi_{m} V_{km} + w_{0,km}  \\
\mu_{1,km}  &= \Xi_{1,m} + w_{1,km}
\end{align}

Let \(\Xi_{0,m}\) be the grand mean outcome under no treatment across
all districts. Without loss of generality, we will set \(\Xi_{0,m} = 0\)
for all \(m\). \(\Xi_{1,km}\) is the grand mean impact across districts.

We now consider the distributions of random effects and impacts
\(w_{0,km}\) and \(w_{1,km}\). First, we have \(E(w_{0,km}) = 0\),
\(Var(w_{0,km}) = \eta_{0,m}^2\), and correlation between outcomes
\(M \times M\) matrix \(\boldsymbol{\rho}^{w_0}\). \[ \left(
\begin{array}{c}
w_{0,k1}    \\
\vdots  \\
w_{0,kM}
\end{array}\right)
\sim
N\left[\left(
\begin{array}{c}
0       \\
\vdots  \\
0
\end{array}\right),\left(
\begin{array}{ccc}
\eta_{0,1}^2                        & \cdots        & \rho^{w_0}_{1M} \eta_{0,1} \eta_{0,M} \\
\vdots                              & \ddots        & \vdots    \\
\rho^{w_0}_{M1} \eta_{0,M} \eta_{0,1}   & \cdots        & \eta_{0,M}^2
\end{array}
\right)\right].\]

Similarly, we have \(E(w_{1,km}) = 0\),
\(Var(w_{1,km}) = \eta_{1,m}^2\), and correlation between outcomes
\(M \times M\) matrix \(\boldsymbol{\rho}^{w_1}\).

We now consider the joint distribution, \((w_{0,km}, w_{1,km})\) as
bivariate normal on the margin with correlation \(\kappa^{w}_{mm}\):
\begin{equation}
\left(
\begin{array}{c}
w_{0,km}\\
w_{1,km}
\end{array} \right)
\sim
N\left[\left(
\begin{array}{c}
0\\
0\end{array}
\right),\left(
\begin{array}{cc}
\eta_{0,m}^2                                    & \kappa^{w}_{mm} \eta_{0,m} \eta_{1,m} \\
\kappa^{w}_{mm}  \eta_{1,m} \eta_{0,m}              & \eta_{1,m}^2
\end{array}
\right)\right], \label{eq:pairwise_corr_level3}
\end{equation}

But we want these values to be correlated across outcome (i.e., district
average math test will be correlated with district average reading
test). We therefore generate the full set of district \(k\)'s random
effects across all outcomes as a \(2M\) vector of multivariate normal
residuals:
\[ (w_{0,k1}, \ldots, w_{0,kM}, w_{1,k1}, \ldots, w_{1,km} ) \sim MVNorm( \vec{0}, \Sigma_{full}^{w} ) \]

with \[ \Sigma_{full}^{u} =
\left(
\begin{array}{cc}
\Sigma_{w_0}        & \Sigma_{w} \\
\Sigma_{w}^\prime   & \Sigma_{w_1}
\end{array}
\right)
\]

and \[ \Sigma_{w_0} =
\left(
\begin{array}{ccc}
\eta_{0,1}^2                                & \cdots & \rho^{w_0}_{1M} \eta_{0,1} \eta_{0,M}  \\
\vdots                                      & \ddots & \vdots \\
\rho^{w_1}_{M1} \eta_{0,M} \tau_{0,1}       & \cdots & \eta_{0,M}^2
\end{array}
\right)
\]

\[ \Sigma_{w_1} =
\left(
\begin{array}{ccc}
\eta_{1,1}^2                            & \cdots & \rho^{w_1}_{1M} \eta_{1,1} \eta_{1,M}  \\
\vdots                                  & \ddots & \vdots \\
\rho^{w_1}_{M1} \eta_{1,M} \eta_{1,1}   & \cdots & \eta_{1,M}^2
\end{array}
\right)
\]

For the form of the off-diagonal blocks, \(\Sigma_{w}\), we first
construct \(M \times M\) matrix \(\boldsymbol{\kappa}^{w}\) with entries
\(\{\kappa^{w}_{mm}\}\). The diagonals of this matrix are
\(\kappa^{w}_{mm}\), which have been previously defined as the
correlation of the intercept and impact for outcome \(m\). The
off-diagonals are, for \(m \neq m^\prime\),
\[ \kappa^{w}_{m m^\prime} = cor( w_{0,km}, w_{1,km^\prime} ) .\]

We assume these are fixed across different values of \(k\), i.e.~that
the correlations are constant across districts.

For example, consider a \(3 \times 3\) case:
\[ \boldsymbol{\kappa}^{w} =
\left(
\begin{array}{ccc}
cor( w_{0,k1}, w_{1,k1} )   & cor( w_{0,k1}, w_{1,k2} ) & cor( w_{0,k1}, w_{1,k3} ) \\
cor( w_{0,k2}, w_{1,k1} )   & cor( w_{0,k2}, w_{1,k2} ) & cor( w_{0,k2}, w_{1,k3} ) \\
cor( w_{0,k3}, w_{1,k1} )   & cor( w_{0,k3}, w_{1,k2} ) & cor( w_{0,k3}, w_{1,k3} ) \\
\end{array}
\right)
\]

We note that this matrix does not necessarily have to be symmetric.

This gives \[ \Sigma_{w} =
\left(
\begin{array}{ccc}
\kappa^{w}_{11} \eta_{0,1} \eta_{1,1}   & \cdots & \kappa^{w}_{1M} \eta_{0,1}\eta_{1,M} \\
\vdots                                  & \ddots & \vdots                           \\
\kappa^{w}_{M1} \eta_{0,M} \eta_{1,1}   & \cdots & \kappa^{w}_{MM} \eta_{0,M}\eta_{1,M}
\end{array}
\right)
\] Note how the diagonals correspond to the off-diagonal in
Eq\textasciitilde{}\ref{eq:pairwise_corr_level3}.

\subsection{Generate level 2 (school) data}

\subsubsection{Level 2 covariates}

Each outcome has its own school-level covariate. For example, school
average reading and math pre-tests, used for adjusting reading and math
outcomes (in practice we might imagine adjusting each outcome with both,
but in the case of few clusters this might not be a good idea due to
degrees of freedom issues).

Index covariates as \(X_{jkm}\) with \(j = 1, \ldots, J\),
\(k = 1, \ldots, K\), and \(m = 1, \ldots, M\). As with the
district-level covariates, we have \(E(X_{jkm}) = 0\) and
\(Var(X_{jkm}) = 1\), and \(\boldsymbol{\rho}^X\) is a \(M \times M\)
symmetric correlation matrix.

\subsubsection{Level 2 outcomes}
\label{sec:level2_outcomes}

Each school \(j\) in district \(k\) has its average outcome under no
treatment \(\theta_{0,jkm}\) and its average impacts \(\psi_{1,jkm}\).
The mean outcome and average impact for school \(j\) in district \(k\)
for outcome \(m\) is \begin{align}
\theta_{0,jkm}  &= \mu_{0,km} + \delta_{m} X_{jkm} + u_{0,jkm}  \\
\psi_{1,jkm}        &= \mu_{1,km} + u_{1,jkm}
\end{align}

We can easily convert from three-level to two-level models. If there are
no districts, then \(\mu_{0,km} = \Xi_{0,m}\) and
\(\mu_{1,km} = \Xi_{1,m}\) for all \(k\). Essentially, we set
\(w_{km} = 0\), \(w_{km} = 0\), and \(\xi_m = 0\) for all \(k\).

The \((u_{0,jkm}, u_{1,jkm})\) follow a multivariate Normal structure as
in Section\textasciitilde{}\ref{sec:level3_outcomes}. We have
\(Var(u_{0,jkm}) = \tau^2_{0,m}\) and \(Var(u_{1,jkm}) = \tau^2_{1,m}\).
Also \(Cov(\boldsymbol{u}_{0,jk\cdot}) = \boldsymbol{\rho}^{u_0}\) and
\(Cov(\boldsymbol{u}_{1,jk\cdot}) = \boldsymbol{\rho}^{u_1}\). Finally,
they relate to each other with
\(Corr(u_{0,jkm}, u_{1,jkm^\prime}) = \kappa^{u}_{mm^\prime}\).

\subsection{Generate level 1 (individual) data}

\subsubsection{Level 1 covariates}

Individuals have individual level covariates, one for each outcome
\(C_{ijkm}\). For example, group-mean centered reading and math scores.
We assume these are homoskedastic and have the same mean across sites.
As with previous covariates, we have \(E(C_{ijkm}) = 0\) and
\(Var(C_{ijkm}) = 1\), and \(\boldsymbol{\rho}^C\) is a \(M \times M\)
symmetric correlation matrix.

\subsubsection{Level 1 outcomes}

For each outcome, the outcome model for the individual is \begin{align}
Y_{ijkm}(0) &= \theta_{0,jkm} + \gamma_m C_{ijkm} + r_{ijkm} \\
Y_{ijkm}(1) &= Y_{ijkm}(0) + \psi_{1,ijkm}
\end{align}

where \(Y_{ijkm}(0)\) is potential outcome \(m\) under no treatment for
individual \(i\) in school \(j\) in district \(k\), and \(\psi_{ijkm}\)
is the unit's individual causal effect.

We assume constant treatment effects for individuals in the same school,
\(\psi_{ijkm} = \psi_{1,jkm}\), but this assumption could be relaxed to
allow for individual treatment-level heterogeneity.

As with previous covariates, we have \(E(C_{ijkm}) = 0\) and
\(Var(C_{ijkm}) = 1\), and \(\boldsymbol{\rho}^C\) is a \(M \times M\)
symmetric correlation matrix.

Finally, individual-level residuals are distributed \(E(r_{ijkm}) = 0\)
and \(Var(r_{ijkm}) = 1\), and \(\boldsymbol{\rho}^r\) is a
\(M \times M\) symmetric correlation matrix.

\subsubsection{Reduced form}

Putting the levels together, we have: \begin{align}\label{eq:reduced}
Y_{ijkm}(0) &= \Xi_{0,m} + \xi_{m} V_{km} + \delta_{m} X_{jkm} + \gamma_m C_{ijkm} + w_{km} + u_{0,jkm} + r_{ijkm} \\
Y_{ijkm}(1) &= Y_{ijkm}(0) + \Xi_{1,m} + z_{km} + u_{1,jkm}
\end{align}

\subsection{Summary: Generating the full table of potential outcomes}

\begin{enumerate}
    \item For $k = 1, \ldots K$, and $m = 1, \ldots M$:
    \begin{enumerate}
        \item Generate district covariates $V_{km}$.
        \item Generate district residuals $w_{0,km}$ and $w_{1,km}$.
        \item Calculate district grand means $\mu_{0,km}$ and $\mu_{1,km}$.
    \end{enumerate}
    \item For $j = 1, \ldots J$, and $m = 1, \ldots M$:
    \begin{enumerate}
        \item Generate school covariates $X_{jkm}$.
        \item Generate school residuals $u_{0,jkm}$ and $u_{1,jkm}$.
        \item Calculate school grand means $\theta_{0,jkm}$ and $\psi_{1,jkm}$.
    \end{enumerate}
    \item For $i = 1, \ldots N$ and for $m = 1, \ldots M$:
    \begin{enumerate}
        \item Generate individual covariates, $C_{ijkm}$ 
        \item Generate individual residuals $r_{ijkm}$.
        \item Generate predicted baseline outcomes ($Y_{ijkm}(0)$ without residuals).
        \item Add residuals to the predicted outcomes to get $Y_{ijkm}(0)$ and calculate $Y_{ijkm}(1)$.
    \end{enumerate}
    
\end{enumerate}

\subsection{Generate observed data}

Once we have our full set of potential outcomes, we generate treatment
assignments to generate the observed outcomes. We generate our treatment
assignment, \(T_{ijk}\) for all \(i = 1, \ldots, n_j\) and
\(j = 1, \ldots, J\) and \(k = 1, \ldots, K\). Once we have our set of
\(T_{ijk}\) (no matter how they were obtained) we calculate the observed
outcomes \begin{equation}
Y_{ijkm}^{obs} = Y_{ijkm}(0) (1-T_{ijk}) + Y_{ijkm}(1) T_{ijk}
\end{equation}

\subsubsection{Randomization schemes}

We can assign at the district, school, or individual level depending on
the design we are generating data for.

\begin{itemize}
\item Blocked individual randomization: simple random sampling occurs within each school, with $\bar{n}\bar{T}$ units assigned to treatment in each school.
\item Cluster 2-level randomization: simple random sampling occurs across schools, with $J \bar{T}$ schools assigned to treatment.
\item Blocked cluster 2-level randomization: school level assignment occurs within each district, with $J\bar{T}$ schools assigned to treatment in each district.
\item Cluster 3-level randomization: simple random sampling occurs across districts, with $K \bar{T}$ districts assigned to treatment.
\end{itemize}

\section{Tuning the DGP parameters}
\label{sec:tune}

We define two main types of parameters. First, model parameters are
those defined in Tables\textasciitilde{}\ref{tab:latent_param} and
\ref{tab:corr_param}, and define the DGP. Second, control or derived
parameters, defined in Table\textasciitilde{}\ref{tab:derived_param},
indirectly tune model parameters. Control parameters are set by the
user, which then given the model parameters fed into the DGP. The
mapping of control parameters to model parameters is in
Table\textasciitilde{}\ref{tab:derived_param}.

We break our model parameters into sets:

\begin{itemize}
\item Set 1: $\{M, J, K, n_{jk}, \Xi_{0,m}, \boldsymbol{\rho}^D, \boldsymbol{\rho}^w, \boldsymbol{\rho}^z, \boldsymbol{\rho}^X, \boldsymbol{\rho}^u, \boldsymbol{\rho}^v, \boldsymbol{\rho}^C, \boldsymbol{\kappa}^{wz}, \boldsymbol{\kappa}^{uv}, p_j\}$ are set directly.
\item Set 2: $\{N, \mu_{0,m}, \mu_{1,m}, \theta_{0,jkm}, \psi_{1,jkm}, Y_{ijkm}(0), Y_{ijkm}(1) \}$ are functions of parameters that are set directly.
\item Set 3: $\{\Xi_{1,m}, \eta^2_{0,m}, \eta^2_{1,m}, \tau^2_{0,m}, \tau^2_{1,m}, \xi_m, \delta_m, \gamma_m\}$ are tuned through control parameters.
\end{itemize}

To translate from our control parameters to the model parameters we
derive several relationships in the following.

\subsection{Calculating the variation in random effects and impacts}

We have variation at the individual, school, and district level. We want
to be able to tune the proportion of variation in each of these levels.
We are interested in the unconditional (covariate-free) ICC.

We have for the variance of the control side: \begin{align*}
Var_m( Y_{ijkm}(0) ) &= \xi_m^2 Var_m(V_{km}) + \delta_m^2 Var_m(X_{jkm}) + \gamma_m^2 Var_m(C_{ijkm}) + \eta^2_{0,m} + \tau^2_{0,m} + \sigma^2_m \\
&=  \xi_m^2 +  \eta^2_{0,m} + \delta_m^2  + \tau^2_{0,m} + \gamma_m^2 + \sigma^2_m.
\end{align*}

Looking at Equation\textasciitilde{}\ref{eq:reduced}, we see:

\[ \text{ICC}_{3,m} = \frac{Var(\theta_{0,km})}{ Var(Y_{ijkm}(0))} = \frac{\xi^2_m + \eta_{0,m}^2}{\left(\xi_m^2 +  \eta^2_{0,m}\right) + \left(\delta_m^2  + \tau^2_{0,m}\right) + \left(\gamma_m^2 + \sigma^2_m\right)}.\]

\[ \text{ICC}_{2,m} = \frac{Var(\mu_{0,jkm})}{ Var(Y_{ijkm}(0))} = \frac{\delta_m^2  + \tau_{0,m}^2}{\left(\xi_m^2 +  \eta^2_{0,m}\right) + \left(\delta_m^2  + \tau^2_{0,m}\right) + \left(\gamma_m^2 + \sigma^2_m\right)} .\]

\subsection{Calculating the covariate coefficients}

\subsubsection{Calculating the level 3 covariate coefficient $\xi_m$}

The regression coefficients for the level 3 covariates, \(\xi_m\), is
dictated by the desired \(R^2\) values. Thus, we would like to find
\(\xi_m\) as a function of the level-3 \(R^2\). We define \(R^2_{3,m}\)
as the proportion of variance between level 3 districts predicted by
level 3 covariates.

We start with \begin{align*}
R_{3,m}^2
&= 1 - \frac{Var(w_{0,km})}{Var(\mu_{0,km})} \\
&= 1 - \frac{\eta^2_{0,m}}{\xi_m^2 Var(V_{km}) + \eta^2_{0,m}},
\end{align*} leading to \begin{align*}
\xi_m  &= \sqrt{\frac{\eta^2_{0,m}R_{3,m}^2}{Var(V_{km})(1 - R_{3,m}^2)}} \\
&= \sqrt{\frac{\eta^2_{0,m}R_{3,m}^2}{1 - R_{3,m}^2}}.
\end{align*}

\subsubsection{Calculating the level 2 covariate coefficient $\delta_m$}

We start with our level-2 \(R^2\) being defined as the proportion of
variance in level-2 schools explained by level-2 covariates:

\[
R_{2,m}^2 = 1 - \frac{Var(u_{0,jkm})}{Var(\theta_{0,jkm} \mid S_{id})}
\]

where the conditioning \(Var(\theta_{0,jkm} \mid S_{id})\) denotes the
variance of outcomes \emph{within} a particular district. This expands
to

\[
R_{2,m}^2 = 1 - \frac{\tau^2_{0,m}}{\delta_m^2Var(X_{jkm} \mid S_{id}) + \tau^2_{0,m}}.
\]

Since our \(X_{jkm}\) are generated independent of district, the
conditional variance is the same as overall. This gives

\[
R_{2,m}^2 = 1 - \frac{\tau^2_{0,m}}{\delta_m^2Var(X_{jkm}) + \tau^2_{0,m}},
\]

Leading to \begin{align*}
\delta_m  &= \sqrt{\frac{\tau^2_{0,m}R_{1,m}^2}{Var(X_{jkm})(1 - R_{2,m}^2)}} \\
&= \sqrt{\frac{\tau^2_{0,m}R_{2,m}^2}{1 - R_{2,m}^2}}.
\end{align*}

\subsubsection{Calculating the coefficient for the Level 1 variable ($\gamma_m$)}

Similar to level 2, we start with our level 1 \(R^2\) being defined as
the proportion of level 1 variance in individuals explained by level 1
covariates:
\[ R^2_{1,m} = 1 - \frac{ \sigma^2_m }{ var( Y_{ijkm}(0) \mid S_{id})},  \]
where the conditioning denotes the variance of outcomes \emph{within} a
particular school.

We find

\begin{align*}
R^2_{1,m} &= 1 - \frac{ \sigma^2_m }{ \gamma_m^2 var( C_{ijkm} \mid S_{id} ) + \sigma^2_m }\\
&= 1 - \frac{ \sigma^2_m }{ \gamma_m^2 var( C_{ijkm}) + \sigma^2_m }\\
\gamma_m &= \sqrt{\frac{\sigma^2_m R_{1,m}^2}{var( C_{ijkm})(1 - R_{1,m}^2)}}\\
&= \sqrt{\frac{R_{1,m}^2}{1 - R_{1,m}^2}}\\
\end{align*}

\subsection{Calculating the grand mean impacts $\Xi_{1,m}$}

This is a function of effect size. The effect size is simply the overall
impact in standard deviation units, with the standard deviation usually
being the marginal standard deviation of the control side:

\[ \Xi_{1,m} = ES_m \cdot SD_m( Y_{ijkm}(0) ) \]

where \(SD_m(Y_{ijkm}(0))\) denotes the standard deviation over \(i\),
\(j\), and \(k\) for fixed outcome \(m\). We have already noted
\(Var_m( Y_{ijkm}(0) ) = \xi_m^2 + \gamma_m^2 + \delta_m^2 + \eta^2_{0,m} + \tau^2_{0,m} + \sigma^2_m\).

\subsection{Final results}

We have produced a system of equations:

\begin{align*}
\text{ICC}_{3,m} &= \frac{\xi^2_m + \eta_{0,m}^2}{\xi_m^2 +  \eta^2_{0,m} + \delta_m^2  + \tau^2_{0,m} + \gamma_m^2 + 1}\\
\text{ICC}_{2,m} &= \frac{\delta_m^2  + \tau_{0,m}^2}{\xi_m^2 +  \eta^2_{0,m} + \delta_m^2  + \tau^2_{0,m} + \gamma_m^2 + 1}\\
\xi_m  &= \sqrt{\frac{\eta^2_{0,m}R_{3,m}^2}{1 - R_{3,m}^2}}\\
\delta_m &= \sqrt{\frac{\tau^2_{0,m}R_{2,m}^2}{1 - R_{2,m}^2}}\\
\gamma_m &= \sqrt{\frac{R_{1,m}^2}{1 - R_{1,m}^2}}
\end{align*}

We solve the system to find our model parameters: \begin{align*}
\gamma_m^2 &= \frac{R_{1,m}^2}{1 - R_{1,m}^2}\\
\delta_m^2 &= \frac{R_{2,m}^2}{1-R_{1,m}^2} \frac{\text{ICC}_{2,m}}{1 - \text{ICC}_{3,m}- \text{ICC}_{2,m}}\\
\xi_m^2 &= \frac{R_{3,m}^2}{1-R_{1,m}^2} \frac{\text{ICC}_{3,m}}{1 - \text{ICC}_{3,m}- \text{ICC}_{2,m}}\\
\tau^2_{0,m}  &= \frac{1-R_{2,m}^2}{1-R_{1,m}^2} \frac{\text{ICC}_{2,m}}{1 - \text{ICC}_{3,m}- \text{ICC}_{2,m}}\\ 
\eta^2_{0,m} &= \frac{1-R_{3,m}^2}{1-R_{1,m}^2}\frac{\text{ICC}_{3,m}}{1 - \text{ICC}_{3,m}- \text{ICC}_{2,m}}
\end{align*}

For details on the algebra, see Section\textasciitilde{}\ref{sec:alg}.

And finally we set: \begin{align*}
\eta^2_{1,m} &= \omega_{3,m} \left(\eta^2_{0,m} + \xi^2_m\right) \\
\tau^2_{1,m} &= \omega_{2,m} \left(\tau^2_{0,m} + \delta_m^2\right) \\
\end{align*}

\section{Appendix: Derivations of parameter formulae}
\label{sec:alg}

Let's start off with expressions we will later use:

\begin{align*}
\tau^2_{0,m}  &= \frac{\delta_m^2(1 - R_{2,m}^2)}{R_{2,m}^2}\\
\delta_m^2 + \tau^2_{0,m}  &= \delta_m^2 + \frac{\delta_m^2(1 - R_{2,m}^2)}{R_{2,m}^2}\\
&= \frac{\delta_m^2R_{2,m}^2 + \delta_m^2 - \delta_m^2R_{2,m}^2}{R_{2,m}^2}\\
\delta_m^2 + \tau^2_{0,m} &= \frac{\delta_m^2}{R_{2,m}^2}\\
\end{align*}

We also note:

\begin{align*}
\frac{\text{ICC}_{3,m}}{\text{ICC}_{2,m}} &= \frac{\xi^2_m + \eta_{0,m}^2}{\delta_m^2  + \tau_{0,m}^2}\\
\xi^2_m + \eta_{0,m}^2 &= \frac{\text{ICC}_{3,m}(\delta_m^2  + \tau_{0,m}^2)}{\text{ICC}_{2,m}}\\
&= \frac{\text{ICC}_{3,m}\delta_m^2}{R_{2,m}^2\text{ICC}_{2,m}}\\
\end{align*}

And finally it's easy to re-express \(\gamma_m^2 + 1\): \begin{align*}
\gamma_m &= \sqrt{\frac{R_{1,m}^2}{1 - R_{1,m}^2}}\\
\gamma_m^2 + 1 &= \frac{R_{1,m}^2}{1 - R_{1,m}^2} + 1\\
&= \frac{1}{1 - R_{1,m}^2}
\end{align*}

Let's start by plugging some of these into our expression for \(ICC_2\)
to find \(\delta_m\):

\begin{align*}
\text{ICC}_{2,m} &= \frac{\delta_m^2  + \tau_{0,m}^2}{\xi_m^2 +  \eta^2_{0,m} + \delta_m^2  + \tau^2_{0,m} + \gamma_m^2 + 1}\\
&= \frac{\frac{\delta_m^2}{R_{2,m}^2}}{\frac{\text{ICC}_{3,m}\delta_m^2}{R_{2,m}^2\text{ICC}_{2,m}} + \frac{\delta_m^2}{R_{2,m}^2} + \gamma_m^2 + 1}\\
\frac{\delta_m^2}{R_{2,m}^2} &= \text{ICC}_{2,m} \left(\frac{\text{ICC}_{3,m}\delta_m^2}{R_{2,m}^2\text{ICC}_{2,m}} + \frac{\delta_m^2}{R_{2,m}^2} + \gamma_m^2 + 1\right) \\
\delta_m^2 &=  \text{ICC}_{3,m}\delta_m^2 + \text{ICC}_{2,m} \delta_m^2 + \text{ICC}_{2,m}R_{2,m}^2(\gamma_m^2 + 1)\\
\delta_m^2 &= \frac{\text{ICC}_{2,m}R_{2,m}^2(\gamma_m^2 + 1)}{1 - \text{ICC}_{3,m}- \text{ICC}_{2,m}}\\
&= \frac{\text{ICC}_{2,m}R_{2,m}^2}{(1 - \text{ICC}_{3,m}- \text{ICC}_{2,m})(1-R_{1,m}^2)}
\end{align*}

Proceeding by a similar method, we can use \(ICC_3\) to find \(\xi_m\):

\begin{align*}
\xi_m^2 &= \frac{\text{ICC}_{3,m}R_{3,m}^2}{(1 - \text{ICC}_{3,m}- \text{ICC}_{2,m})(1-R_{1,m}^2)}
\end{align*}

Now we can plug in to find \(\tau^2_{0,m}\):

\begin{align*}
\tau^2_{0,m}  &= \frac{\delta_m^2(1 - R_{2,m}^2)}{R_{2,m}^2}\\
&= \frac{\text{ICC}_{2,m}R_{2,m}^2}{(1 - \text{ICC}_{3,m}- \text{ICC}_{2,m})(1-R_{1,m}^2)}\frac{(1 - R_{2,m}^2)}{R_{2,m}^2}\\
&= \frac{\text{ICC}_{2,m}(1-R_{2,m}^2)}{(1 - \text{ICC}_{3,m}- \text{ICC}_{2,m})(1-R_{1,m}^2)}
\end{align*}

And similarly:

\begin{align*}
\eta^2_{0,m} &= \frac{\text{ICC}_{3,m}(1-R_{3,m}^2)}{(1 - \text{ICC}_{3,m}- \text{ICC}_{2,m})(1-R_{1,m}^2)}
\end{align*}

\end{document}